%% file: main_arXiv.tex
\documentclass{article}
\input{preamble}

\title{Option Volume Imbalance as a predictor for equity market returns}

\begin{document}

\maketitle

\begin{abstract}
We investigate the use of the normalized imbalance between option volumes corresponding to positive and negative market views, as a predictor for directional price movements in the spot market. Via a nonlinear analysis, and using a decomposition of aggregated volumes into five distinct market participant classes, we find strong signs of predictability of excess market overnight returns. The strongest signals come from Market-Maker volumes. Among other findings, we demonstrate that most of the predictability stems from high-implied-volatility option contracts, and that the informational content of put option volumes is greater than that of call options.  
\end{abstract}

\textbf{Keywords}: 
Options; Implied Volatility; Informational Role of Option Volume; Market Prediction; Market Microstructure; Signal Analysis; Quantitative Trading Strategies; Options Applications

 {
   \hypersetup{linkcolor=blue}
   \tableofcontents
 }

\input{input_main_quant}


\FloatBarrier

\appendix
\input{input_app_ab}

\input{input_app_cde}

\end{document}

%% file: preamble.tex
\pdfoutput=1
\usepackage{caption}
\usepackage[list=true]{subcaption}

\usepackage{arxiv}

\usepackage[utf8]{inputenc} 
\usepackage{url}            
\usepackage{booktabs}       
\usepackage{amsfonts}       
\usepackage{nicefrac}       
\usepackage{microtype}      
\usepackage{lipsum}
\usepackage{amsmath}
\usepackage{bbm}
\usepackage[ruled,vlined]{algorithm2e}
\usepackage{float}
\usepackage{mathtools}  
\usepackage{amssymb}
\usepackage{tabulary}
\usepackage{amsthm}
\usepackage[titletoc]{appendix}
\usepackage[pdftex]{graphicx}

\usepackage[export]{adjustbox}
\usepackage{graphicx}  
\usepackage{chngcntr}
\usepackage{pdfpages}
\usepackage{lineno}

\usepackage{xcolor}
\usepackage{authblk}
\usepackage[colorlinks=true,linkcolor=red,filecolor=green,citecolor=blue]{hyperref}
\usepackage{cleveref}

\usepackage{authblk}
\usepackage{tcolorbox}

\usepackage{placeins}

\usepackage{array,xcolor,colortbl}
\usepackage{multirow}

\newcolumntype{G}{>{\centering\columncolor{gray!20!white}}p{0.2\textwidth}}
\newcolumntype{C}{>{\centering\arraybackslash}p{0.2\textwidth}}
\usepackage{verbatim}

\usepackage[utf8]{inputenc}
\usepackage{titletoc}
    \titlecontents{chapter}
    [0em] %
    {\medskip\bfseries}
    {\thecontentslabel\quad}
    {}
    {\hfill\contentspage}

    \titlecontents{section}[1.72em]{\smallskip}%
    {\thecontentslabel.\enspace}
    {}
    {\titlerule*[1.2pc]{.}\contentspage}%

\newcounter{noteMCctr} 

\author{Nikolas Michael$^{\dag \ast}$, Mihai Cucuringu$^{\dag \ddag \odot}$ and Sam Howison$^{\ddag}$} 

\affil{
$\dag$Department of Statistics, University of Oxford, 24-29 St Giles', Oxford OX1 3LB, UK. \\
        $\ddag$Mathematical Institute, University of Oxford, Andrew Wiles Building, Woodstock Rd, Oxford OX2 6GG. \\
        $\odot$ The Alan Turing Institute, 96 Euston Rd, London NW1 2DB. \\$\ast$Corresponding author.  Email: nikolas.michael@kellogg.ox.ac.uk 
        }

\date{January 2022}

%% file: input_main_quant.tex
\bibliographystyle{apalike}
\section{Introduction}

Since the opening of the CBOE in 1973, derivative markets have been a prominent presence in the financial system, and have grown rapidly ever since. According to BIS estimates~\cite{BIS}, the total global nominal amount outstanding for derivatives for the second half of 2020 amounts to USD 582 tn of which 7 tn are equity-linked Options\footnote{By comparison the global total GDP was estimated by the World Bank to be USD 88 tn~\cite{worldbank}.}. The latter is accounted for by multiple option exchanges across the world, including CBOE, the CME group, Nasdaq, and Eurex. Equity options are appealing to traders due to the higher leverage they offer, as well as the absence of restrictions such as the uptick rule. The limited downside risk of a long put position as opposed to a short futures position in the stock, and the hedging opportunities, are also attractive to investors.

Naturally, derivative markets are strongly intertwined with the spot markets. It has been argued that, due to informed trading, option markets lead the spot market (e.g \cite{black}). In particular, an interesting line of research is concerned with the information content of signed option volumes. A trader who (either from private information or by using predictive models) expects a spot-price rise can take a long call position or a short put position, while a trader who takes a short call position or a long put position may expect a price fall instead. Therefore, aggregating the signed volumes into these two categories of positive and negative signals yields a metric of the discrepancy in trader anticipation. Our main question is whether this discrepancy can act as a predictor for the direction of future price movements in the spot market.

To this end, we consider a decomposition of the trade flow into positive and negative ``signals'' (defined below), and use their normalized difference as a predictor for future stock returns. We refer to this metric as the \textit{Option Volume Imbalance (OVI)}. We employ two data sets, from the \textsc{NOM} and \textsc{PHLX} option exchanges, which dis-aggregate the daily option volume into 10-minute buckets, distinguishing between five different market participants and between call and put options. Although our analysis simply makes use of the aggregate volumes rather than, say, order book data, it still yields significant results.

Similar research usually uses either customer data or generally non-market maker data, without distinguishing between classes of market participants. In this paper, a special emphasis is placed into the decomposition between different classes of traders (or market participant classes), and especially that of Market Makers. While this may create an instance of double counting orders for both sides of a trade, it provides an insight on how the same transactions can be used from different angles, to provide trading signals of varying performances. Furthermore, this shows that on a purely speculative level, only certain classes can produce a positive signal under this analysis. If for example Customers are at least partially successful in their speculative ability, this should translate into a financial signal that produces positive returns. On the other hand, Market Makers who profit in other ways (e.g spreads) would not be expected to provide a positive signal, but rather a negative one, reflecting the speculative ability of all those who trade with them.

Furthermore, we empirically assess how the predictive power of OVI varies based on different values of the options' Greeks, and other characteristics. We compare the two different exchanges, and provide a qualitative description of certain intraday patterns we observe. Lastly, we investigate the cross-sectional effects of this metric, by modeling the interactions across the various assets using a directed network.

Throughout the paper, we use the cumulative Profit and Loss (P\&L) and Sharpe Ratio as measures of predictability in future returns. However, as we consider multiple types of OVI (e.g different market participants), we also wish to consider their total predictive power. This motivates us to formulate a novel regression approach which directly maximizes the cumulative P\&L, which we refer to as \textit{P\&L regression}. Unlike other research of this type, which usually tests the relationship between covariates and future returns by considering $t$ or $R^2$ values, our approach is not limited to linear relationships and can deal effectively with correlated features. We summarize our main contributions as follows. \\

\begin{tcolorbox}[top=1mm, bottom=1mm, left=1mm, right=1mm]
\paragraph{Summary of main contributions} 
The main contribution of our work is an analysis of the Option Volume Imbalance (OVI), and in particular its relationship with the future price of the underlying equity/ETF, as follows. (1) We provide evidence that the OVI metric exhibits predictive power on directional overnight price movements for the underlyings.  
(2) We show, by the analysis of the OVIs of different market participants, that they bring different information to the market; this segmentation is novel in the literature. 
%
(3) We analyze of the effect of various Option Features on the capacity of the OVI to predict future equity returns. (4) Finally, we provide evidence of cross-impact within the universe of underlying equity/ETF assets.
\end{tcolorbox}

As a secondary theme, we also consider a short high-level overview of the two markets, showing how most of the activity comes from Market Makers and Customers. We also provide an approximation of how option volume flows between the different market participants.

\paragraph{Paper outline} The remainder of the paper is organized as follows. In \Cref{section:OVI} we give an overview of related literature, and define the Option Volume Imbalance metric (OVI). In \Cref{section:datasets}, we describe the data sets used in our analysis. In \Cref{section:evaluation}, we discuss the evaluation methods used subsequently. In \Cref{section:MarketMakers}, using volumes from different market participants, we show the predictive power of the OVI for the future returns in the spot market. In \Cref{section:greeks}, we see the effects of various option features, such as the Greeks, on the predictive power of the OVI metric. In \Cref{section:intraday}, we extend our analysis by comparing the two data sets (from different exchanges), we consider results for longer time horizons as well as some intraday variants of the OVI-based strategies. In \Cref{section:networks}, we check for cross-impact effects, using a network model. Lastly, we summarize our conclusions and detail future research directions in \Cref{sec:conclusion}. Additional numerical results are deferred to the Supplementary Material. 

\section{Motivation and the Option Volume Imbalance}
\label{section:OVI}

\subsection{Related literature}


Traditionally, derivative markets have been used for hedging purposes, and have been shown to play a role in reducing the risk of portfolios. However, the high leverage offered by options has led to the suggestion that they play an additional role as an avenue for informed trading (\cite{black} and \cite{grossman}). A theoretical model was provided by \cite{ohara}, showing that under certain reasonable conditions, informed traders choose to capitalize on their private information by trading in both the stock and the options market. Under this model, informed trading is more prevalent when the liquidity and leverage of the option markets are sufficiently high. Furthermore, under \cite{Biais}'s model, the introduction of options reduces the informational efficiency of the market, making it harder for market makers to distinguish the information content of trades, as the option market becomes a more attractive avenue for informed traders.

The early empirical investigation \cite{mandle}, focused on the opportunity signals appearing whenever implied option prices from the Black-Scholes (BS) model differed significantly from observed prices, speculated that there was information in the option prices not reflected in the stock markets. This was extended by~\cite{Batcha} to intraday prices, with a pseudo-American BS model for the implied options, supporting and extending the results of \cite{mandle}. They used a number of different holding periods for equities, with most intraday strategies resulting in losses due to transactions costs, even though they do report profits in the overnight holding case.

Under the sequential information arrival model of \cite{copeland1976}, volumes can be used as a proxy for the rate of information arrival. Based on that, \cite{anthony} showed that call volumes lead stock volumes, which indicates the presence of informed traders in the options market. Challenging these results, \cite{stephan_whaley} showed empirically that the stock market leads the option market by at least 15 minutes, rather than vice versa. It was observed by \cite{ohara} that, while price changes in the stock market lead the option volumes, as articulated by \cite{stephan_whaley}, the results are quite different if one aggregates volumes into positive (call buys and put sells) and negative (call sells and put buys) flows. They rejected the hypothesis that these aggregated volumes hold no information about the spot market. This led to research based on option flow volume or option imbalance, i.e. features proportional to the difference between volumes corresponding to positive and negative flow.

One of the most important studies focusing on the decomposition to positive and negative flows, was \cite{poteshman2005}, who showed that the signed option volume, (which they referred to as \textit{demand for volatility}) was correlated with the realized volatility, and that market makers responded to imbalances by raising the spreads, in order to shield themselves from informed traders. According to the findings of \cite{SCHLAG200569}, signed option volumes have some predictive power of short-term returns, which however reverse at longer timescales, in contrast to the predictive power they found in futures. Later on, \cite{Pan_Poteshman} used put/call ratios from the opening of new buyer-initiated options, and showed their predictive power for future stock returns. They concluded that the source of predictability was primarily informed trading, by differentiating between publicly and non-publicly available data. Furthermore, they showed that deep out-of-the-money options had the highest predictive power, as well as data that came from customers who traded using full-service brokerage houses; in contrast, firm proprietary trades had no predictive power. 

Another option volume flow metric was defined by \cite{Hu2014}, who used delta-adjusted volumes and showed its significant predictive power for stock returns. The sign of the resulting coefficients did not reverse for longer horizons, as also observed earlier by \cite{Pan_Poteshman}. Furthermore, they found that the predictability of future returns is stronger for assets with a higher ratio of informed traders, low analyst coverage, large bid-ask spreads, small market capitalization, and low institutional ownership. 

Another promising metric proposed was the Option to Stock volume ratio (O/S), first constructed by \cite{ROLL20101}, who concluded that the cross-sectional and time-series variation of this metric reflected the presence of informed trading. The predictive power of this metric for future returns was empirically verified by \cite{JOHNSON2012262}, and was shown to be stronger when short-sale constraints were higher. 
\cite{GE2016601} compared the different signed option volumes by using the O/S ratio and concluded that there was no significant difference in predictive power between long and short positions, thereby speculating that the source of predictability was the embedded leverage common to the two positions, while the short-sale costs played no significant role. Later, \cite{Lin2015} decomposed the O/S metric into eight distinct ratios, differentiating between open and close, buy and sell, and calls and puts. Noticing how the informational content in both long and short positions was similar, they concluded that the most important channel of predictability was the embedded leverage in the options market.  

A more recent study was \cite{kalok}, regarding the interplay between returns and net volume (buyer-initiated minus seller-initiated) across stock, call and put markets. They found that it was the stock net volume that led the spot and option quote-price revisions, while the net option volumes had no effect. However, both net flows seemed to be negatively auto-correlated, perhaps indicating the effect of inventory control. 

The hypothesis that informed trading is the source of predictability deriving from the option market is strengthened by studies focusing on eventful dates. For example, \cite{Amin&Lee} found that around announcement days, both the profitability and the effective bid-ask spread of options increased. \cite{lin2013} provided conclusive evidence that, prior to any type of corporate analyst events, both the skew and spread of implied volatility become significant predictors, especially in liquid markets. It has been conjectured that this is due to analyst tipping (analysts give traders tips on upcoming information). For example,~\cite{cao} investigated the effect of option volumes prior to takeover and merger announcements. They found that the lagged call option imbalance (the difference between buyer and seller initiated volumes) is not a significant predictor for next day returns, but that this was not the case for pre-announcement periods. They also showed that short-term out-of-the-money options experienced the largest increase in trading volume during these periods.

Lastly, \cite{GoncalvesPinto2020WhyDO}, contrary to previous research, argued that the main driving force behind the predictive power of option prices stems from price pressure and not from informed trading, as the predictive power was not lowered when considering option contracts with zero volumes for a specific day.

\vspace{-2mm}
\subsection{Option Volume Imbalance}
\label{subsec:OVI}
Denote by $N_d$ the number of assets whose options were traded on financial day $d$, and $C_{i,d}$ denote the number of different option contracts corresponding to the $i^{th}$ asset. Given day $d$ and asset $i$, let $T_{i,j,d,m}$ and $V_{i,j,d,m}$ denote the number of trades and volume for option contract $j$ by investors of market participant class $m$. "Number of trades" refers to the total number of transactions (regardless of their size), while volume
refers to the total number of option contracts traded. Also let $P^{option}_{i,j,d}$ denote the average of opening and closing price of option contract $j$ for day $d$. 

As \cite{ohara} observed, while the option volumes do not reliably predict movements in the stock market, this changes when one considers the volumes of positive and negative signals separately. Positive signals include call buys and put sells, while negative signals include call sells and put buys. If a trader (with or without private information) predicts a positive movement in the price of the underlying, they can opt to enter into either or both of a long call position and a short put position. Conversely, if predicting a downward movement, they can take either or both of a long put position and a short call position\footnote{This excludes customers who trade in the options market to hedge an existing stock position, and customers who trade option straddles and other more complicated strategies. The latter poses no issue as the two types of volumes cancel each other out, while, unfortunately, we are unable to identify customers purchasing options for hedging purposes.}. Thus, we define 
\[ V^{\textrm{Up}, \mathcal{F}}_{i,j,d,m} = V^{\textrm{Call\_Buy}, \mathcal{F}}_{i,j,d,m} + V^{\textrm{Put\_Sell}, \mathcal{F}}_{i,j,d,m} \textrm{\ \ and \ \ } 
 V^{\textrm{Down}, \mathcal{F}}_{i,j,d,m} = V^{\textrm{Call\_Sell}, \mathcal{F}}_{i,j,d,m} + V^{\textrm{Put\_Buy}, \mathcal{F}}_{i,j,d,m}, \]
to be the volumes corresponding to speculation for upwards and downwards movement respectively. 
We similarly define 
\[ T^{\textrm{Up}, \mathcal{F}}_{i,j,d,m} = T^{\textrm{Call\_Buy}, \mathcal{F}}_{i,j,d,m} + T^{\textrm{Put\_Sell}, \mathcal{F}}_{i,j,d,m} \textrm{\ \ and \ \ } 
 T^{\textrm{Down}, \mathcal{F}}_{i,j,d,m} = T^{\textrm{Call\_Sell}, \mathcal{F}}_{i,j,d,m} + T^{\textrm{Put\_Buy}, \mathcal{F}}_{i,j,d,m}, \]
to be the number of trades corresponding to speculation for upwards and downwards movement respectively. Note that $\mathcal{F}$ will be used to encode different filters we apply to the transaction data before running different experiments (deliberately left general). 

Therefore, using the volume of Call-Buy/Sell and Put-Sell/Buy transactions as positive and negative directional signals respectively, and normalizing by the total volume, we define the Option Volume Imbalance (\textsc{OVI}) for asset $i$ and market participant class $m$ (for day $d$) as
\begin{equation} \label{eq:OVI}
\textrm{OVI}_{i,d,m}^{[\mathcal{F} ]} =  \frac{\sum_{j=1}^{C_{i,d}} \left( X^{\textrm{Up}, \mathcal{F}}_{i,j,d,m} -  X^{\textrm{Down},\mathcal{F}}_{i,j,d,m} \right)}{\sum_{j=1}^{C_{i,d}} \left( X^{\textrm{Up},\mathcal{F}}_{i,j,d,m} +  X^{\textrm{Down},\mathcal{F}}_{i,j,d,m} \right)},
\end{equation}  
where X denotes a generic flow variable \footnote{Note that for this definition we define $0/0 = 0$, i.e when the total volume is equal to $0$, the OVI is also equal to $0$.}. 
We will mostly be using the volume-based OVI, which is defined by setting $X = V$, i.e $X^{\textrm{Up}, \mathcal{F}}_{i,j,d,m} = V^{\textrm{Up}, \mathcal{F}}_{i,j,d,m}$ and $X^{\textrm{Down}, \mathcal{F}}_{i,j,d,m} = V^{\textrm{Down}, \mathcal{F}}_{i,j,d,m}$. 
The normalization ensures that OVI lies in the interval $\left[ -1,1 \right]$, so that the OVI of different assets can still be directly compared, independent of the asset's liquidity or market cap.

Note that each transaction is double-counted, from the perspective of both parties. For example, if a Customer buys a call option from a Market Maker, with the intention of creating a long position, it is counted both as a Customer-Buy and a Market Maker-Sell transaction. More details regarding the data set are given in \Cref{section:datasets}.   

One potential shortcoming of volume-based OVI, is that more weight could be given to option contracts with lower prices (e.g deep OTM options) as they are sometimes transacted in larger volumes. To ameliorate this concern, we can define alternative OVI features by using the total number of trades (which we refer to as \textit{trade-based} OVI) by setting $X$ = $T$, i.e  $X^{\textrm{Up}, \mathcal{F}}_{i,j,d,m} = T^{\textrm{Up}, \mathcal{F}}_{i,j,d,m}$ and $X^{\textrm{Down}, \mathcal{F}}_{i,j,d,m} = T^{\textrm{Down}, \mathcal{F}}_{i,j,d,m}$. We also define the \textit{nominal volume-based} OVI, by weighting the volume by the option price, i.e set $X^{\textrm{Up}, \mathcal{F}}_{i,j,d,m} = P^\textrm{Option}_{i,j,d} V^{\textrm{Up}, \mathcal{F}}_{i,j,d,m}$ and $X^{\textrm{Down}, \mathcal{F}}_{i,j,d,m} = P^\textrm{Option}_{i,j,d} V^{\textrm{Down}, \mathcal{F}}_{i,j,d,m}$. Nominal-volume based OVI explicitly gives more weight to higher cost trades, thus preventing low-cost options from having a disproportionate influence \footnote{Note that the choice of sign in \eqref{eq:OVI} is arbitrary; depending on the market participant, the OVI can be either positively or negatively correlated with future spot returns. The choice of signs, was made in accordance to the intuition of positive and negative news, as characterizations of Call-Buys,Put-Sells and Call-Sells,Put-Buys respectively.}.

Our OVI-based approach is similar in spirit to, but slightly different from the put-to-call volume ratio, that was used by \cite{Pan_Poteshman}; those authors restricted their analysis to Open-Buy transactions (see appendix for definition) and showed how the put-to-call ratio is a negative predictor for the price movement of the underlying.
Since our data set splits transactions based on Market Participant, we do not need to restrict to Open-Buy transactions. Instead, by using \eqref{eq:OVI}, all of the available types of transaction are used to construct a metric, for each Market Participant Class.

Importantly, unlike other similar studies in this area, we do not make a linearity assumption between OVI and future returns. In contrast, our analysis uses the sign of the OVI as a directional predictor for subsequent price movements, by making a hypothetical transaction in the corresponding direction. If a significant profit (or loss) is made from this transaction, then we conclude that the OVI contains significant information about the future spot market returns.
 
\FloatBarrier

\vspace{-2mm}
\section{NASDAQ data sets}
\label{section:datasets}

\paragraph{Data sets Description}
We consider two data sets, from the Nasdaq Option Market (NOTO) and Nasdaq PHLX (PHOTO) exchanges. The data sets provide information about the aggregate option volumes for the period 02 Jan 2015--31 Dec 2019. For each day in this range, we are given a time series of intraday updates from the exchange, disseminated at 10-minute intervals, detailing the total cumulative volume of options transacted up to that intraday time point. As the market opens at 9:30 ET, the first report is disseminated at 9:40 ET, and the last report at 16:00 ET (a total of 39 intraday time points). We are also given a daily summary report containing the total daily option volume for each contract. In the daily report, we are also given its low, high, opening and closing price, its open interest, and total electronic volume, as well as the total volume for the specific exchange. Aggregate volumes are further broken down into five different market participant classes (hereafter MPCs) which are: Firm (Proprietary Trades), Brokers, Market Makers, Customers, and Professional Customers. Definitions for these are given in \Cref{appendix:definitions}. 

Furthermore, for each MPC other than Market Makers, volumes are further decomposed into four parts based on the type of transaction and intention of the agent: Opening Buy, Opening Sell, Closing Buy and Closing Sell (definitions are given in \Cref{appendix:definitions}) . In \Cref{table:notation}, we give a summary of the main notation used throughout the paper, some of which is defined later on.

\begin{table}[h]
\begin{center} 
\begin{tabular}{r c p{14cm}}
\toprule
$D$ & : & Number of days covered by the data set.\\
$N_d$ & : & Number of assets, whose option contracts were traded on  day $d$.\\
$C_{i,d}$  & : & Number of option contracts traded for asset $i$ on day $d$. \\
$I_{i,d}$ & : & Number of option contracts corresponding to asset $i$ traded on day $d$.\\
$P_{i,d}^{\textrm{open}}$ & : & Opening price of underlying asset $i$, on day $d$. \\  
$P_{i,d}^{\textrm{close}}$ & : & Closing price of underlying asset $i$, on day $d$. \\ 
$P_{i,j,d}^{\textrm{option}}$ & : & Average of opening and closing price of option contract $j$ for underlying asset $i$, on day $d$. \\ 
$\textrm{OVI}_{i,d,m}^{[\mathcal{F} ]}$ & : &  OVI of MPC $m$ for asset $i$ on day $d$. \\
$f_{i,d}$ & : & Raw return of asset $i$ on day $d$ (overnight returns, unless otherwise specified). \\
$\tilde{f}_{i,d}$ & : &  Excess Market Return, of asset $i$ on day $d$ (overnight returns, unless otherwise specified). \\
$s_{i,d}$ & : &  Signal observed on day $d$ for asset $i$'s price movement. \\
$P\&L_d$ & : & Profit and Loss for portfolio at day $d$ \\
$\textrm{SR}$ & : & Sharpe Ratio performance metric of the portfolio for the time span of $D$ days \\
$\textrm{PPD}$ & : & Profit Per Dollar performance metric of the portfolio for the time span of $D$ days \\
$V_{i,j,d,m}$ & : & Option volume (total number of contracts traded) of $j^{th}$ option contract of the $i^{th}$ asset on day $d$, by investors of MPC $m$ \\
$T_{i,j,d,m}$ & : & Option trades (total number of transactions) of $j^{th}$ option contract of the $i^{th}$ asset on day $d$, by investors of MPC $m$ \\
$g$ & : & activation function for P\&L regression \\
$L$ & : &  Objective function for P\&L regression. \\
$h$ & : &  Smooth proxy for the absolute value (used for P\%L regression).\\
$l$ & : &  Window Length for sliding window.\\
\bottomrule
\end{tabular}
\end{center}
\caption{Main Notation Used}
\label{table:notation}
\end{table}

\begin{figure}[h]
   \includegraphics[width=1\textwidth]{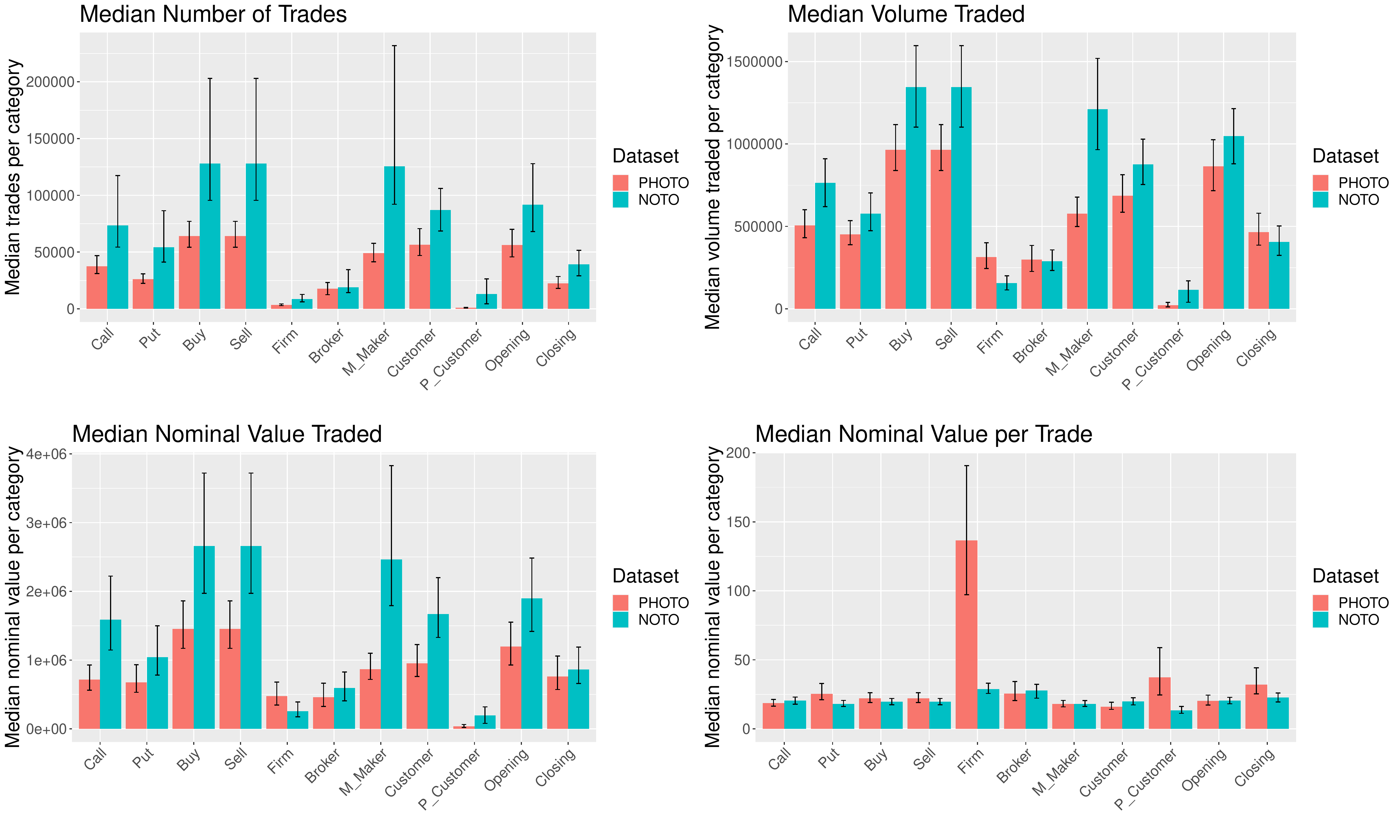}
   \caption{Breakdown of the two data sets into their various constituents: The plots show the median daily number of trades (top-left), median daily volume traded (top-right) and median nominal value traded (bottom-left) and the average \textit{nominal size per trade} (bottom-right) which is the ratio between the values of the first and third quantities. The whiskers indicate the Inter-quartile Range across days for each constituent. 
   }
   \label{fig:breakdown}
\end{figure} 

\subsection{Descriptive Analysis of the Market.}

In the rest of this Section, we show some preliminary analysis for the two data sets, including an analysis of the interaction between different MPCs. The purpose of this section is twofold; first the analysis in itself is novel and some interesting observations are noted. Second, these observations will help us interpret the results of the subsequent sections.

In \Cref{fig:breakdown}, we show summary statistics (on a daily basis) for the trades in the two data sets, disaggregated by trade type and MPC as defined in \cref{appendix:definitions}. Although NOTO has a higher average trade count across MPCs, the aggregates of the two data sets are qualitatively similar. In particular, for both data sets call options have a higher median volume than put options and the most prevalent MPCs are Market Makers and Customers. Firms and Professional Customers have a very small median volume. However, when it comes to nominal value per trade for the PHOTO data set, Firms have by far the highest ratio compared with other MPCs. Overall, we expect the OVI of Market Makers and Customers to be the most informative, as the OVI of the other MPCs is constructed from fewer transactions and thus more likely to be noisy (this is confirmed in \Cref{section:MarketMakers}, where we observe higher performance for Market Maker and Customers OVI).  
\begin{figure}[h]
\centering
\includegraphics[width=1\textwidth]{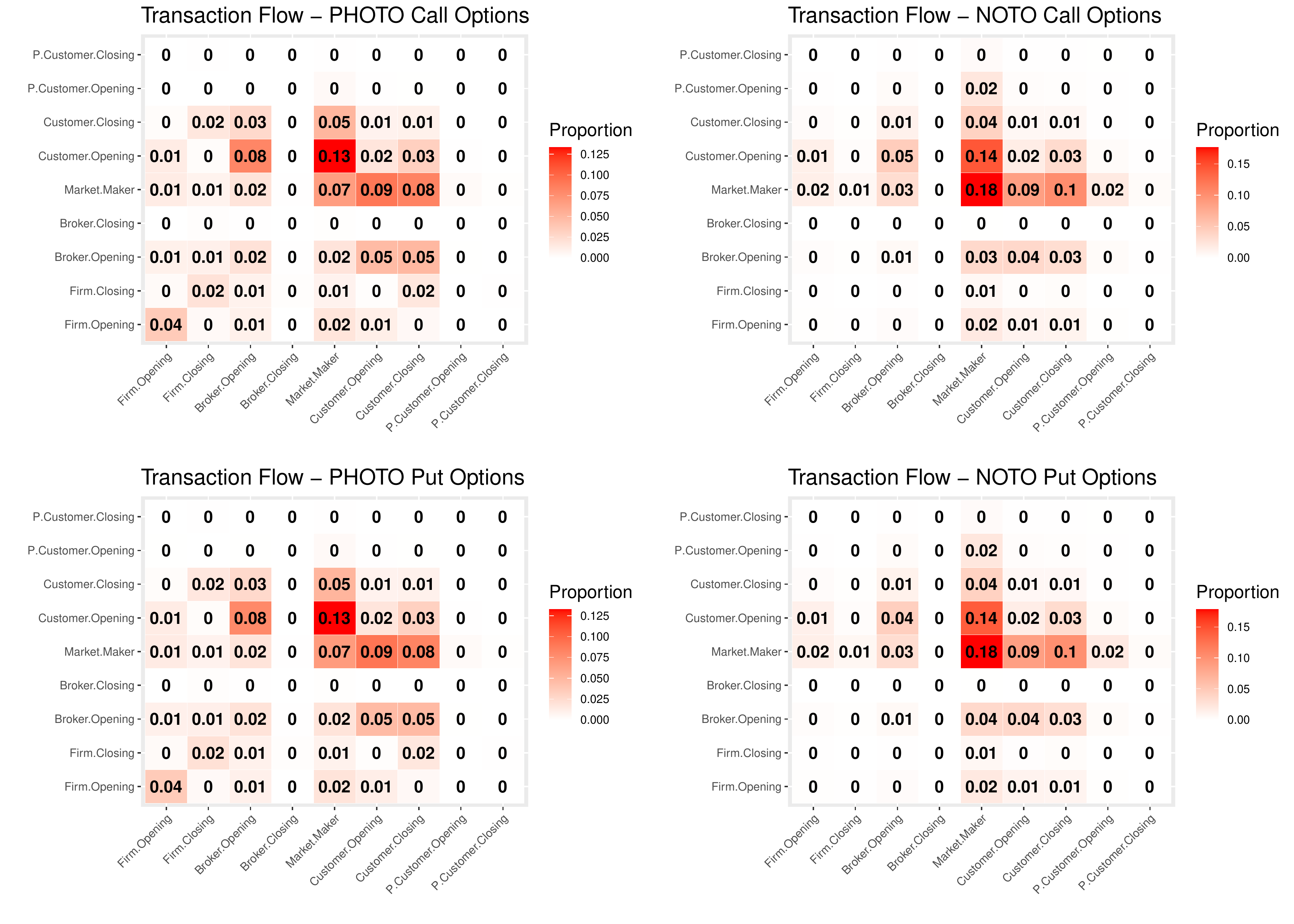}
\caption{An approximation of the volume flow between MPCs (see~\eqref{flow_fraction} for the approximation). The rows/columns correspond to the Buyer/Seller side of transactions respectively. Cell values show the median of the approximated flow proportion between the two participants (to 2 dp), with red highlighting indicating high proportions.}
\label{fig:flow_breakdown}
\end{figure} 

To complement the overview of \Cref{fig:breakdown}, we also show the flow between different MPCs, i.e how much they trade with each other. Denote by $\widetilde{\textrm{NF}}_{m_1,m_2}$, the average trade flow from MPC $m_1$ (as a seller) to MPC $m_2$ (as a buyer), which represents the median percentage of trades that correspond to these MPCs. The full transaction details are not available, thus we can approximate it by inspecting the 10-minute intraday aggregate volume reports, and then interpreting and matching the flow between different MPCs. For details see \Cref{appendix:tradeflow}. The heatmap of the median flow for the two data sets is shown in \Cref{fig:flow_breakdown}, differentiating between call and put options. Once again, the two data sets display many similarities in their flow structure. Firstly, in both cases the call and put flow structures are very similar. Second, whenever a flow from MPC $m_1$ to $m_2$ ($\widetilde{\textrm{NF}}_{m_1,m_2}$) is relatively high, the reverse flow ($\widetilde{\textrm{NF}}_{m_2,m_1}$) also tends to be relatively high. In addition, a very large proportion of the flow includes either an (Ordinary) Customer and/or a Market Maker in the transaction. In both cases, the flow between Brokers and Customers is much higher than between Brokers and Market Makers, but in the case of PHOTO the difference is much higher. 

The proportion of trades involving the Firms MPC is much higher for PHOTO in comparison to NOTO. The opposite is true for Market Makers, where higher proportion for NOTO, and sharing a significant proportion of the flow with Professional Customers Opening.

In both data sets, Firms and Professional Customers trade mostly with Market Makers, while for Brokers their flow with Customers is larger than for Market Makers. 
\begin{figure}[h]
\centering
\includegraphics[width=1.15\textwidth,center,trim=0cm 4cm 0cm 4cm,clip]{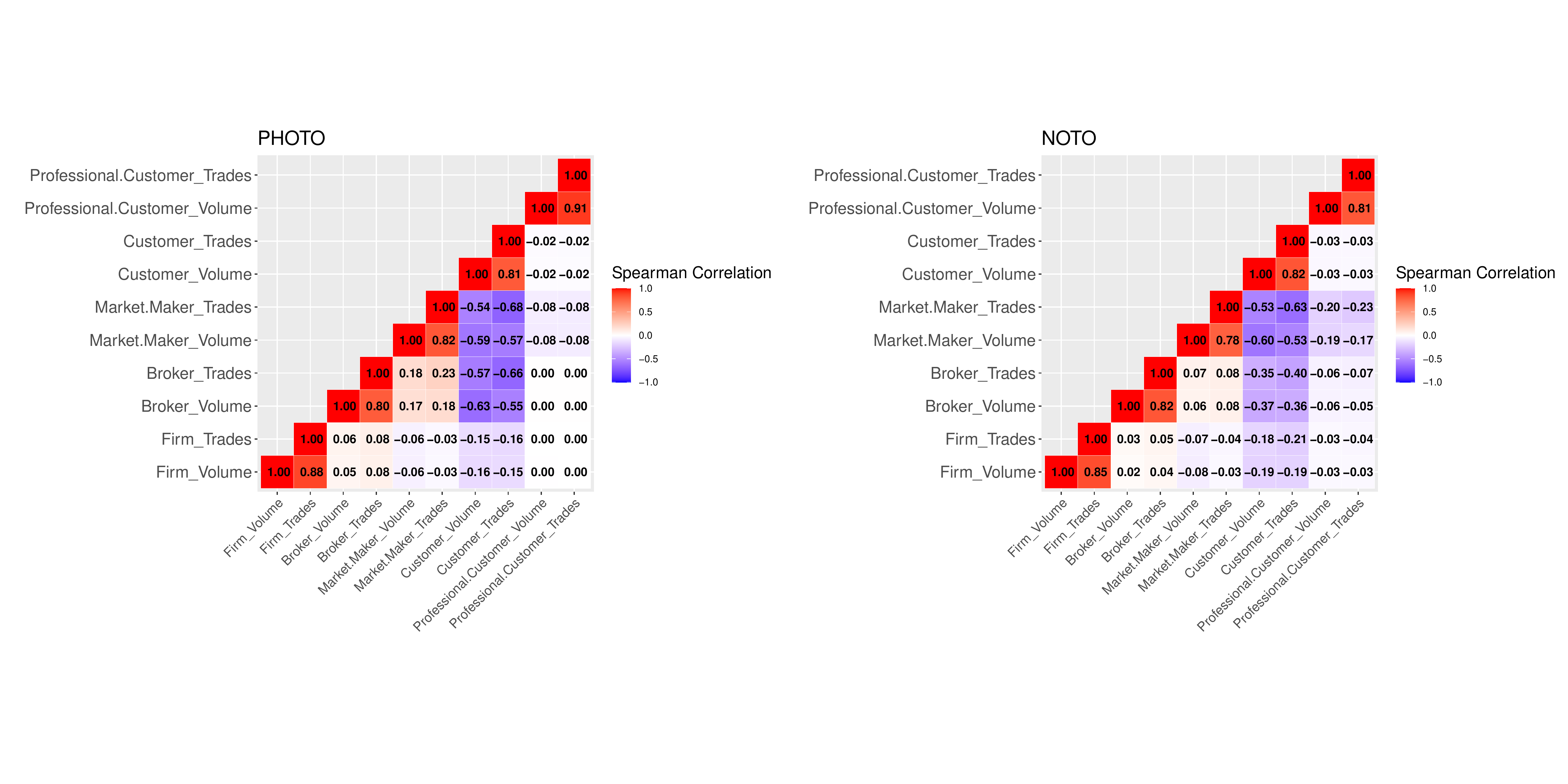}
\caption{OVI Correlation Map: Plot of the median (across days) Spearman correlation between the OVI (trades and volume based) of different MPCs.}
\label{fig:imb_cor}
\end{figure}

\paragraph{Correlation of Volume and OVI} 
The correlation between PHOTO and NOTO option volume ($\sum_m \sum_j V^{\textrm{PHOTO}}_{i,j,d,m}$ and $\sum_m \sum_j V^{\textrm{NOTO}}_{i,j,d,m}$) ranges between 0.00 to 0.93, for different assets. The correlation between the total volume for different MPCs ($\sum_i \sum_j V_{i,j,d,m}$ for different $m$) are also naturally quite high, ranging between $0.8-1.0$. More interestingly, in \Cref{fig:imb_cor} we plot the median (across days) correlations between the OVI features built using the data from the five MPCs \footnote{Note that the correlation map is very similar across days, with each entry having a standard deviation of 0.05 or less, aross days. Also note that for this analysis the Spearman correlation was chosen for its simplicity and ability to identify nonlinear relationships (the Pearson correlations were, however, very similar).}. Unsurprisingly (as trades are double-counted), there is a strong correlation between the volume- and trade-based OVI when they correspond to the same MPC. More importantly, for both data sets (Ordinary) Customers have a high negative correlation, of around $-$0.6, with Market Makers and Brokers, and a low negative correlation, of around $-0.20$ with Firms. We also observe a weak correlation, of around $0.2$, between Market Makers and Brokers for the PHOTO data set, as well as a weak correlation, of around $-0.2$ between Market Makers and Professional Customers for the NOTO data set. The rest of the correlations are below 0.1 in magnitude. Therefore, we should keep in mind the strong multicollinearity present in the OVI features (which later on brings up the need for a regularization term in the P\&L regression \Cref{sec:pnl}).


\paragraph{Principal Component Analysis}
\begin{figure}[h]
\centering
\includegraphics[width=0.9\textwidth,center]{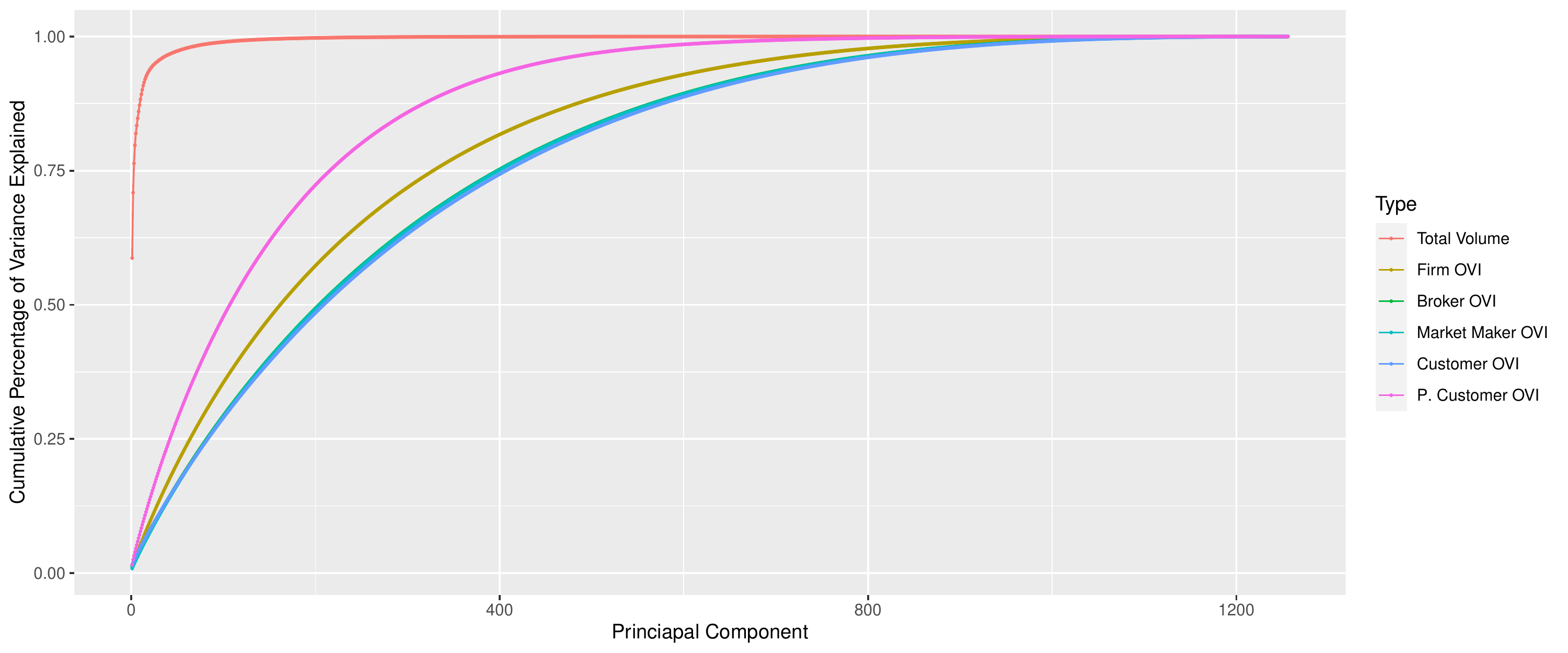}
\caption{Volume and OVI Principal Components: This plot shows the cumulative percentage of variance captured by principal components for Volume and the different OVI of each MPC.}
\label{fig:pca_cumsum}
\end{figure}

\begin{figure}[h]
\centering
\includegraphics[width=0.9\textwidth,center]{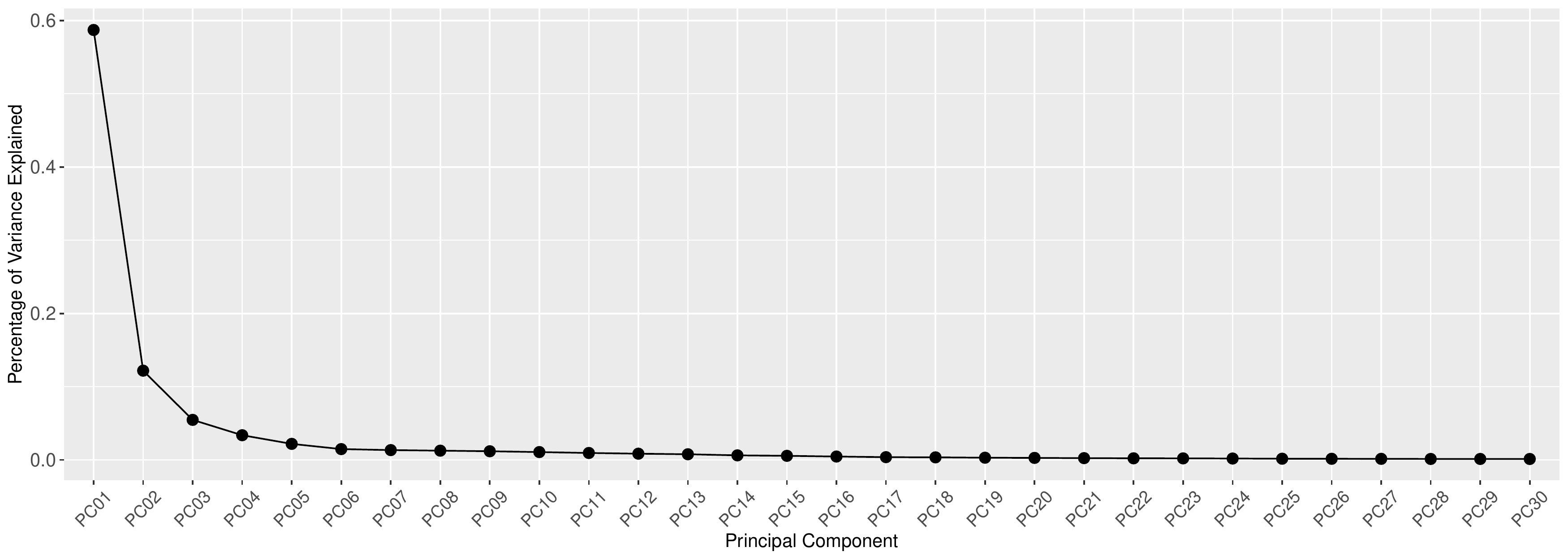}
\caption{Volume Scree Plot: This plot shows the percentage of variance captured by the first few principal components for total option volume (for the PHOTO data set).}
\label{fig:screeplot}
\end{figure}
We carry out PCA to investigate the intrinsic dimensionality of the option volumes and OVI observed for the PHOTO data set. In \Cref{fig:pca_cumsum}, we plot the cumulative percentage of variance explained for volumes and each type of OVI. For volumes, the first principal component explains 59\% of variance, with the 12 first components explaining more than 90\%. In \Cref{fig:screeplot}, we show a scree plot for the first few principal components corresponding to PHOTO option volumes. For all OVIs, more than 100 principal components are required to explain 50\% of the variance. MPCs with lower daily volumes require fewer principal components, with Professional Customers requiring just 107 components, while Customers require 211 components. Thus, using option volumes to construct OVI significantly increases the intrinsic dimensionality of the data set. This could potentially be an indication that OVI is a better metric in determining asset-specific returns, when compared to option volumes (which have a very low intrinsic dimensionality). 

Overall, we see that PHOTO and NOTO are similar in terms of the trade breakdown, while the breakdowns between call and put options are almost identical. We see that the major MPCs are Customers and Market Makers, while all the OVIs are highly interconnected. On the other hand, the higher intrinsic dimensionality observed from the OVI is further motivation for its use instead of the option volumes themselves.

\vspace{-3mm}
\section{Evaluation methods}
\label{section:evaluation}
In the subsequent sections, we investigate how the different OVI features can be used as predictors for the daily returns of the corresponding underlying assets. Underlying assets include mostly stocks, but also a selection of 11 ETFs. We obtain stock and ETF return data for NYSE from the Center for Research in Security Prices (CRSP) database.

\paragraph{Profit and Loss (P\&L) and related performance measures}
Let $P^{\textrm{open}}_{i,d}$ and $P^{\textrm{close}}_{i,d}$ denote the opening and closing price for asset $i$ at day $d$. We define the following return types \footnote{Note that all returns considered are adjusted for splits and dividends} for asset $i$: 
\[ \textrm{Daily Close to Close Returns (CL\_tmCL):} \ \ \ f_{i,d}^{\textrm{CL}\_\textrm{tmCL}} = \frac{P^{\textrm{close}}_{i,d+1}-P^{\textrm{close}}_{i,d}}{P^{\textrm{close}}_{i,d}},  \] 
\[ \textrm{Daily Open to Close Returns  (tmOP\_tmCL):} \ \ \  f_{i,d}^{\textrm{tmOP}\_\textrm{tmCL}} = \frac{P^{\textrm{close}}_{i,d+1}-P^{\textrm{open}}_{i,d+1}}{P^{\textrm{open}}_{i,d+1}},  \] 
\[ \textrm{Daily Overnight Returns (CL\_tmOP):} \ \ \ f_{i,d}^{\textrm{CL}\_\textrm{tmOP}} = \frac{P^{\textrm{open}}_{i,d+1}-P^{\textrm{close}}_{i,d}}{P^{\textrm{close}}_{i,d}}. \] 

Whenever considering returns for any of the above types, unless specified otherwise, we use the overnight CL\_tmOP returns. In this case we drop the superscripts and simply write $f_{i,d}$. Also denote the market-excess returns for asset $i$ at day $d$ \footnote{Using Risk-Adjusted return using the Fama and French 3-factor model (\cite{FF:1992}) does not significantly change the results.}
     \begin{equation*}
     \widetilde{f}_{i,d} = f_{i,d} - f_{\textrm{SPY},d},     
     \end{equation*}
    where $f_{\textrm{SPY},d}$ denotes the future return of the S\&P-500 index. The S\&P-500 index is a widely-used proxy for market returns, due to its wide range of large-cap firms and quarterly updates. Considering market excess returns, allows us to consider a market neutral study, by considering price movements relative to the market price movement. We will refer to market excess returns as EMR, and to raw returns as RR. 



Generally, we are interested in predicting directional movement of future prices. We denote by $s_{i,d}\in\mathbb{R}$ the observed signal, on day $d$, for the future price of asset $i$. Note that $s_{i,d}$ as defined here is deliberately left general and could be either some observed variable or the output of a predictive model. However, in the context of this paper, this will be either a single OVI metric or a linear combination of several OVI metrics (as seen later in this section). Then, the sign of this signal is used to select the direction in which to make a hypothetical transaction in the underlying, with $+$ and $-$ corresponding to buy and sell respectively. We also specify the size of the trade (the 'bet size'), denoted by $b_{i,d}$ in one of various ways described below; this quantity may be used to express one's degree of confidence in the predictive power of the signal. A sequence of such trades constitutes a strategy. To assess the performance obtained using a series of signals, we consider the (excess-returns-adjusted) Profit and Loss (P\&L) metric defined by
\begin{equation} \label{eq:PnL}
\textrm{P}\&\textrm{L}_{d} = \sum_{i=1}^{N_d}  b_{i,d} \times \widetilde{f}_{i,d} \times \textrm{sign} \left(  s_{i,d} \right),
\end{equation} 
\vspace{-3mm}


The P\&L is the difference in profit or loss that would have been earned, had we invested $b_{i,d}$ units of currency (in our case dollars) in comparison to making an equivalent transaction in the benchmark SPY ETF (which is tracking the S\&P-500 index). Note that the idea of P\&L described here may not derive from a strategy that is executable in practice (for example, there may be liquidity or timing constraints); nor, given the existence of transaction costs (which we do not take into account here), does it show the realistic return obtained from the strategy. 
Instead, this metric is a measure of predictability across a number of stocks based on the predictor $s$.

The P\&L is defined above for each day $d$, and this gives a vector  
$\textbf{P\&L} = \left( \textrm{P\&L}_1,\textrm{P\&L}_2,\dots,\textrm{P\&L}_{D-1}\right)$ of daily P\&L's which measures the outcome of the hypothetical strategy over the whole time horizon. 
A measure of the long-term performance of a strategy is the annualized Sharpe Ratio (hereafter SR) \cite{Shapiro1994}, defined as

\vspace{-4mm}
\begin{equation}
\textrm{SR} = \frac{\textrm{mean}(\textbf{P\&L}) \times \sqrt{252}}{\textrm{sd}(\textbf{P\&L})}.
\end{equation}   
Lastly, we define the P\&L per dollar traded (PPD) as 

\vspace{-4mm}
\begin{equation}
    \textrm{PPD} = \frac{\sum_{d=1}^{D-1} \textrm{P\&L}_d}{\sum_{d=1}^{D-1} B_d}, 
\end{equation} 
\vspace{-4mm}

\noindent  
where $B_d = \sum_{i=1}^{N_d} b_{i,d}$ is the total bet size for day $d$. 

When assessing whether a hypothetical strategy is significantly profitable, we use a test with the null hypothesis $H_0: \textrm{SR} = 0$. We make use of this test throughout the paper, by conducting a test in the same fashion as \cite{bailey2014}, which takes into account the non-normality of returns. In particular, we use the test statistic 

\vspace{-5mm}
\[ \frac{\textrm{SR(\textbf{P\&L})}}{\sqrt{\frac{1-\textrm{skewness}(\textbf{P\&L}) * \textrm{SR}(\textbf{P\&L}) + (\textrm{kurtosis}(\textbf{P\&L})-1)*(\textrm{SR}(\textbf{P\&L})*2)/4}{T-1}}}, \]
\vspace{-4mm}

\noindent where the denominator term is based on the estimate provided by \cite{bao}, for the variance of the Sharpe Ratio.

Whenever comparing the SRs of two different strategies, we need to test the hypothesis of the form of $H_0:  \textrm{SR}_1 = \textrm{SR}_2$. For that purpose we use the hypothesis test of \cite{wolf2008}. To adjust for the non-normality of returns, they use a studentized bootstrap test on the residuals of a VAR model (ia a circular block approach) to adjust for the non-normality of returns. 

\paragraph{Quantile rank portfolios}
The P\&L \eqref{PnL} only uses the sign of the observed signal (for example, the OVI), and does not take into account its magnitude. Naturally, a larger magnitude for the signal should entail a higher confidence in the prediction. To that end, for each MPC and each day, we split underlying assets into 5 groups, in two different ways.

For each financial day, we divide the underlying assets into five quantile buckets on the basis of the magnitude (strength) of the corresponding option-market signal, $|s_{i,d}|$. We denote by QB1 the group of assets with the 20\% of weakest signals, and so on up to QB5, which corresponds to the 20\% strongest signals. Our first split of the equities is simply into the five quantile buckets QB1--QB5, and we refer to this as Quantile-Ranked buckets. For the second and main split, all groups contain equities corresponding to different percentages of the strongest signals, with the first corresponding to all 100\% of signals and the fifth to just the strongest 20\% signals. In particular we define Q1 = QB1$\cup$QB2$\cup$QB3$\cup$QB4$\cup$QB5, Q2 = QB1$\cup$QB2$\cup$QB3$\cup$QB4, and so on; we refer to it as a quantile ranked group. 



We assess performance of the P\&Ls of strategies based on OVI-based signals using a number of  different bet weighting schemes, where the bet size (ie., capital) allocated to each equity (equation \eqref{PnL}) depends either on the liquidity of the asset in the option market for that financial day (as measured by traded volume) or on the magnitude of its signal (OVI). 

\paragraph{Bet-size weighting schemes}
We now discuss the bet-size term $b_{i,d}$ of \eqref{eq:PnL}, by discussing a number of different betting schemes. Naturally, we can naively allocate equal capital to all the equities traded (we refer to this as the Uniform Weighting Scheme). However, we could take into account our confidence in the used signals, by basing the bet scheme either on the magnitude of the signal or the liquidity of the option contract (as for more liquid assets, the OVI is calculated using more transaction data). In each case, we first define a ``raw'' bet size $b'_{i,d}$ for each equity on the basis of an appropriate measure of the market, and set 

\vspace{-6mm}
\[
b_{i,d} = \frac{b'_{i,d}}{\sum_j b'_{j,d}}.
\]
\vspace{-3mm}

\noindent
In this way we ensure comparability of schemes by betting a total of $1$ dollar, distributed across all the underlying assets for which trading is indicated, each day. This way, the cumulative P\&L \eqref{PnL} has the interpretation of being the cumulative percentage return.
\begin{itemize}
     \item \textbf{Uniform Weighting Scheme}. Here $b'_{i,d}=1, \ \forall i,d$. In this case we do not differentiate between the different underlying assets, allocating equal capital to each asset $i$ with $s_{i,d}\neq 0$.
     \item \textbf{Imbalance Weighting Scheme. } Here $b'_{i,d}= |s_{i,d}^h|$. Using this betting scheme is equivalent to using the signal $s_{i,d}$ instead of just its sign as in~\eqref{PnL}. As higher-magnitude signals (for example, higher OVI magnitude) indicate a higher confidence level, this weighting scheme allows us to incorporate our prediction confidence in the bet size. Splitting equities into quantile rank portfolios is an alternative way to differentiate based on the magnitude of the signals, but using this scheme allows us to use the whole portfolio, without splitting into groups.
    \item \textbf{(Nominal) Volume Weighting Scheme}. Here the bet size is set to $b'_{i,d}= log(1 + \sum_{j=1}^{C_{i,d}} V_{i,j,d,m})$, or $b'_{i,t}= log(1 + \sum_{j=1}^{C_{i,d}} P_{i,j,d} V_{i,j,d,m})$ for the nominal volume case (depending on the MPC $m$). The (nominal) volume is used as a proxy for option liquidity in order to de-emphasize illiquid cases. The log transformation is applied to soften the significant differences in volumes between different contracts.
    \item \textbf{Relative Volume Weighting Scheme}. Here the bet size is set to $b'_{i,d}= log(1 + \sum_{j=1}^{C_{i,d}} \frac{V_{i,j,d,m}}{OI_{i,j,d}})$ (depending on the MPC $m$), where $OI_{i,j,d}$ is the open interest for option contract $j$ (of asset $i$). One issue with the volume weighting scheme, is that equities with lower volumes in the option market, will tend to get a lower weight, even when their liquidity is relatively high. The relative volume weighting scheme, takes this shortcoming into account, by normalizing volumes by the open interest of each contract. 
\end{itemize}
Unless specified otherwise, we will be using a Uniform Weighting Scheme.

\vspace{-4mm}

\subsection{P\&L Regression.}
\label{sec:pnl}
In our analysis, we consider several variants of OVI, based on different option transaction types (Buys, Sells or both), data sets (PHOTO, NOTO or both) and MPCs. We are 
also interested in analyzing the joint predictive power of such OVI features, in explaining future returns. Usually a linear method is employed (OLS or LASSO Regression), and the corresponding $t$ or $p$-values are used to compare the explanatory power of different covariates (e.g \cite{Pan_Poteshman}). 
 
However, a linear model is only suitable for finding linear relationships.
Furthermore, the objective function of OLS regression is based on the euclidean distance between observed and fitted values. Such an objective function can be unsuitable for directional predictions (as predicting the direction is an inherently nonlinear problem). A predictor that is successful in capturing the signs of future returns, may be deemed uninformative with respect to the euclidean distance. In the case of linear models, while a predictor's sign may match the observations perfectly, we may still observe a small corresponding $R^2$ value. Finally, some of the options corresponding to an asset may not be traded at all on a given day, meaning that the OVI can be zero for a large number of days. Linear methods are also unable to deal with feature sparsity, as they are unable to ignore zero-valued features. Indeed, while OLS regression is able to show that all of the OVIs are significant predictors for future returns, the corresponding $R^2$ values are quite low (check section 1 in the Supplementary Material) and it is hard to distinguish between the explanatory power of the different OVIs. 
 
For these reasons, we suggest a novel method for regressing the future returns against different variables (in our case, different variants of OVI), where we directly set an objective function equal to the cumulative P\&L \eqref{PnL}, thus searching a combination of covariates which maximizes the net profit. Equivalently, we can minimize the negative P\&L. Inspired by the popular LASSO regression \cite{LASSO}, we also impose an $\ell_1$ penalty term on the objective function. This allows us to a distinguish between significant and non-significant covariates, by penalizing large coefficient values for the latter. Similar previous research has optimized the Sharpe Ratio by including it in the objective function for portfolio optimization problems such as \cite{Zhang8} and \cite{cao_SR_opt}, but in this case we are looking to combine multiple features for the forecasting of future returns for different assets. 

Suppose that we have a total of $K$ variants of OVI. Denote by $A_{i,d,k}$ the $k^{th}$ type of feature/OVI, for asset $i$ at day $d$ (e.g $A_{i,d,k} = \textrm{OVI}^{[\mathcal{F}]}_{i,d,m}$ for some MPC $m$ and some restrictions $\mathcal{F}$). In this case, we define a signal which is a linear combination of the different OVI

\vspace{-5mm}
\begin{equation} \label{lasso_s}
s_{i,d}(\beta;\mathbf{A}) = \beta_0 + \sum_{k=1}^K \beta_{k}A_{i,d,k} ,
\end{equation}
\vspace{-4mm}

\noindent  
for some vector $\beta = \left( \beta_0,\dots,\beta_K \right)$. The goal is to choose $\beta$, in a way which maximizes the profit of a strategy based on the signal in \eqref{lasso_s}. In particular, we look for a vector $\beta$, which maximizes the total P\&L (denoted as $\textrm{PL}$) defined as
\begin{equation} \label{eq:lasso_pnl}
\textrm{PL}(\beta;\mathbf{A},f) = \sum_{d=1}^{D-1} \textrm{P\&L}_d = \sum_{d=1}^{D-1} \left( \sum_{i=1}^N b_{i,d}(\beta; \mathbf{A}) f_{i,d} \textrm{sign} \left( s_{i,d}(\beta; \mathbf{A}) \right)  \right).
\end{equation} 
Thus, we look for an objective function which incorporates the total P\&L. 
In order to use gradient methods, we require a smooth objective function. The sign function used in \eqref{eq:lasso_pnl} is not smooth, therefore we wish to substitute it with a differentiable activation function $g:\mathbb{R} \to \mathbb{R}$, such as 
\begin{equation} \label{lasso_sigma}
g(x;\alpha_1) = 2 \left( \frac{1}{1+e^{- \alpha_1 x}} - \frac{1}{2} \right), \ \ \alpha_1 > 0. 
\end{equation} 
This is equivalent to the \emph{tanh} function for $\alpha_1=2$, and for large $\alpha_1$ it converges pointwise to the sign function.
We can get rid of the $\textrm{sign}$ function by choosing a bet size, similar to the uniform scheme  
\begin{equation} \label{lasso_b}
b_{i,d}(\beta; \mathbf{A}) = \frac{|g \left( s_{i,d}(\beta; \mathbf{A}) \right)|}{\sum_{j} |g \left( s_{j,d}(\beta; \mathbf{A}) \right)|}.
\end{equation} 

Therefore, \Cref{eq:lasso_pnl} can be rewritten as 
\begin{equation} \label{eq:lasso_pnl_2}
\textrm{PL}(\beta;\mathbf{A},f) = \sum_{d=1}^{D-1} \textrm{P\&L}_d = \sum_{d=1}^{D-1} \left( \sum_{i=1}^N \frac{1}{\sum_{j} |g \left( s_{j,d}(\beta; \mathbf{A}) \right)|} f_{i,d} g \left( s_{i,d}(\beta; \mathbf{A}) \right) \right).
\end{equation}

This brings another differentiability issue with the absolute value function, which is used twice, one for the bet sizes \eqref{lasso_b}, and for the $\ell_1$ regularization term, which is non-differentiable at $0$. There exist many methods to circumvent this, but we shall make use of one of the simpler ones, which is to replace these functions with a smooth approximation, in the same spirit as how we used the modified sigmoid. We use the function $h$, defined as 
 \begin{equation} \label{lasso_h} h(x;\alpha_2) = \sqrt{x^2 + \alpha_2}, \alpha_2 > 0, \end{equation}  for some small value of $\alpha_2$.
In summary, the objective function to be minimized, takes the form
\begin{equation} \label{eq:lasso_Lpnl} L(\beta;\mathbf{A},f,q,\alpha_1,\alpha_2,\lambda) = -\sum_{d=1}^{D-1} \left( \sum_{i=1}^N \frac{1}{\sum_{j} h \left( g \left( s_{j,d}(\beta; \mathbf{A}) \right) \right)} f_{i,d} g \left( s_{i,d}(\beta; \mathbf{A}) \right) \right) + \lambda \sum_{k=1}^K h( \beta_k;\alpha_2).
\end{equation} 

To minimize \eqref{eq:lasso_Lpnl}, we employ the ADAM algorithm \cite{ADAM}\footnote{Using the ADAM optimizer, requires choosing two parameters, the so-called moment parameters. For this paper, we use the recommended values given in the original paper ($0.9$ and $0.999$).}. The derivative of the objective function is given by 
\begin{align*} \label{lasso_lpnl2} 
\frac{\partial L}{\partial \beta_k} = 
& - \sum_{d=1}^{D-1} \sum_{i=1}^N f_{i,d} \left( \frac{1}{\sum_{j} h \left( g \left( s_{j,d}(\beta; \mathbf{A}) \right) \right)}  \frac{\partial g \left( s_{i,d} (\beta) \right)}{\partial \beta_k} - \right. \\ & \left. g \left(s_{i,d}(\beta) \right) \left( \frac{1}{\sum_{j} h \left( g \left( s_{j,d}(\beta; \mathbf{A}) \right) \right)} \right)^2 \sum_{j=1}^N h'(g\left(s_{j,d}(\beta)\right)) \frac{\partial g \left( s_{i,d} (\beta) \right)}{\partial \beta_k} \right) + \lambda \left(1-\mathbbm{1}_{(k=0)}\right)  h'(\beta_k), 
\end{align*} 
where 
\begin{equation} \label{lasso_lpnl3} \frac{\partial g\left( s_{i,d} (\beta) \right)}{\partial \beta_k} =  g' \left( \beta_0 + \sum_{k=1}^K A_{i,d,k} \beta_{k} \right) A_{i,d,k}. \end{equation} 
Equivalently, written in vector form, this amounts to  
\begin{align*} \label{lasso_lpnl4} \frac{\partial L}{\partial \beta} = - &\sum_{d=1}^{D-1} \sum_{i=1}^N f_{i,d} \frac{1}{\sum_{j} h \left( g \left( s_{j,d}(\beta; \mathbf{A}) \right) \right)} \frac{\partial g\left( s_{i,d} (\beta)  \right)}{\partial \beta}  \\ & + 
\sum_{d=1}^{D-1} \left( \sum_{i=1}^N f_{i,d} g \left(s_{i,d}(\beta) \right) \left( \frac{1}{\sum_{j} h \left( g \left( s_{j,d}(\beta; \mathbf{A}) \right) \right)} \right)^2 \right)   \left( \sum_{j=1}^N h'(g\left(s_{j,d}(\beta)\right)) \frac{\partial s_{j,d} (\beta)}{\partial \beta} \right)\\ &  + 
\lambda \left[0,  h'(\beta_1) , \dots, h'(\beta_K) \right]^T. 
\end{align*} 

Lastly, once the algorithm converges and a vector $\hat{\beta}$ is obtained, we rescale the  coefficients, so that the non-intercept coefficients have a total absolute value of one, i.e 
\[  \tilde{\beta} = \frac{\hat{\beta}}{\sum_{k=1}^K | \hat{\beta_k} |}.
\] 
Since $\beta$ only affects the P\&L only through the signal $s$ (see (\ref{lasso_s})), rescaling all of the coefficients, does not affect the objective function if we are using the sign function. Note however, that due to using the proxy $tanh$ function, there is in fact a small difference that is caused by the rescaling. 

As already mentioned, this method has various advantages over the standard linear regression, as it is for example, better able to uncover monotonic relationships than standard linear regression. The disadvantage is that unlike OLS, the minimization of \Cref{eq:lasso_Lpnl} does not have a closed-form solution nor is it convex, and we are also burdened with the choice of $\lambda$. On the other hand, the magnitude of the fitted coefficient $\beta$ has a direct interpretation as the weighting factor that contributes in making the buy/sell/neutral decision. 

Throughout the rest of the paper, signals $s_{i,d}$ will either be equal to a particular type of OVI or take the form \eqref{lasso_s}. In the first case the strategy does not require any parameters, and so remains the same throughout the whole evaluation period. In the second case, the model could be trained on more recent data, via a sliding-window approach.

In particular, we set a window size $l$ of 500 trading days (about 2 years), a testing window size of $T= 100$ days\footnote{Note that choosing different $l$ and $T$, does not significantly alter results.}. A window at day $d$, contains the range of days $\{d-l+1,\dots,d\}$. For each window, we regress the future returns against the OVI features, using the P\&L regression described above, to obtain coefficients $\{ \beta_k^{(d-l+1):d} \}_{k=1}^K$. After obtaining the fitted coefficients $\{ \beta_k^{(d-l+1):d} \}_{k=1}^K$, we calculate the P\&L in \eqref{eq:lasso_pnl_2} using the days in the testing set for the window at day d \[ \textrm{Test}_d = \{ d+1,\dots,d+T \}, \]
i.e $\text{PL}(\{ \beta_k^{(d-l+1):d} \}_{k=1}^K;A_{\left[ \cdot,\textrm{Test},\cdot  \right]},f_{\left[ \cdot,\textrm{Test}  \right]},q,\alpha_1,\alpha_2)$. This is an out-of-sample metric, with the direct interpretation of being the hypothetical profit one would make in 100 days of trading by following such a strategy (in the absence of trading costs).

\FloatBarrier 
\section{Main Empirical Results} 
\label{section:MarketMakers}
\FloatBarrier 

\begin{figure}[h]
\includegraphics[width=1\textwidth,left]{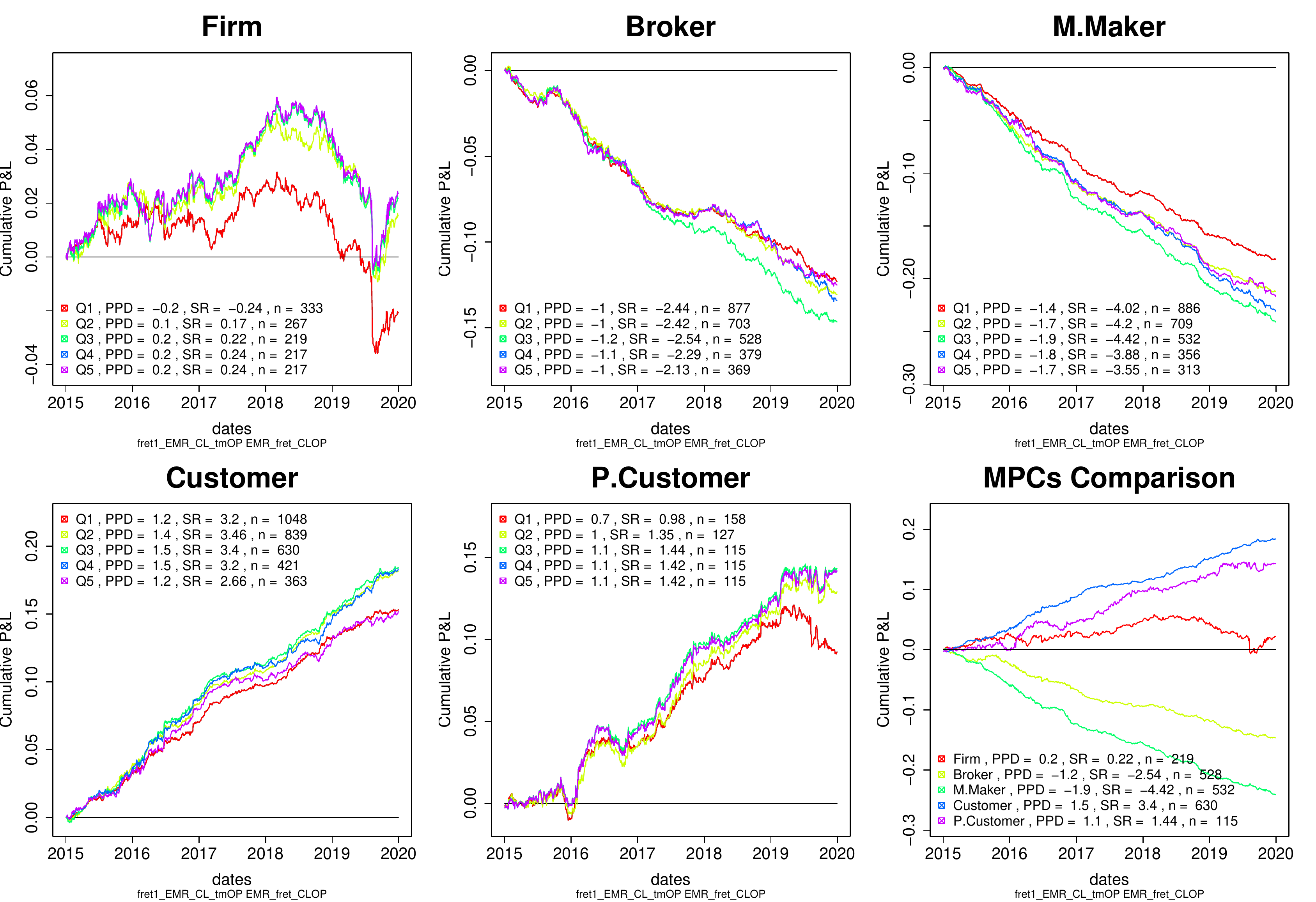}
\caption{Cumulative P\&L plot - Overnight Returns: This figure shows the cumulative P\&L by following a uniform scheme - one-day holding period strategy, for the five QR groups Q1-Q5 and each MPC, for overnight returns (EMR\_CL\_tmOP). Also given, are SR (the Annualized Sharpe Ratio), daily PPD (Profit per dollar in basis points), and n (the average number of assets in the portfolios). The first five subplots show the 5 QR groups for each MPC, while the last subplot corresponds to using the 3rd QR group for each MPC. }
\label{fig:cumulative_PnL_overnight}
\end{figure}

In this section, we assess the predictive power of OVI against the future returns across different MPCs, across differet return types (overnight,intraday and close to close) for the PHOTO data set, and by using different betting schemes.

\paragraph{Comparison across market participant classes}
We consider the OVI for each of the 5 MPCs seperately by using their sign as the signal for future returns ($s_{i,d}$ = $\textrm{OVI}_{i,d,m}$ for each $m$). We then evaluate their corresponding P\&L using \Cref{PnL}, for their excess market overnight returns (EMR CL\_tmOP), and for different quantile ranked groups (Q1-Q5) built as described in \Cref{section:evaluation}. In particular, we plot the time series for the cumulative P\&L incurred over the period 2015-2019 in \Cref{fig:cumulative_PnL_overnight}.

The highest performance is observed for Market Makers, where the cumulative returns are between 17\% and 25\% and annualized SRs lie between 3.5 and 4.4 depending on the quantile rank portfolio used. The second best performance is observed for Customers, whose cumulative P\&L display a clear upwards trend, and SRs with magnitudes in the range 2.7--3.5. Strong positive signals are also observed for Brokers, with SRs between 2.1 and2.5. Lastly, Professional Customers display a downwards trend with SRs of magnitude 1--1.4. For Firms, while the total P\&L is slightly positive, there does not appear to be a clear upwards trend, with SRs very close to 0. Note that there seems to be a trend shift happening after 2018 for Firms, from a somewhat upward trend to a downward trend (a trend shift can also be noticed for other MPCs). 

\begin{figure}[h]
\centering
\includegraphics[width=0.9\textwidth]{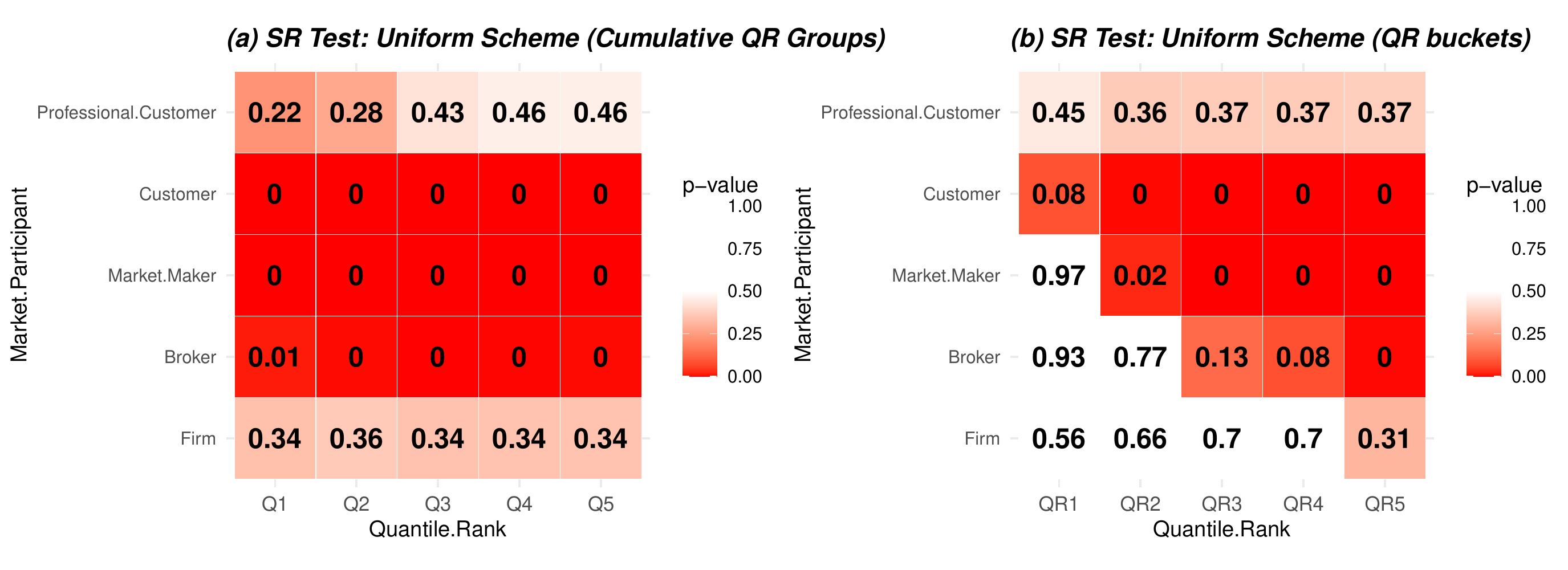}
\caption{Sharpe Ratio Significance tests: The grid shows the p-value for the significance of the SR for different OVIs. The coloring shows the significance level for the strategies, where red denotes statistically significant returns associated with that strategy (either positive or negative). Sub-figure(a) uses the cumulative QR groups where as sub-figure(b) uses the (distinct) quantile buckets.}
\label{fig:single_tests}
\end{figure}

In \Cref{fig:single_tests}(a), we report the $p$-values obtained when running a significance test for the SRs corresponding to excess market CL\_tmOP returns. The tests for \textit{Brokers, Market Makers, and Customers} are rejected with a $p$-value of 0 (up to 2 decimal places) across all quantile ranks. \textit{Firm} and \textit{Professional Customers} have $p$-values close to 0.5 across all ranks. Thus, in the span of five years, we observe a clear signal from three out of five of the MPCs, with the \textit{Market Maker} achieving a return of about 30\% in a period of 5 years. The two participants without a clear signal are also the MPCs with the lowest mean volume per day (see \Cref{fig:breakdown}), which partly accounts for the reduced performance. The clearest signals come from \textit{Market Makers}, followed by \textit{Customers}, which have on average the highest volume per day. These observations clearly mark the OVI as holding predictive power for future returns. Note that this verifies the findings of the earlier study \cite{poteshman2005}, who report that the informational context of Customer volumes is higher than Firm volumes.

\paragraph{Quantile rank portfolios}  
For all MPCs except Firms, the best performing portfolio is Q3 both in terms of PPD and in terms of SR. The worst performing portfolio in terms of PPD is Q1, which in most cases has a noticeably worse performance than other QR groups. In terms of SR, the worst performing portfolio is either Q1 (all assets) or Q5 (top 20\% of strongest signals) depending on the MPC. Therefore, the strategies of taking into account all the non-zero OVIs, or only considering a very small subset of the strongest signals, are both outclassed by following a portfolio consisting of the top 60\% or 80\% of the highest-magnitude OVIs. 

The same analysis is carried out for quantile ranked buckets. In \Cref{fig:single_tests}(b), it becomes apparent that across MPCs, lower QR Buckets highly outperform the higher buckets, with QB1 being almost uninformative. For both \textit{Market Makers} and \textit{Customers}, QB2-QB5 appear to be the most significant. Overall, this confirms that not only is the sign of the OVI informative on future returns, but that the predictability observed is stronger for higher magnitudes of the OVI, which justifies using the OVIs as signals.

\paragraph{Comparison across Return Modes}
\begin{figure}[h]
\includegraphics[width=1.05\textwidth,left]{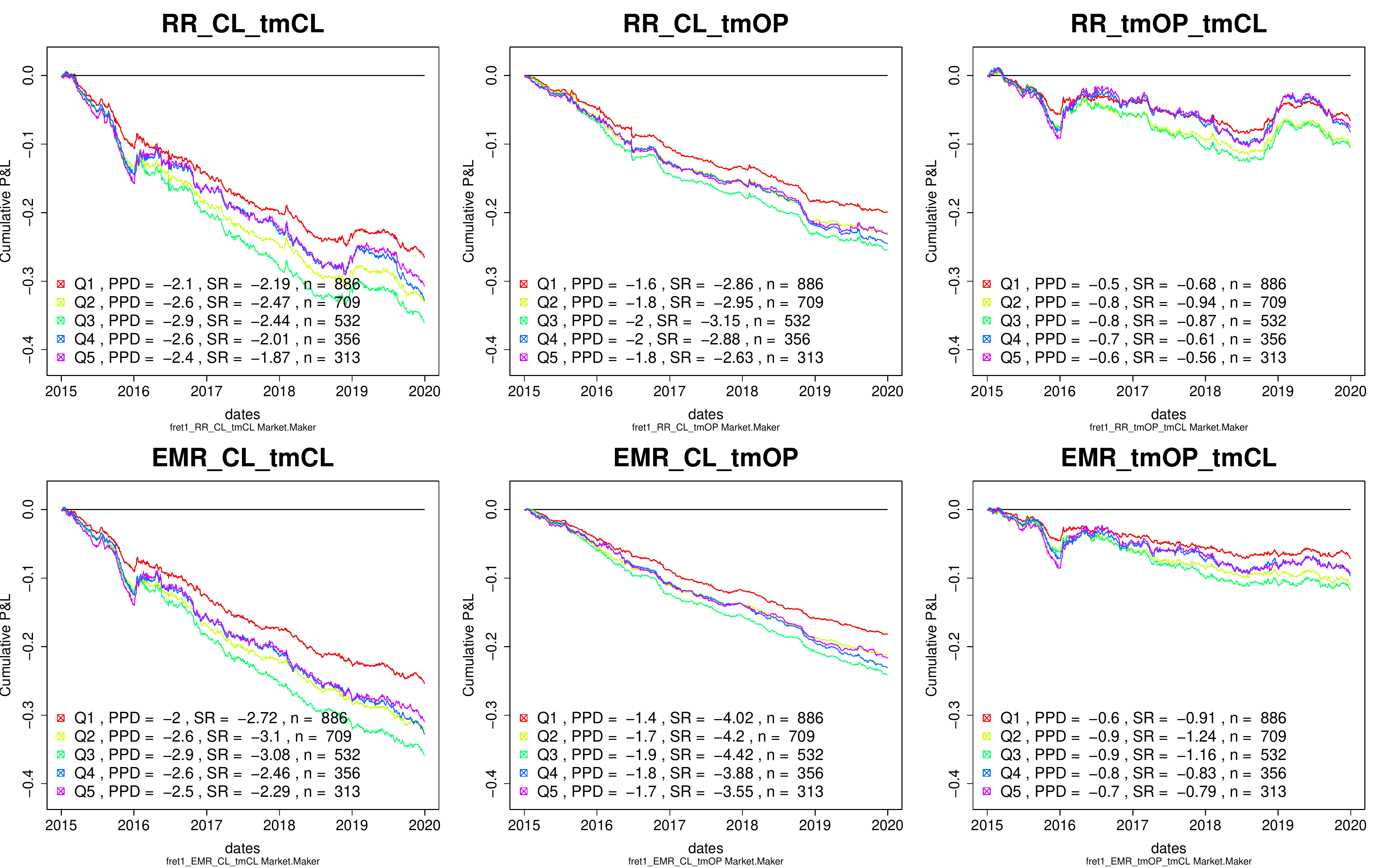}
\caption{Cumulative P\&L plot - Market Maker: This shows the cumulative P\&L from following a Uniform Weighting Scheme - one-day holding period strategy, for the five quantile rank groups Q1-Q5, using the OVI of Market Makers for six different types of returns. Also given, are SR (the Annualized Sharpe Ratio), PPD (Profit per dollar in basis points), and $n$(the average number of assets in the portfolios).}
\label{fig:cumulative_PnL_mm}
\end{figure}

In \Cref{fig:cumulative_PnL_mm}, the cumulative P\&L corresponding to each type of return is shown, when using the Market Maker OVI. 

The highest predictability is clearly observed for the the overnight (CL\_tmOP) returns, followed by the close-to-close (CL\_tmCL), and then by the open-to-close (tmOP\_tmCL) returns. Close-to-close return is a combination of the other two types of returns, with most of its predictability stemming from the overnight returns rather than the next-day returns. 
Using market excess returns, results in higher returns as shown by higher SRs, but raw returns also give significant results across MPCs and quantile rank groups. 
\begin{figure}
\includegraphics[width=1\textwidth,left]{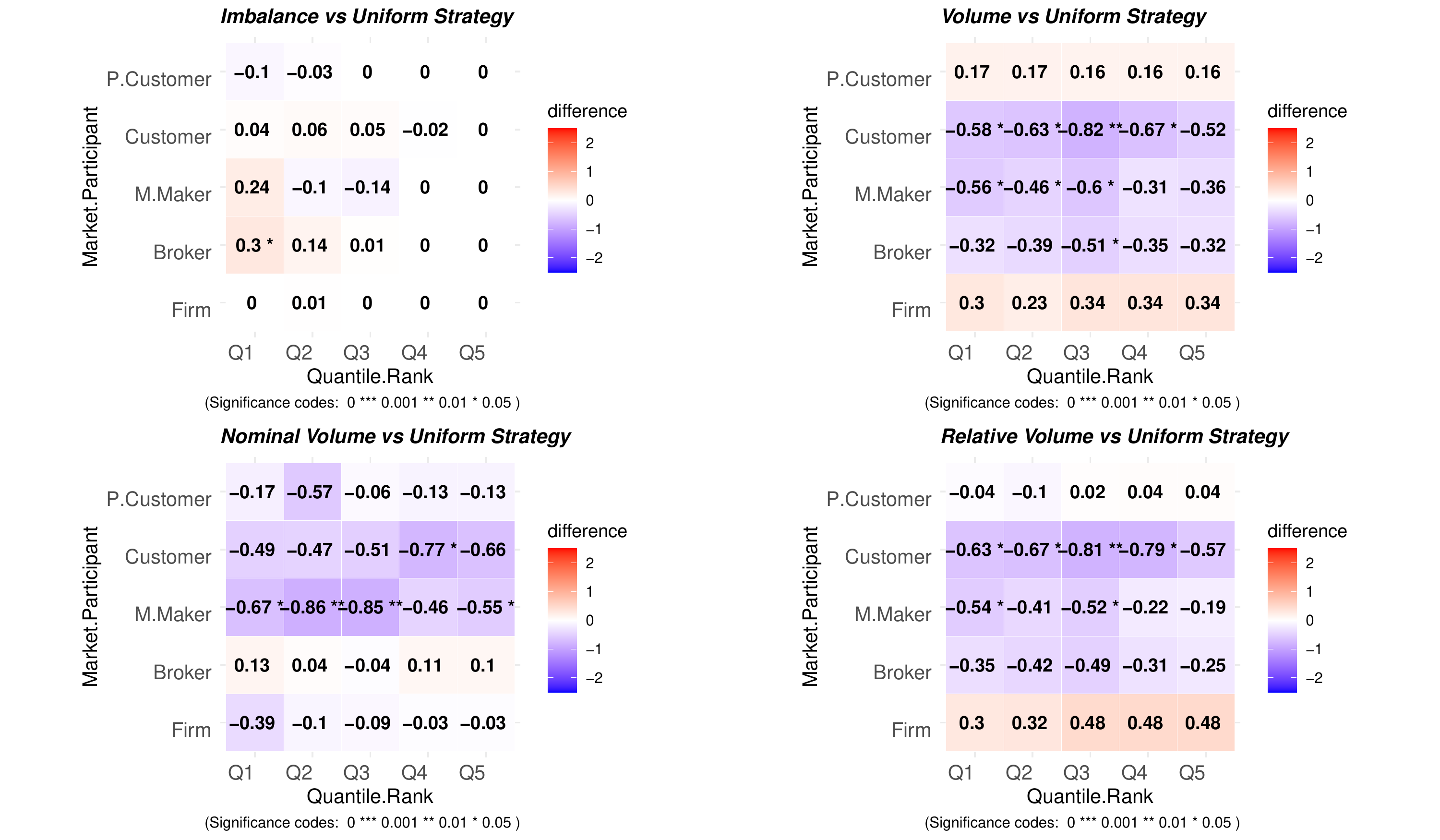}
\caption{Sharpe Ratio Difference Tests for different betting scheme: Cells colored red (resp. blue) denote that the scheme considered outperforms (resp. is outperformed) by the Uniform Weighting Scheme. The significance level is also given after the number of stars (1-3), if applicable.}
\label{fig:betting_strategies_sharpe_differences}
\end{figure}

\begin{figure}
\includegraphics[width=1\textwidth,left]{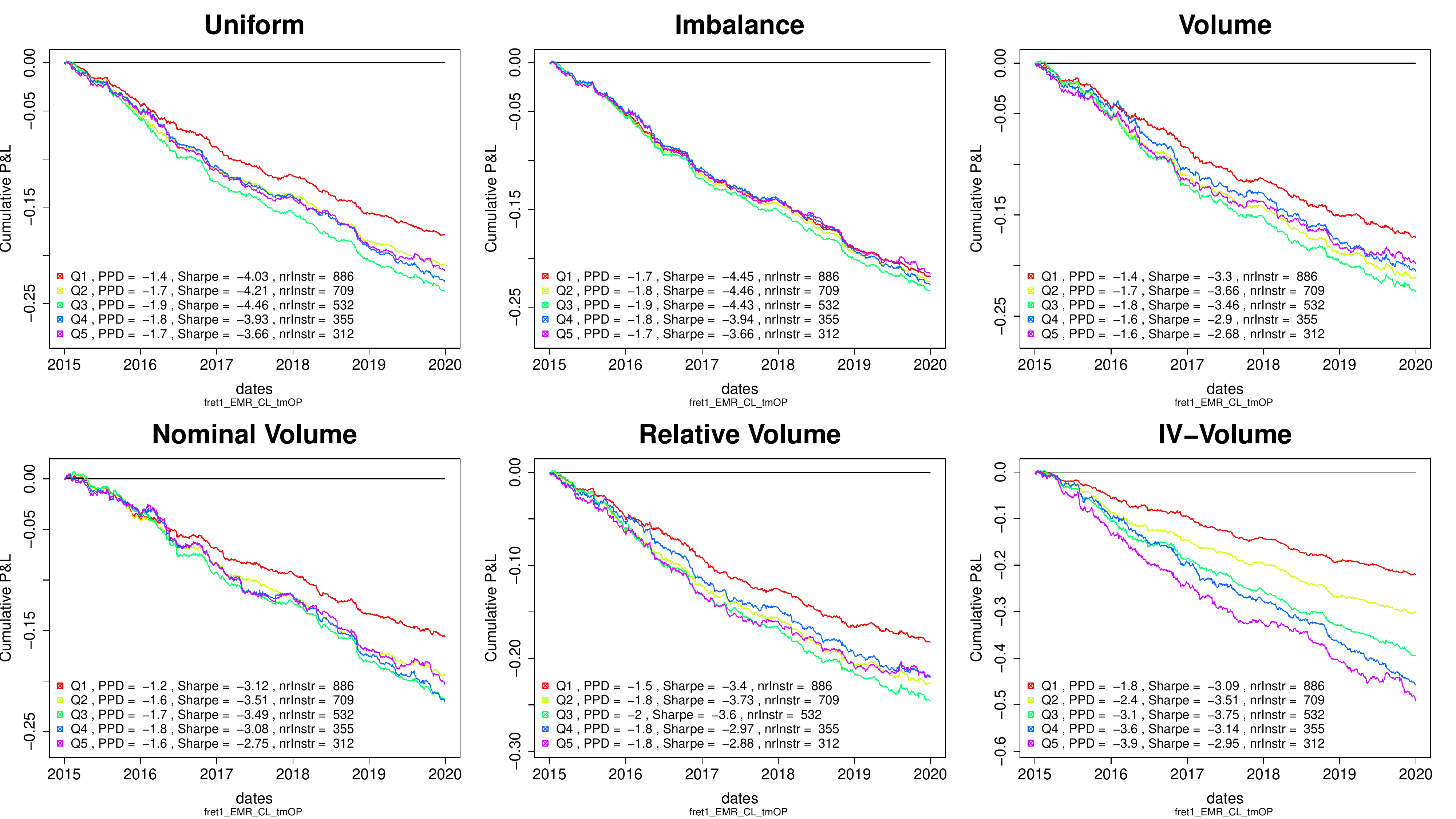}
\caption{Cumulative Returns for Different Betting Schemes: This shows the cumulative EMR CL\_tmOP returns P\&L by following the different betting schemes with one-day holding period. The OVI used is for Market Makers, and the five quantile rank portfolios Q1-Q5. Each portfolio Q-k corresponds to considering the assets corresponding to the ($20k$) \% highest OVIs in absolute value, for six different target returns. Also given, are Sharpe (the Annualized Sharpe Ratio), PPD (Profit per dollar in basis points), and nrInstr (the average number of assets used on the portfolio).}
\label{fig:betting_strategies_cumsum}
\end{figure}

\paragraph{Comparison across OVI types}
\begin{figure}[h]
\includegraphics[width=1.05\textwidth,left]{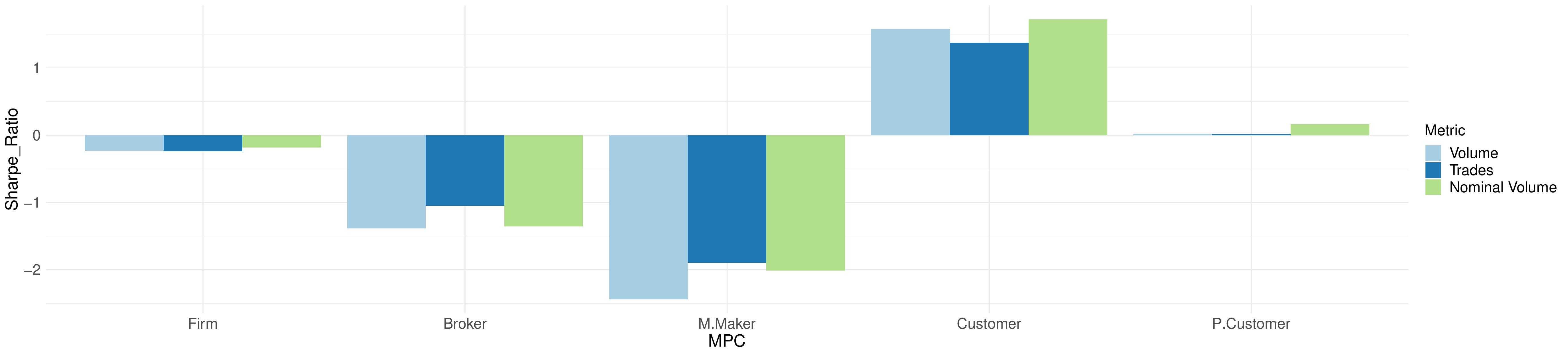}
\caption{Comparison across OVI types: For each MPC, the SR corresponding to EMR CL\_tmOP returns are shown, for each OVI type (as defined in \Cref{subsec:OVI}). We use a uniform betting scheme, Q3 strategy for each are shown for each OVI type.}
\label{fig:sharpe_ratio}
\end{figure}
Throughout this section we have been using the Volume-based OVI. In \Cref{fig:sharpe_ratio}, we compare the different types of OVI (for Q3 quantiles, Uniform Weighting Scheme) across MPCs. Trade-based OVI has the lowest performance for Brokers (significantly lower than both volume and nominal volume based OVI), Market Makers (significantly lower than volume based OVI) and Customers (significantly lower than nominal volume based OVI), and therefore seems as a less suitable feature, than the others. This is expected as it is naturally the least informative of the three types of features, compared to volume and nominal volume. Volume-based OVI scores the highest (in magnitude) SR for Market Makers and Brokers, and the difference with the other two types for Market Makers is significant when running a pairwise SR Difference test. 

\vspace{-2mm}
\subsection{Alternative Weighting Schemes}
In this subsection, we compare the Uniform Weighting Scheme with the alternative schemes defined in the previous section. We use the Market Maker OVI, and record the P\&L corresponding to excess market overnight returns \Cref{PnL}. In \Cref{fig:betting_strategies_sharpe_differences}, we carry out a SR difference test to compare the performance of different weighting schemes for the overnight returns with the uniform scheme. In \Cref{fig:betting_strategies_cumsum} we plot the cumulative P\&L for excess market overnight returns obtained by following a strategy for each of the different weighting schemes, for the Market Maker OVI.

\paragraph{Imbalance Weighting Scheme}  
In most cases across MPCs and quantile rank groups, when performing pairwise SR difference tests between the Uniform and Imbalance Weighting Schemes, the hypothesis is not rejected, and if we correct for multiple testing, none of the hypothesis' is rejected. However, as seen in \Cref{fig:betting_strategies_sharpe_differences}, the SRs corresponding to the Imbalance Scheme are slightly higher than those for the Uniform Weighting Scheme, for all quantile rank groups for Brokers and Customers, as well as Q2, Q4 and Q5 for Market Makers. Therefore, this suggests that the Imbalance Weighting Scheme is a better way to incorporate the information embedded in the OVI, than a simple uniform scheme.

Furthermore, as seen in the cumulative plot \Cref{fig:betting_strategies_cumsum}(e), we can clearly see that the quantile rank groups are more indistinguishable (both in terms of SR and PPD) in comparison with the Uniform Scheme, especially for Q1-Q3. In addition, the $5^{\textrm{th}}$ QR group is the worst performing, despite being one of the highest performing groups for the other weighting schemes.

\paragraph{Volume-Weighting Scheme}  
In \Cref{fig:betting_strategies_sharpe_differences}, it can be seen that in most cases, both the Volume and Nominal Volume Weighting Schemes are outperformed by the Uniform Weighting Scheme both in terms of SR as well as PPD. The Relative Volume Scheme, is also outperformed by the Uniform Scheme in terms of SR, but slightly outclasses it in terms of PPD. In addition it outclasses the other two schemes in terms of both SR and PPD, for most of the QR Groups. 
The lower SRs are a result of higher variability introduced when incorporating volume into the betting scheme. However, none of the volume betting schemes seems to be suitable in replacing the uniform betting scheme. Note, that defining a weighting scheme based on the raw option volume (without the log transform), the sqrt-transformed volume, or the liquidity of the underlying, also fails to outperform the uniform scheme, even though the raw-Volume Weighting Scheme does outperform it in terms of PPD (but has a very high variability).  

\begin{figure}
\begin{center}
\includegraphics[width=1\textwidth]{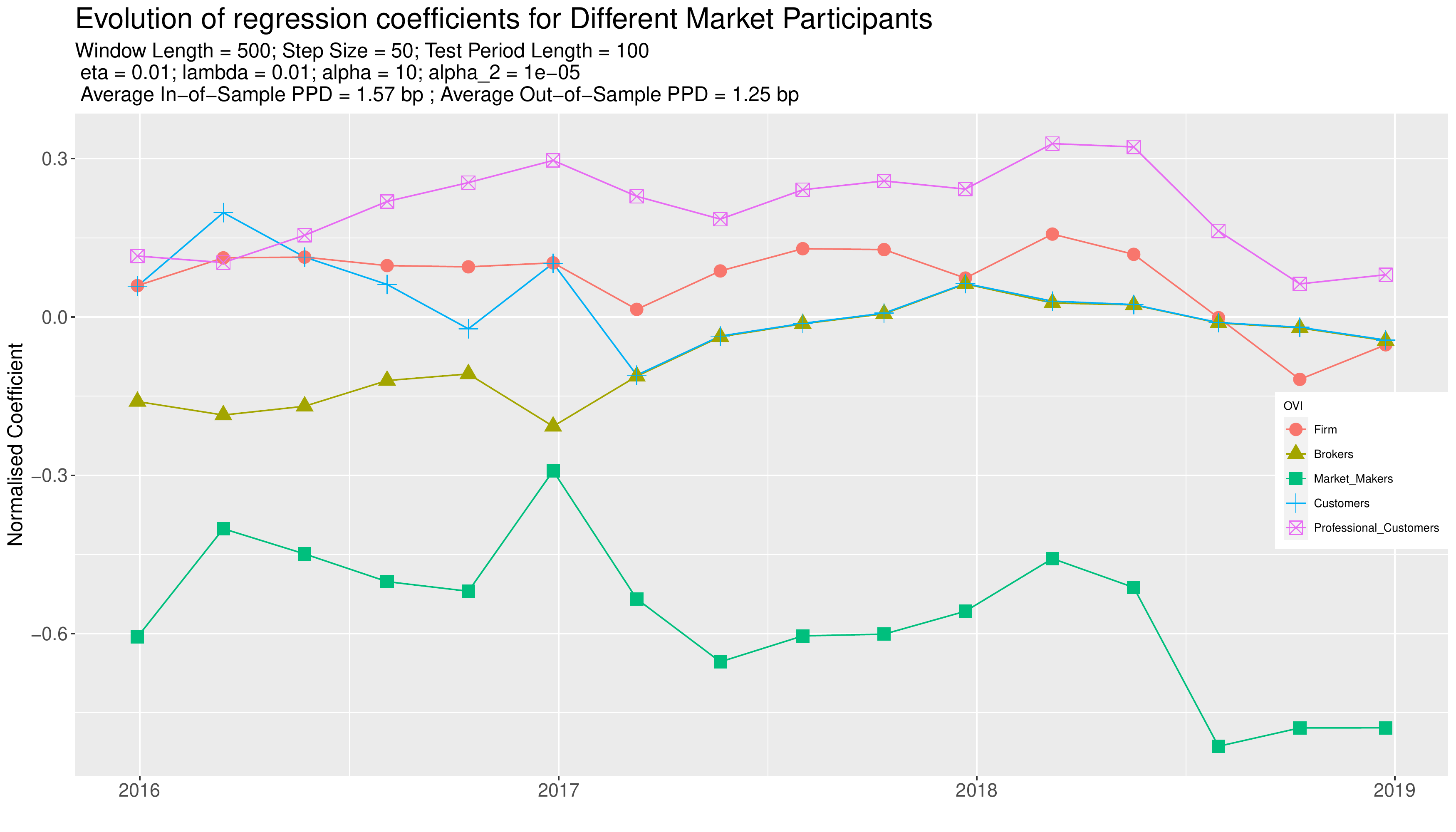}
\end{center}
\caption{Multiple-factor regression for different MPCs: The line graphs, show the evolution of the coefficients in equation~\eqref{eq:model2}, fitted using the P\&L regression for M2 across time, as indicated by the plotted points. Note that coefficients are normalized to have a total $\ell_1$ norm equal to 1 (excluding the intercept term).}
\label{fig:regression_coefficients_five}
\end{figure}

\subsection{Multi-factor analysis and stability analysis across time}
In predicting future overnight returns, using the signal given by the sign of the Market Maker OVI, has reulted in the highest performance. The Market Maker signal can be written as 
\begin{equation} \label{eq:model1}
        s^1_{i,d} = \textrm{OVI}^{\textrm{PHOTO}}_{i,d,\textrm{M.Maker}}.
\end{equation} 
In this subsection we evaluate the joint predictive power of the OVI corresponding to the five different market participants (K=5) using the P\&L regression method described in \Cref{section:evaluation}, to predict the future EMR CL\_tmOP return, by using a signal of the form 
\begin{equation} \label{eq:model2}
        s^2_{i,d}\left( \mathbf{\beta} \right) = \beta_0 + \beta_{1} \textrm{OVI}^{\textrm{PHOTO}}_{i,d,\textrm{Firm}} + \beta_{2} \textrm{OVI}^{\textrm{PHOTO}}_{i,d,\textrm{Broker}}  + \beta_{3} \textrm{OVI}^{\textrm{PHOTO}}_{i,d,\textrm{MarketMaker}} + \beta_{4} \textrm{OVI}^{\textrm{PHOTO}}_{i,d,\textrm{Customer}} + \beta_{5} \textrm{OVI}^{\textrm{PHOTO}}_{i,d,\textrm{P.Customer}}.
\end{equation}  

We regress in a sliding window manner and record the evolution of the normalized coefficients during 2016-2019, as shown in \Cref{fig:regression_coefficients_five}. Note that at each time point, the coefficient of \textit{Market Maker} OVI dominates, by having a magnitude of 0.3-0.75. Generally, the Market Maker OVI seems to have a higher weight for the later years, which is partly due to the coefficients of Broker and Customer OVI shrinking in magnitude. Firms show mostly negative coefficients, with low magnitude for the entire 5-year period. Interestingly, the OVI for Professional Customers has a consistently negative coefficient throughout the 5-year period, with a relatively high magnitude. Since the volumes of Professional Customers are quite low (i.e the OVI is equal to zero in most cases), the high coefficient does not disrupt the model (which would have not been possible if we used a linear regression model). On the other hand, as seen in the correlation map in \Cref{fig:imb_cor}, the OVI for Professional Customers is not correlated with the other features, which potentially shows that additional information is brought when it is used in conjunction with Market Makers' OVI.

Since such a method results in a coefficient vector dominated by the Market Maker's OVI, it is worth comparing the performance of $s^1$ and $s^2$, when using the bet sizes \Cref{lasso_b}. The two strategies give similar mean in-sample PPDs with 1.50bps and 1.57bps respectively. However, using $s_1$ results in a slightly higher out-of-sample mean PPD of 1.33bps compared to 1.25bps for $s_2$, indicating some overfitting in the second case.  

\FloatBarrier 

\vspace{-2mm}
\section{Influence of options features on OVI predictability}
\label{section:greeks}

 In the previous section, we showed how the OVI can be used as a predictor for next day returns. In this section, we investigate which options carry most of the predictability, by segmenting options according to the value of certain option features, namely a subset of the ``Greeks'': \textit{moneyness}, \textit{maturity}, and \textit{implied volatility}; we calculate all of these features in a standard Black--Scholes model without dividends. One might naturally expect option moneyness, being a measure of the relative sizes of the spot and strike prices, to influence OVI predictability. Indeed, \cite{Pan_Poteshman} and \cite{GE2016601} both find that the highest predictability over comparatively long timescales  stems from volumes of OTM options, which is attributed to the relatively high leverage that they offer speculative traders targeting (or hedging) large price moves over these timescales. Here, however, we consider next-day returns, and we expect ATM options returns to be more informative, as they offer a more geared return from small price moves than OTM options. On the other hand, ITM options' intrinic value, results in a higher price, hindering the high leverage advantage offered by the option market, and resulting in relatively low volumes. 


Note that there is not a one-to-one map between moneyness and other Greeks; while the Greeks are related in various ways, different Greeks can serve to segment the universe of options in different ways. For example, $\Gamma$ is highly correlated with the absolute value of moneyness, while $\Delta$ depends monotonically on moneyness. As we see below, this enables us to consider a range of perspectives.

For each day, we divide the option contracts into four disjoint buckets, based on the quantiles of the feature in question. The first bucket contains options corresponding to the lowest 25\% of values for that feature, and so on up to the $4^{th}$ bucket, which contains options corresponding to the highest 25\%. The OVI pipeline described in the previous sections is then applied to each bucket separately, and the performance of the resulting buckets is compared. For each feature, we run the pairwise SR difference test between the four buckets (at the 5\% level). We only follow this process for Brokers, Market Makers and Customers, as they are the MPCs whose OVI was concluded to be a significant predictor of future returns.

Below we show some interesting results based on the $\Delta$, as well as the implied volatility of options, calculated (for simplicity and ease of computation) using the Black--Scholes (BS) model\footnote{Note that since equity options are mostly American, the Black--Scholes Model may systematically undervalue them. However, as these differences are typically of the order of 1--2\%, we accept the inaccuracy and avoid large-scale computations.} for European Options without dividends \cite{BS}. We also provide a short discussion on moneyness and maturity, to interpret the results. 


Under the BS model, a call option has a price $ P^{\textrm{Call}}$ equal to 
\begin{equation}\label{call-option}
    P^{\textrm{Call}}(S,K,\tau,r,\sigma) =   S e^{-r \tau}  \Phi \left( d_1 \right) - K e^{-r \tau} \Phi \left( d_2 \right), 
\end{equation}
and a put option has a price $ P^{\textrm{Put}}$ equal to
\begin{equation}\label{put-option} 
    P^{\textrm{Put}}(S,K,\tau,r,\sigma) =  K e^{-r \tau} \Phi \left( -d_2 \right) -S e^{-r \tau}  \Phi \left( -d_1 \right),
\end{equation}
where

\vspace{-6mm}
\[ d_1(S,K,\tau,r,\sigma) = \frac{ \textrm{log} \left( S / K \right) + \left( r + \sigma^2/2  \right) \tau }{\sigma \sqrt{\tau}} \textrm{  and} \] 

\vspace{-5mm}
\[ d_2(S,K,\tau,r,\sigma)  = d_1(K,S,\tau,r,\sigma) - \sigma \sqrt{\tau}. \] 
\vspace{-4mm}

Here $S$ refers to the current price of the underlying asset (in our case, we use the average of close and open), $r$ refers to the risk free interest rate, $\tau$ to the time until maturity, $K$ to the strike price, $\sigma$ the unobserved volatility of the underlying and $\Phi$ the cumulative distribution function of a normal distribution. Lastly, $P^{\textrm{Call}}$ / $P^{\textrm{Put}}$ denote the average of the opening and closing prices. The implied volatility $\sigma_I$, is the value of $\sigma$ that solves \eqref{call-option} / \eqref{put-option} (with $P^{\textrm{Call}}$ / $P^{\textrm{Put}}$ being the average of the opening and closing prices), and can be obtained using a simple iterative method. Importantly, note that all these quantities describe an option from the buyer-perspective (e.g all call options will have a positive $\Delta$).

\begin{figure}[h]
   \centering
   \includegraphics[width=1\textwidth]{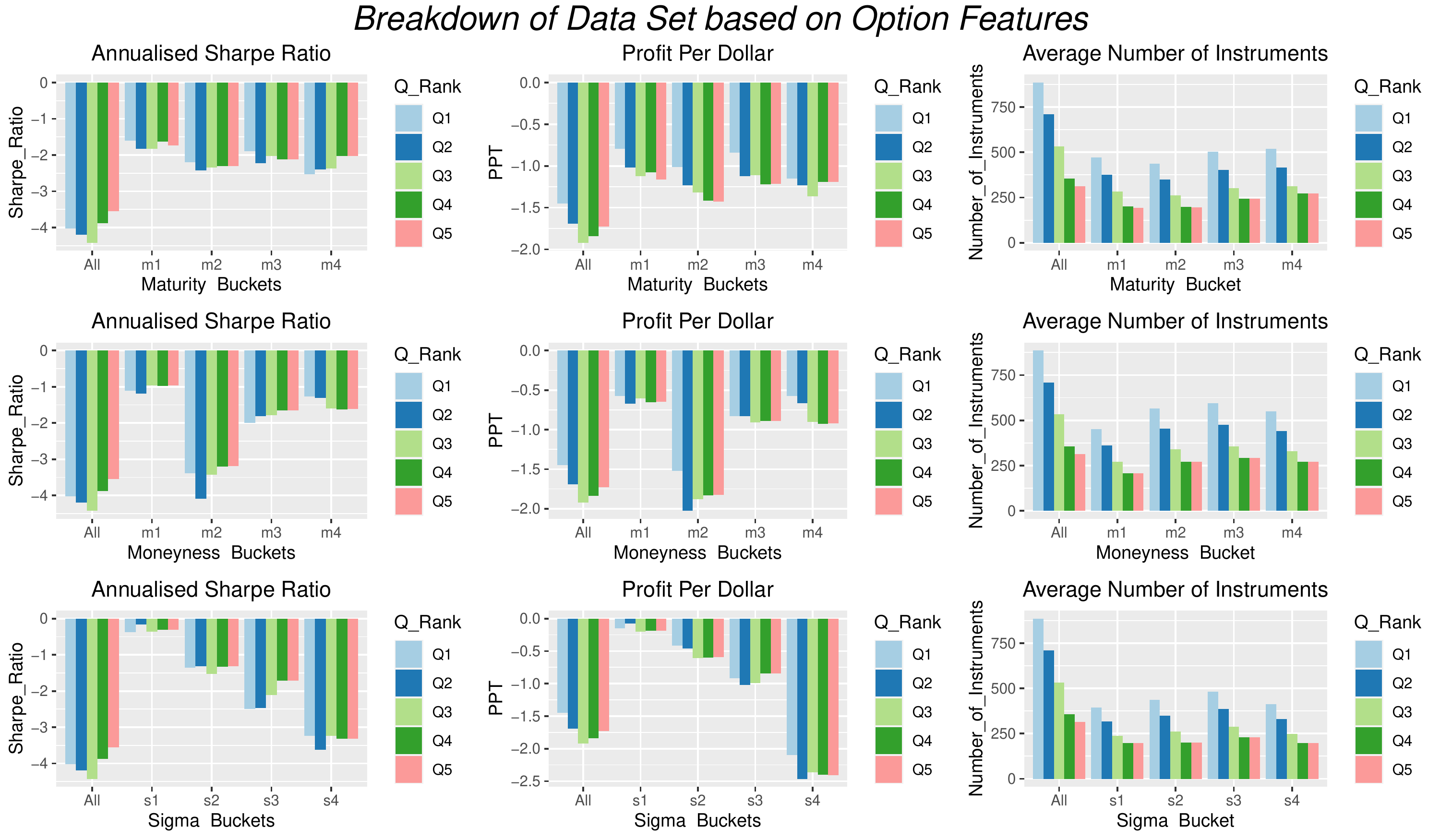}
   \vspace{-2mm}
   \caption{Summary statistics from the breakdown of the PHOTO option data set for Market Makers into four buckets based on the value of some option feature. For each feature, the first bucket corresponds to option contracts whose features have the lowest values, whereas the last bucket corresponds to option contracts whose features have the highest values. The label ``All'' corresponds to using the complete data set, without conditioning on the values of a Greek. Quantile Groups are defined as in \Cref{section:evaluation}, while buckets are based on the quantiles of the Greek corresponding to each option contract. The sub-figures for each Greek correspond to the Annualized Sharpe, Profit Per Dollar, and Average Number of Insturment used in the portfolio.}
   \label{fig:features_mm}
\end{figure}

\begin{figure}[h]
   \centering
   \includegraphics[width=0.9\textwidth]{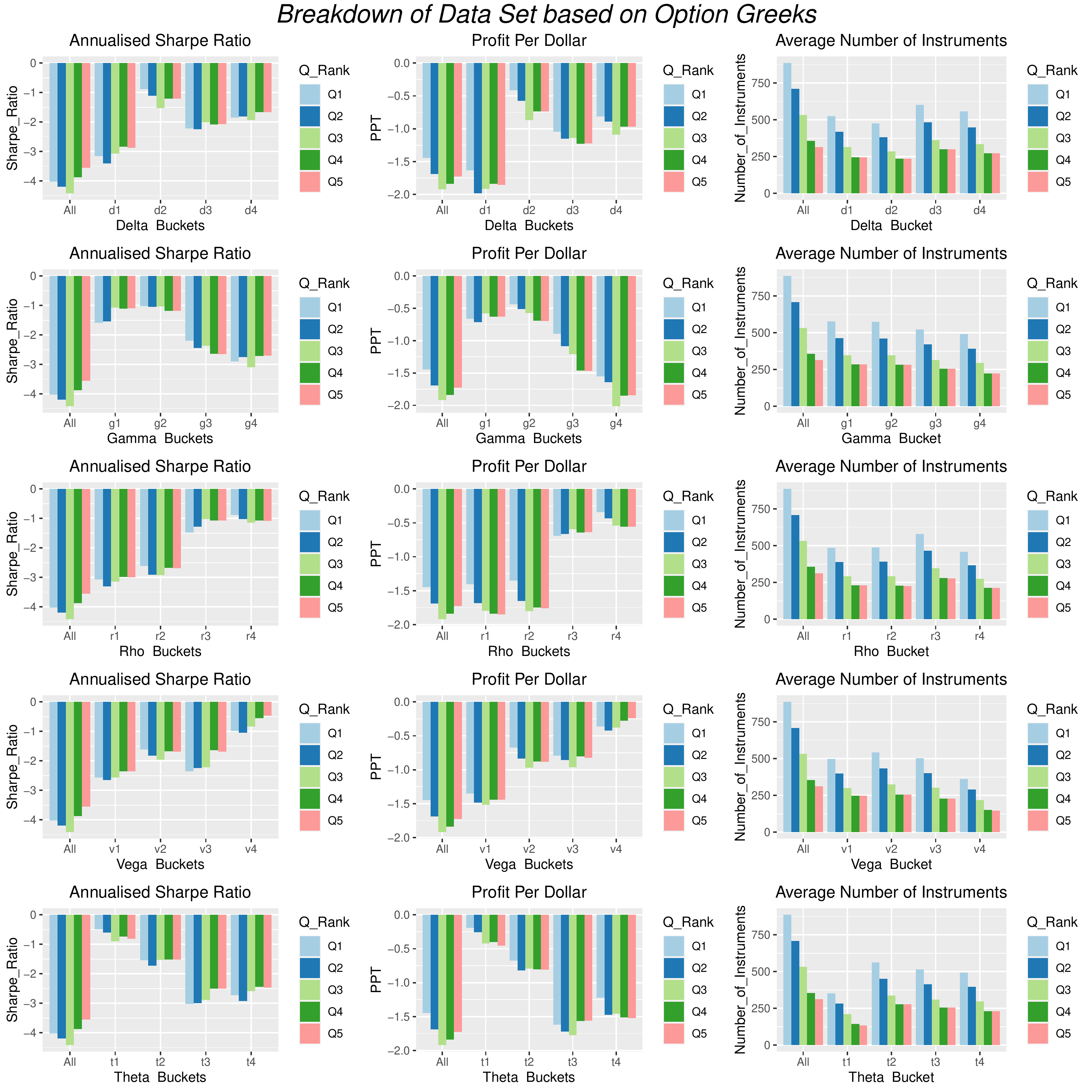}
   \vspace{-4mm}
   \caption{Summary statistics from the breakdown of the PHOTO option data set for Market Makers into four buckets based on the value of some Greek. For each feature, the first bucket corresponds to option contracts whose features have the lowest values, whereas the last bucket corresponds to option contracts whose features have the highest values. The label 'All' corresponds to using the complete data set, without conditioning on the values of a Greek. Quantile Groups are defined as in \Cref{section:evaluation}, while buckets are based on the quantiles of the Greek corresponding to each option contract. The sub-figures for each Greek correspond to the Annualized Sharpe, Profit Per Dollar, and Average Number of Instruments used in the portfolio.}
   \label{fig:buckets_mm}
\end{figure}

In \Cref{fig:features_mm} and \Cref{fig:buckets_mm} we plot, for Market Makers, the results for the option Maturity~$\tau$, standardised Moneyness defined as~\[\textrm{Mon} = \left\{
\begin{array}{ll}
      +\log \left(S  e^{r \tau}/K \right)/(\sigma  \sqrt{\tau}) & \textrm{, for call options} \\
      -\log \left( S  e^{r \tau}/K \right)/(\sigma  \sqrt{\tau})& \textrm{, for put options,}\\ 
\end{array}
\right. \] and implied volatility~$\sigma_I$, as well as the five main Greeks (Delta, $\Delta = \partial P/\partial S$; Gamma, $\Gamma = \partial^2 P/\partial S^2$; Theta, $\Theta = \partial P/\partial t$; Rho, $\rho=\partial P/\partial r$; and Vega, $\mathcal{V}=\partial P/\partial \sigma$).  

\begin{figure}[h]
   \centering
   \includegraphics[width=0.9\textwidth]{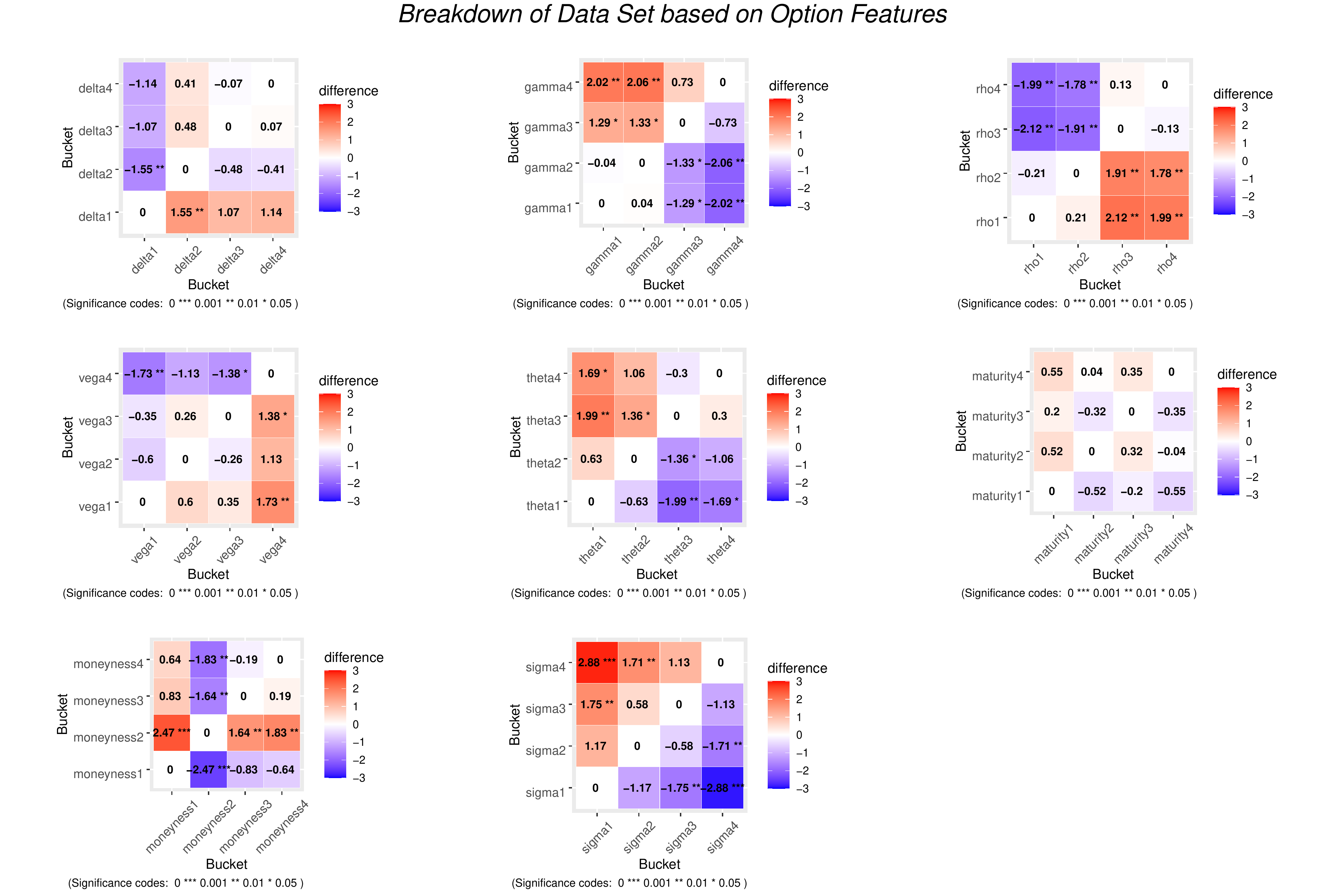}
   \caption{This figure shows the pairwise SR difference tests. In each of the grids depicted in the above figure, the results of running the pairwise SR difference tests for different buckets are shown. The difference shown in each cell is the difference between the row and column bucket and is colored red (resp. blue) if the row bucket has a higher (resp. lower) SR than the column bucket. The significance level is also given by the number of stars (1-3), if applicable.}
   \label{fig:buckets_diff}
\end{figure}

\paragraph{Moneyness, Gamma and Theta}
Consistently with our expectation noted above, ATM options show the highest predictability. For example, across MPCs, higher Gamma options/higher Gamma buckets (corresponding to ATM options) tend to perform better. 
For Market Makers, the $4^{th}$ Gamma bucket significantly outperforms the others, clearly indicating how higher $\Gamma$ is linked to higher predictability. Likewise, higher $\Theta$ (also closer to being ATM) have the highest performance, with the $3^{rd}$ and $4^{th}$ buckets significantly outperforming the first two ones. Results are not as clear with moneyness buckets, with no consistent pattern followed by the buckets, except that the $1^{st}$ bucket is outperformed by the rest, hinting that deep OTM options have the lowest informational content\footnote{Replicating this analysis with five buckets still fails to produce any significant results. This is mainly because the five buckets do not exaclty correspond to deep OTM, OTM, ATM, ITM and deep ITM, as the moneyness distribution is skewed towards OTM options.}. 


\paragraph{(Non) Effect of Maturity ($\tau$)}
One could expect maturity to play an important role in the predictability, and as is often done in other works, one might be tempted to filter options with very high maturities. In particular, similar to the moneyness case, one could expect that higher periods to maturity would result in a poor informational content. However, no consistent pattern can be observed across MPCs for the different maturity buckets, indicating that maturity has no effect on predictability. Thus, there is no need to filter option contracts with high or low maturity as is done in other works, when calculating the OVI. 

\paragraph{Difference between call and put options}
By definition, call options have a positive $\Delta$, while put options have a negative $\Delta$. In addition, as the future value of assets rises with interest rates, call options also have a positive $\rho$, while put options have negative $\rho$.

We can clearly observe that put options play a more important role in predicting future returns compared to call options. Across all MPCs, the first bucket for Delta (consisting mostly of ITM put options) outperforms the other buckets (consisting mostly of call and OTM put options). For Market Makers and Customers, the $1^{st}$ bucket significantly outperforms the other buckets. In addition, the first two buckets for $\rho$ (corresponding to the Put Options) are the best performing, with the difference being statistically significant for all MPCs. 

\paragraph{Implied Volatility ($\sigma$) and Vega ($\mathcal{V}$)}
For all of Brokers, Market Makers, and Customers, higher implied volatility results in higher predictability. In particular, the $4^{th}$ bucket is the best performing bucket across all MPCs, and significantly outperforms all other buckets except the $3^{rd}$ bucket for Brokers. The $4^{th}$ bucket also outperforms the complete data set, in terms of PPD for Brokers and Market Makers, as well as in terms of the SR for Brokers. The $3^{rd}$ bucket is also consistently the second best performing, significantly outperforming the first two buckets for Market Makers, as well as the $1^{st}$ bucket for Brokers. Overall, these results clearly indicate that most of the predictability stems from high implied volatility contracts. 

Given that this feature has the most significant and consistent effect on the predictability of OVI, we consider an additional betting weighting scheme. Similarly to the volume weighted betting scheme, we consider a strategy based on the product's volume, multiplied by the implied volatility of the asset. We set the IV-Volume Weighting Scheme as

\vspace{-4mm}
\[ b'_{i,d}= \sum_{i=1}^{N_d} \sum_{j=1}^{C_{i,d}} \sum_{m} \log \left( 1 + \min \{ \sigma_{i,j,d},2 \} \right) V_{i,j,d,m}, \] 
\vspace{-4mm}

\noindent
where $\sigma_{i,j,d}$ is the implied volatility corresponding to option contract $j$ of asset $i$ on day $d$. As can be seen from \Cref{fig:betting_strategies_cumsum}, this scheme results in a higher PPD than all the other schemes, even though its SR is lower than most schemes with the exception of the volume scheme. Therefore, similarly to the Relative Volume Weighting Scheme, this is a more profitable yet less consistent scheme than the Uniform Scheme.

For Vega, no consistent pattern can be observed across MPCs. Thus, while implied volatility plays a big role in determining the informational content of an option's volume, the sensitivity of options to implied volatility does not.

\FloatBarrier 

\section{Extensions}
\label{section:intraday}
In this section, we build on the results of \Cref{section:MarketMakers}, by performing further tests for more general scenarios. These include checking the price persistence for longer holding periods, comparing the PHOTO and NOTO data sets, and a short intraday and cross-sectional analysis.

\subsection{Short-term versus long-term price persistence}
\begin{figure}[h]
\centering
\includegraphics[width=0.8\textwidth]{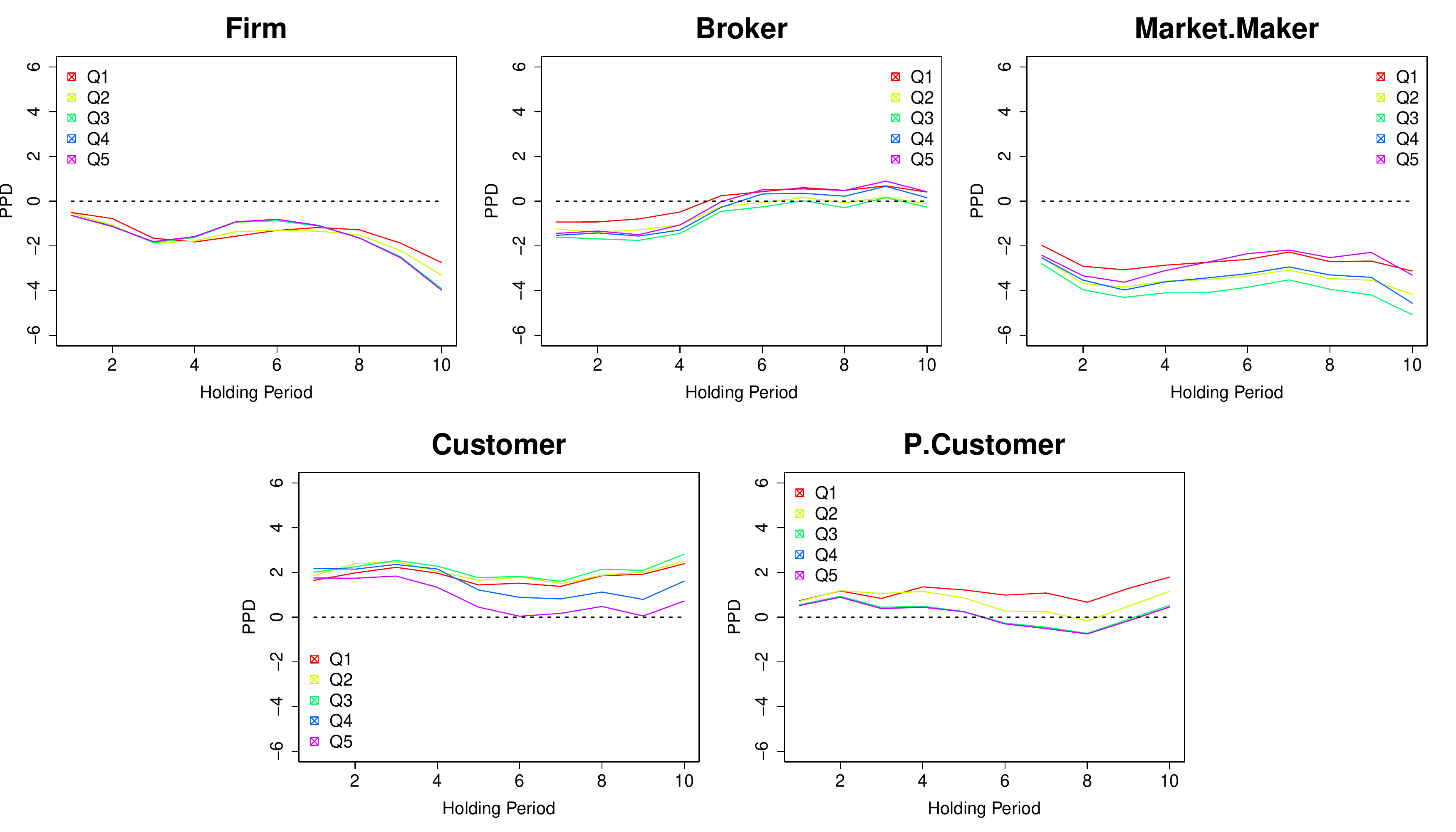}
\caption{Return persistence over longer holding periods:  
This figure shows the cumulative returns obtained for different holding periods ranging between 1 and 10 days. Results are shown for each MPC, when the corresponding OVI strategy is followed, for each QR Group and the Uniform Weighting Scheme.} 
\label{fig:holding_period}
\end{figure}
Previously, we observed that for Market Makers, Customers and Brokers, a strategy based on the OVI with a holding period of one day, results in either significant profits or losses. In a similar fashion to \cite{Pan_Poteshman}, we test how the results change for different holding periods. For each strategy, we record the cumulative returns obtained when using a longer holding period. Therefore, for a holding period of $h$ days we calculate the P\&L as
\begin{equation} \label{PnL}
\textrm{P}\&\textrm{L}_{d}^h = \sum_{i=1}^{N_d} b_{i,d} \times \left( \sum_{k=1}^{h}\widetilde{f}^{\textrm{CL}\_\textrm{tmCL}}_{i,d+(k-1)} \right) \times \textrm{sign} \left(  s_{i,d} \right),
\end{equation} 
and plot the cumulative returns in \Cref{fig:holding_period}, for holding periods of up to 10 days.

For Market Makers, aside from a significant SR displayed during the first day, the SR corresponding to the (non-cumulative) returns of the second day after execution are also statistically significant (with a p-value of 0.00). The cumulative returns do not revert over the following days, clearly demonstrating a persistent return. 

For the rest of the MPCs, only the first day (non-cumulative) returns are significant. There is a long lasting effect shown for Customers' Q4 and Q5, where cumulative returns do not significantly decrease after the first day, not displaying any mean reversion. For Brokers, this is not the case, as the cumulative returns revert signs from around the fifth day onward. The other two MPCs do not produce significant results for the first day.

Overall, as the returns for Market Maker OVI do not revert, the predictive power does not seem to stem from price pressure, but is potentially due to a degree of informed trading, confirming the observations of \cite{Pan_Poteshman}.

\subsection{Difference between Buys and Sells}
\begin{figure}[h]
\centering
\includegraphics[width=0.9\textwidth,trim=0cm 1.3cm 0cm 1.3cm,clip]{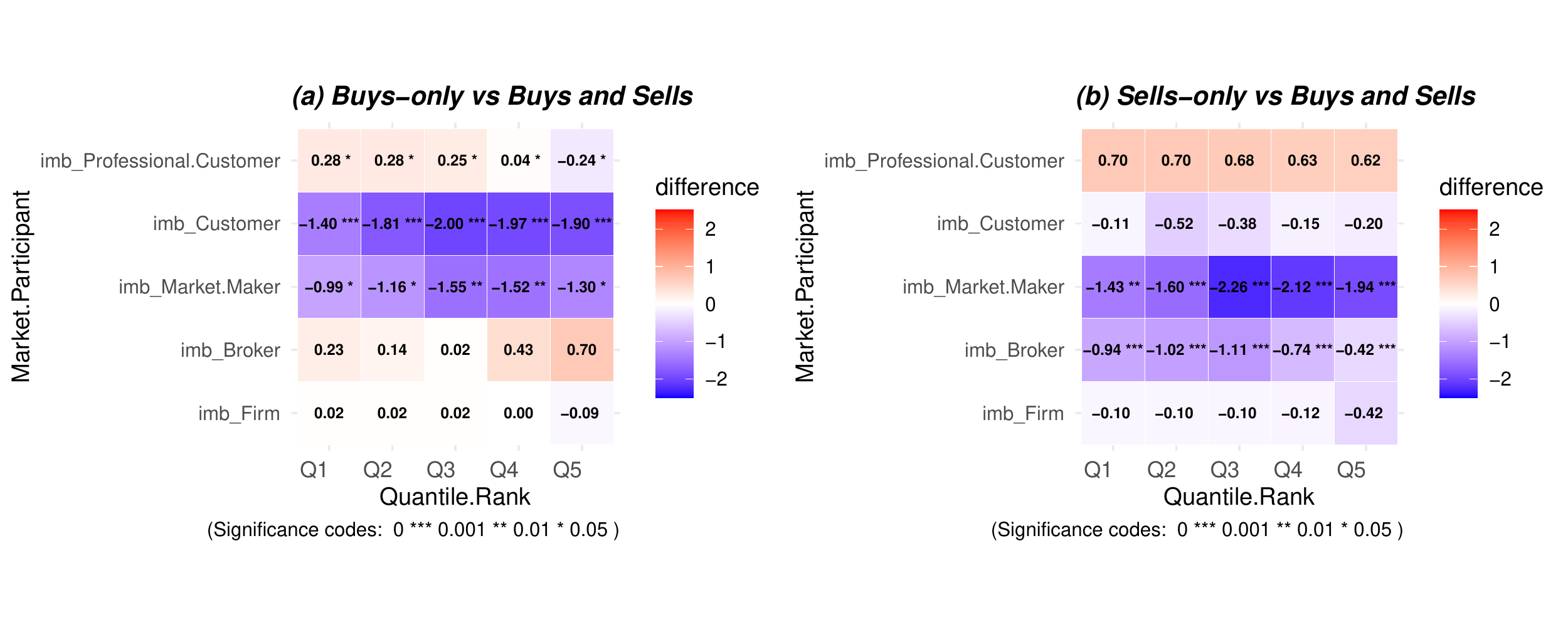}
\caption{SR Difference Tests for Buy vs. Sells: The left (resp. right) figure shows the difference in SRs obtained between using only the buys (resp. sells) vs using the complete data set. Blue cells denote cases where using only the buys (resp. sells) is outperformed by the complete data set, and red when it outperforms the complete data set instead. The significance level is also shown after the numeric value, by 1-3 stars, if applicable.)} 
\label{fig:buy_sells}
\end{figure}

Recalling the definition of OVI \eqref{eq:OVI}, the terms $V^{\textrm{Up}}_{i,j,d,m}$ and  $V^{\textrm{Down}}_{i,j,d,m}$ are defined as using the volumes from both the buying and selling activities of a MPC. In order to gain a better understanding of which type of transaction is the major source of predictive power, we calculate the OVI restricted to only the ``buy'' transactions (resp. sells) denoted as $\textrm{OVI}_{i,d,m}^{\textrm{Buy}}$ (resp. $\textrm{OVI}_{i,d,m}^{\textrm{Sell}}$). 
We compare both of these with the standard volume-based OVI feature ($\textrm{OVI}_{i,d,m}$), 
by using the SR difference test. As can be seen from \Cref{fig:buy_sells}, $\textrm{OVI}_{i,d,\textrm{Customer}}^{\textrm{Buy}}$ performs significantly worse than $\textrm{OVI}_{i,d,\textrm{Customer}}$ across QR Groups, with its OVI being almost uninformative. $\textrm{OVI}_{i,d,\textrm{Customer}}^{\textrm{Sell}}$ also has a lower SR than $\textrm{OVI}_{i,d,\textrm{Customer}}$, but the difference is not statistically significant. This implies that for Customers most of the predictability comes from the selling activity. The opposite is true for Brokers, where $\textrm{OVI}_{i,d,\textrm{Broker}}^{\textrm{Sell}}$ performs significantly worse than $\textrm{OVI}_{i,d,\textrm{Broker}}$, but $\textrm{OVI}_{i,d,\textrm{Broker}}^{\textrm{Buy}} $ has a slightly higher performance than $\textrm{OVI}_{i,d,\textrm{Broker}}$. When it comes to Market Makers, both differences are significant across all quantile rank groups. Recall that the differences in volume for buying and selling activity was small for all of the MPCs (see \Cref{fig:flow_breakdown}), therefore the above results could indicate that Customers utilize their speculative ability (or possibly inside information) mostly by writing option contracts.  

\subsection{Additional features from the NOTO data set} 
\begin{figure}[h]
\centering
\includegraphics[page=1,width=1\textwidth]{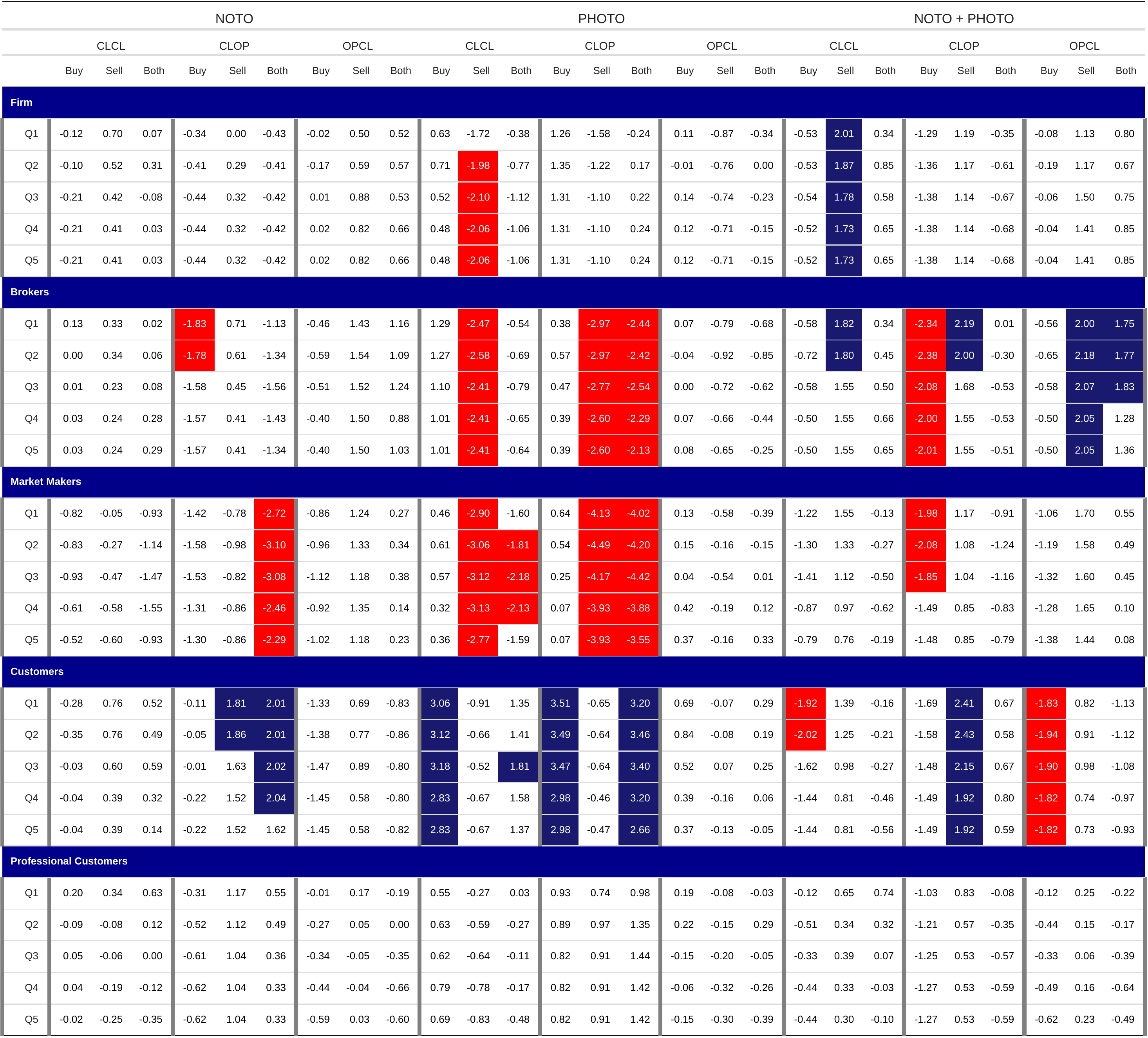}
\caption{Overview of predictability across data sets: This table shows the SRs obtained for each OVI feature. The columns are grouped by the data set, where PHOTO corresponds to the PHLX and NOTO to the NOM option exchange. At a lower level, the columns are grouped by the return type used for their evaluation, and further down by the type of transactions considered (Only Buys / Only Sells / Both). The rows are broken down at the level of MPC and then finally by the QR Group. Cells colored red (respectively blue) when they are significantly negative (respectively positive), at the Bonferonni Corrected level. } 
\label{fig:noto}
\end{figure}

We have so far considered only the PHOTO data set corresponding to the PHLX option exchange. We now consider the NOTO data set, which corresponds to the NASDAQ exchange. In \Cref{fig:noto}, we show a complete overview of the SR corresponding to the three types of returns for the five MPCs, across quantile rank groups, using either buys, sells or both for the calculation of the OVI feature. Furthermore, we show these values when using NOTO, PHOTO or the union of the two data sets. About half of the values in the table have p-values less than 0.05 for the SR significance. In addition, the colored cells denote the cells which have a significant SR when Bonferonni corrected across all tests, i.e using a p-value of $0.05/(5 \times 5 \times 3 \times 3 \times 3) \approx 7.4 \times 10^{-5}$. 

Similarly to the PHOTO data set, the highest performance corresponds to the CL\_tmOP future returns for Customers and Market Makers, while Brokers also have a significantly good performance. Another similarity is that the middle quantile rank groups (Q2-Q4) tend to outperform Q1 and Q5. In the cases of statistical significance, CL\_tmOP and CL\_tmCL future returns have identical signs, with the CL\_tmOP returns consistently scoring higher SRs. For the NOTO data set, tmOP\_tmCL returns display higher SRs than for PHOTO, with some of them being significant (when ignoring the Bonferonni correction) such as Broker sells or Customers buys. Note that in both of these cases, the sign of SR is in the opposite direction of what is displayed by the CL\_tmOP returns, showing a price reversal effect between the overnight and tmOP\_tmCL returns, which is not as strongly present in the PHOTO data set. In fact, one can further notice that no result corresponding to CL\_tmCL returns is significant for the NOTO data set. This seems to indicate a sharper price reversal present in the NOTO data set. Coupled with the the highest SRs scored by the PHOTO data set, this could be indicating that a higher proportion of traders in the PHLX exchange are informed when compared to the NOM exchange, where most of the predictability stems from short term price pressure. 

Due to the higher median daily volume displayed for NOTO (as seen in \Cref{fig:breakdown}), the combined OVI's SR takes the same sign as the SR of the corresponding NOTO value in almost all cases. When a combined OVI is significant, we observe a weak result when using PHOTO's OVIs and a strong result using NOTO's OVIs (that is weaker than the result of using the combined OVI). An interesting example is when using the selling transactions for Firms corresponding to the CL\_tmCL returns, where there is a significantly positive SR for the PHOTO data set, but a significantly negative SR when using the combined data set. 

Most of the predictability comes either from the buying or selling transactions, and in almost no cases do they both appear to result in a significant SR. The one exception is the NOTO OVI of Market Makers for CL\_tmOP returns, where both $\textrm{OVI}_{i,d,\textrm{M. Maker}}^{\textrm{Buy,NOTO}}$ and $\textrm{OVI}_{i,d,\textrm{M. Maker}}^{\textrm{Sell,NOTO}}$ are negative, resulting in a statistically significant performance for $\textrm{OVI}_{i,d,\textrm{M. Maker}}^{\textrm{NOTO}}$. Therefore, using only Buys or Sells outperforms the OVI obtained by using the complete data set in almost all cases. 
\begin{figure}[h]
\centering
\includegraphics[width=1\textwidth]{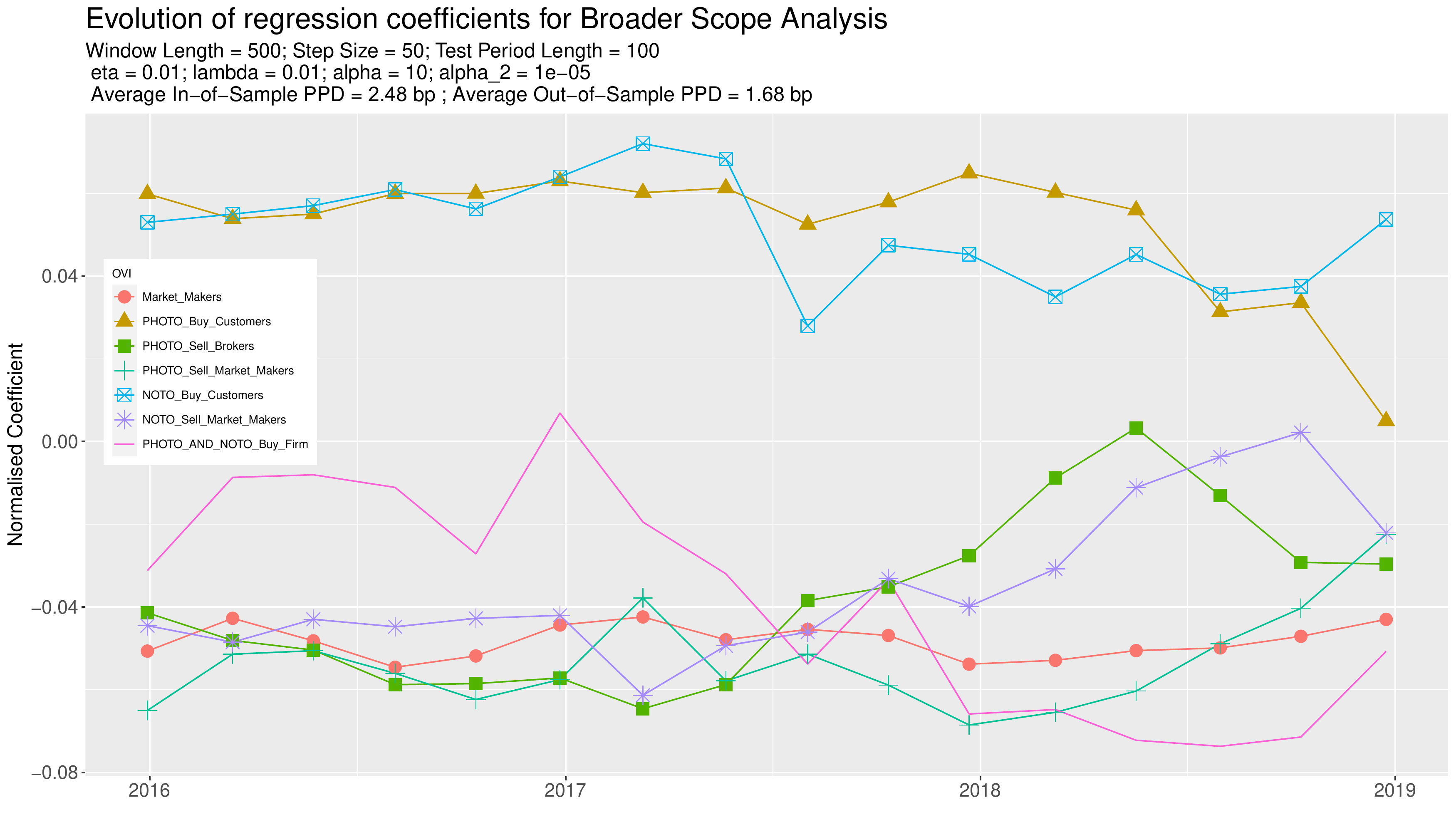}
\caption{Multiple Factors Regression for Broader Scope Analysis: The line graph shows the evolution of the coefficients fitted for model \eqref{eq:model3} using the P\&L regression for QR Bucket 1, across the different windows. The points are shown at the midpoint of their corresponding window. Note that coefficients are normalized to have a total $l_1$ norm equal to 1 (excluding the intercept term). This figure only shows the coefficients with the highest average absolute values, with most of the other factors not being consistently positive or negative.}
\label{fig:regression_coefficients_forty_five}
\end{figure}

Taking into account the plethora of different OVI features, we consider another model, in order to test whether these 45 features can be used to build a better signal (for overnight returns)
\begin{equation}
\label{eq:model3}
        \textrm{M3}: s^4_{i,d} = g \left( \beta_0 + \sum_{M \in \textrm{MPCs}} \sum_{S \in \{\textrm{Buy},\textrm{Sell},\textrm{Both} \}} \sum_{D \in \{ \textrm{NOTO}, \textrm{PHOTO}, \textrm{Both} \} }  \beta_{M,S,D} \textrm{OVI}^{S,D}_{i,d,M}  \right),
\end{equation}
where $\textrm{OVI}^{S,D}_{i,d,m}$ is the OVI corresponding to the $i^{th}$ stock for the $d^{\textrm{th}}$ day, for MPC $M$, data set $D$ and type of transaction $S$ of the data set (Buys, Sells or Both). In \Cref{fig:regression_coefficients_forty_five}, we plot the evolution of some of the coefficients (specifically the ones with the highest median magnitude across time). Note that M3 raises the overall PPD compared to M1-M2, with 2.48bp in-sample and 1.68 out-of-sample mean PPD, which signifies that the predictive power can be increased by considering the NOTO data set and decomposing the data set into Buys and Sells.

Looking at the significant coefficients, we draw the following conclusions
\begin{itemize}
\item $\textrm{OVI}_{i,d,\textrm{M.Maker}}$ gives a consistently negative coefficient (similar to M2), and the same holds true for $\textrm{OVI}^{\textrm{Sell}}_{i,d,\textrm{M. Maker}}$ and $\textrm{OVI}^{\textrm{Sell,NOTO} \cup \textrm{PHOTO}}_{i,d,\textrm{M.Maker}}$.
\item $\textrm{OVI}^{\textrm{Sell}}_{i,d,\textrm{Broker}}$ starts with a significantly positive coefficient, which falls close to zero right after 2018, which is similar to the case of $\textrm{OVI}_{i,d,\textrm{Broker}}$ for M2.
\item $\textrm{OVI}^{\textrm{NOTO,Buy}}_{i,d,\textrm{Customer}}$ yields consistently positive coefficients, similar to  $\textrm{OVI}_{i,d,\textrm{Customer}}$ for M2. 
\item $\textrm{OVI}^{\textrm{Buy}}_{i,d,\textrm{Customer}}$ is also consistently negative, but its coefficient shrinks for latter years similar to $\textrm{OVI}^{\textrm{Sell}}_{i,d,\textrm{Broker}}$.
\item $\textrm{OVI}^{\textrm{NOTO} \cup \textrm{PHOTO, Buy}}_{i,d,\textrm{Firm}}$ has a fluctuating coefficient for the early years, but a positive one for the latter years, reflecting the pattern observed earlier in  \Cref{fig:cumulative_PnL_overnight}. 
\end{itemize}

\FloatBarrier 
\subsection{Cross-Impact analysis}
\label{section:networks}

\begin{figure}[h]
   \includegraphics[width=1\textwidth]{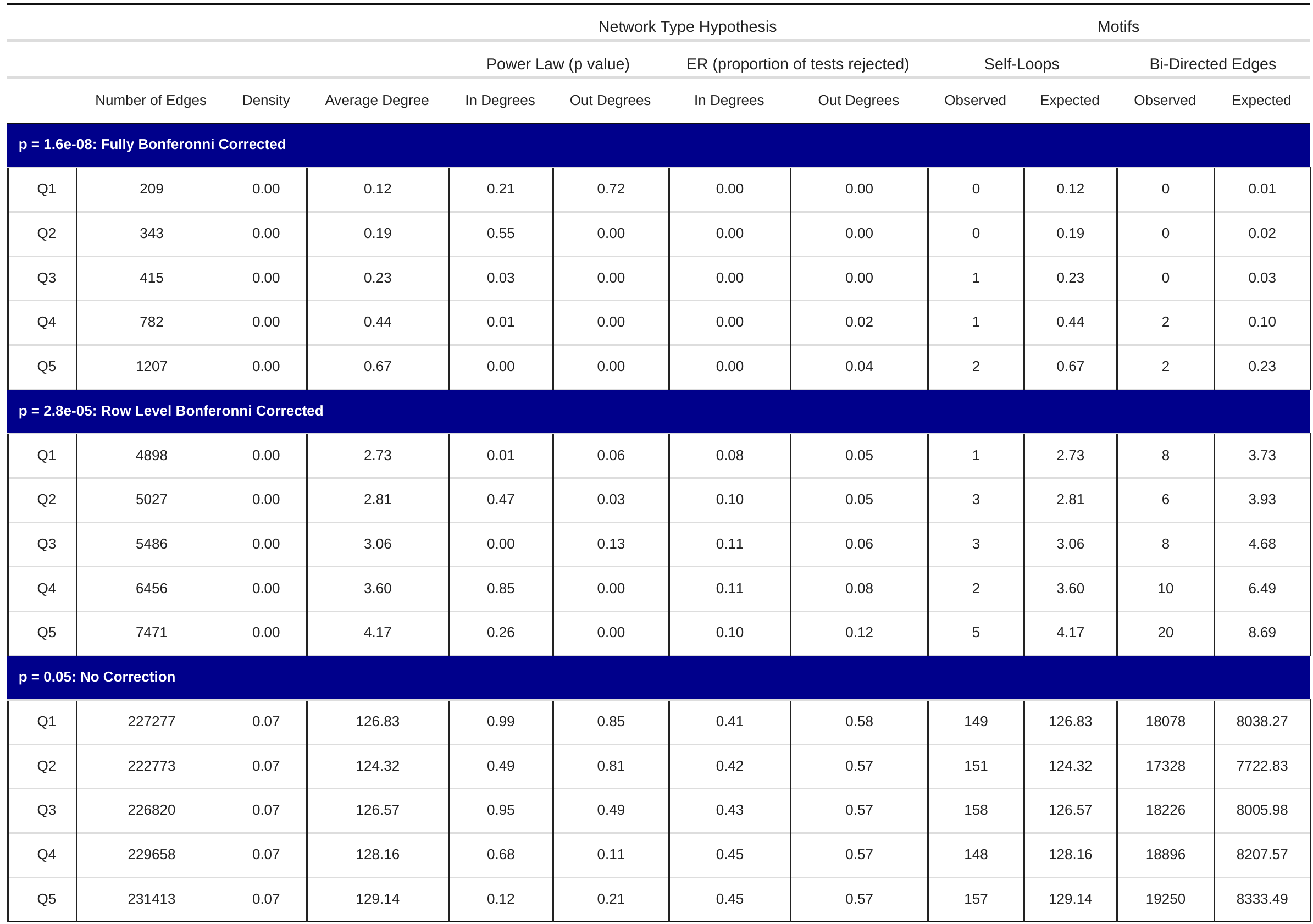}
   \caption{Summary Information for Cross Impact Networks: The above table shows information on 15 cross impact networks based on 3 different p-values (no Bonferonni Correction, row-level Bonferonni Corrected and fully Bonferonni Corrected) as well as five quantile rank groups. The summary information includes the in and out degrees, the p-values based on the null hypothesis that the networks' degrees follow a power-law distribution, and the proportion of tests rejected for pairwise differences between the distributions of the nodes assuming binomial distributions (using bonferonni correction). We also give the number of self-loops and bi-directed edges present in each network as well as the number of each of these motifs that would be expected in a Erdos-Renyi network with the same density.}
   \label{fig:summary_info}
\end{figure}

We perform a cross-sectional analysis, where the OVIs of all the assets are tested for predictive power against the future returns of all the underlying equities. With $N$ stocks in the universe, there are $N^2$ relationships to be tested. These interactions can be represented as an unweighted directed network, where the existence of an out-edge from vertex $i$ to vertex $j$ indicates that the OVI of asset $i$ can be used as a predictor for the equity returns of asset $j$. 

Our approach is similar to the one employing Bonferonni networks (\cite{tum2011}). Such a network was constructed by \cite{curme2014emergence}, to investigate the intraday return correlations across stocks. Instead of correlations, we employ the cumulative P\&L obtained by following a simple strategy of using the Market Maker OVI of asset $i$ as a predictor for the overnight (CL\_tmOP) returns of another asset $j$. An edge is added to the network if the SR differs significantly from 0, according to the \cite{bailey2014} hypothesis test. 

In particular, let $G$ denote the adjacency matrix of the network. For each potential edge $(i,j)$, we iterate across days and trade equity $j=\{1,\ldots,N\}$, for days where asset $i$'s OVI belongs to the Q3 portfolio (top 60\% of the highest OVI's in magnitude). In that case, we buy asset $j$, if the OVI of asset $i$ is positive, and sell if it is negative. Let $Q(i,d,q) = 1$, if the $i^{th}$ asset is in the $q^{th}$ quantile group for day $i$ and 0 otherwise. Then, the daily return for following this simple strategy is equal to $\textrm{PnL}_{i,j,d} = \mathbbm{1}_{Q(i,d,q)} \textrm{sign}(s_{i,d}) \tilde{f}_{j,d}$. We perform the SR significance test on $\{ \textrm{PnL}_{i,j,d} \}_{d=1}^D$, and set $G_{i,j} = 1$, if the obtained p-value is less than a predetermined value $p$ (and $G_{i,j} = 0$ otherwise). By varying $q = 1,2,\dots,5$, we obtain five different networks for each threshold. 

To account for multiple testing, we use the conservative method of a Bonferonni-corrected test by choosing a uniform significance level $\alpha_1 = p/ (\textrm{Number of tests})$ for all individual tests, giving a bound for the Family-wise error rate (FWER) $\le p$ (see for example Appendix A of \cite{curme2014emergence}). We use three different types of p-values corresponding to 0.05, $0.05/N$ (applying the correction to the row-level), and $0.05/N^2$ (applying a full Bonferonni correction). Depending on the p-value and quantile group, we build a different network, resulting in 15 different networks. 

For this analysis, we only use assets listed in CRSP for the whole duration from 01 January 2015-31 December 2019, leaving a reduced universe of 1792 NYSE assets. In \Cref{fig:summary_info}, we summarize the information regarding each of these networks. Obviously, as the p-value grows, the network also grows, with the networks corresponding to a p-value of 0.05 having a density of 7\%. For higher QR groups, the network is larger, with Q5 resulting in the largest networks. The fully Bonferonni-corrected networks have a total number of edges between 209 and 1207, which clearly rejects any hypothesis of no cross-impact effects.
 
Furthermore, out of the 1207 edges of the Q5 network, only two are \textit{self-loops} (i.e using an asset's own OVI feature, to predict its future returns). This is higher than what would be expected under an Erd\H{o}s-R\'{e}nyi network with the same density (which is 0.67 here). \textit{Bi-directed} edges (i.e OVI of two different assets, being good predictors of each other's future returns) are also quite rare, with only two such edges, which is, again, much higher than the expected value of 0.23.

Lastly, we test whether this network is structurally similar to any of the main types of networks usually observed in practice. The degrees of many real-world networks follow a \textit{power-law} distribution; to this end, we employ the test of \cite{clauset_powerlaw}, to test the null hypothesis that in/out-degrees follow a power-law. As shown in \Cref{fig:summary_info}, the network degrees do not appear to follow such a power-law distribution. In particular, five of the networks reject the null hypothesis that in-degrees follow a power-law distribution, while six of the networks reject the equivalent hypothesis for out-degrees. 

Let $d^{\textrm{IN}}_j = \sum_{i=1}^N G_{i,j}$, and $d^{\textrm{OUT}}_j = \sum_{i=1}^N G_{j,i}$, denote the in- and out-degrees of node $j$, respectively.  Furthermore, let 

\vspace{-4mm}
\begin{equation*}
\hat{p}^{\textrm{IN}}_j = \frac{\sum_{l=1}^N G_{l,j}}{N}, \ \ \textrm{and} \ \ \ \hat{p}^{\textrm{OUT}}_j = \frac{\sum_{l=1}^N G_{j,l}}{N},
\end{equation*}
\vspace{-4mm}

\noindent denote the empirical edge density corresponding to node $j$. %
Another question is whether the differences between network degrees are statistically significant. In particular, assume that a node's in- and out-degrees follow a binomial distribution, i.e
\[ d^{\textrm{IN}}_j \sim \textrm{Binomial}(N,p^{\textrm{IN}}_j) \ \ \textrm{and} \ \ d^{\textrm{OUT}}_j \sim \textrm{Binomial}(N,p^{\textrm{OUT}}_j), \forall 1 \le j \le N. \]
Then, we can form the following hypotheses
\[ \textrm{H0: \ \ } p^{\textrm{IN}}_j = p^{\textrm{IN}}_i,  \ \  \forall \ 1 \le i,j \le N, \textrm{\ \ and \ \ }
\widetilde{H0}: \ \  p^{\textrm{OUT}}_j = p^{\textrm{OUT}}_i, \  \ \ \ \forall \ 1 \le i,j \le N. \] 
If the network is a simple Erd\H{o}s-R\'{e}nyi network, then both hypotheses should hold in conjunction. To test whether node $i$ and node $j$ degrees are similar, we use the conservative two-sample binomial t-test, with test statistic

\vspace{-6mm}
\[ 
\textrm{Test}_{i,j} = \frac{\hat{p_i}-\hat{p_j}}{\sqrt{\frac{\hat{p} (1-\hat{p})}{2N}}}. 
\] 
\vspace{-3mm}

Under the null hypothesis, we have $\textrm{Test}_{i,j} \sim N(0,1)$. To account for multiple testing, we once again use a Bonferonni correction by setting the p-value equal to $0.05/N^2$. 

Note that such a test is very conservative, both due to the Bonferonni correction, but also the choice of a t-test for the pairwise difference tests. Indeed, for both in- and out-degrees, smaller networks (fully Bonferonni corrected) do not result in any rejections of the null hypothesis. This is different for the larger networks, as we see a 5-12\% rejection rate for the row-level Bonferonni corrected networks, and a 41-58\% rejection rate for the uncorrected networks. Thus, the networks at hand seem to neither follow a power-law nor form an Erdos-Renyi network, as there are significant differences between node degrees. 

In Section 2 of the Supplementary Material, we provide additional results based on the Q3 row-level Bonferonni corrected network. We show differences in the average degrees of different sectors, and list some of the instruments with the highest centralities, demonstrating a more central role within the network.

\vspace{-2mm}
\section{Conclusion and future research directions}
\label{sec:conclusion}
In this paper, we have defined the OVI feature and showed how it can act as a predictor for future equity returns. Focusing on the PHLX exchange, we compared OVI across MPCs, and found that the Market Maker' OVI consistently provides the highest predictability, yielding annualized Sharpe Ratios of up to 4.5, for a simple betting scheme (without taking into account transaction costs).
In terms of PnL, the tail portfolios corresponding to the strongest signals, can attain up to 4 bpts per day, depending on the sizing scheme employed.   
We have shown that some level of predictability is also present for Customer and Broker OVIs, while no predictability was concluded for Firm Proprietary trades and Professional Customers. We demonstrated how to improve performance, by taking into account the OVI's magnitude. In particular, when using quantile rank groups, we found that the $2^{nd}$-$4^{th}$ quantile rank groups are typically the best performing.

We showed that most of the predictability is present in either Buys or Sells depending on the MPC and the data set used. Decomposing into opening or closing transactions did not significantly affect the observed predictability. However, decomposing transactions based on certain option features significantly affects predictability. In particular, the OVI from high implied volatility contracts is significantly more informative than option contracts with low implied volatility. Other option characteristics that seem to consistently induce higher performance include negative values for Rho, and negative values for Delta (i.e put options). We have also shown that cross-impact effects are present, and in fact, significantly outnumber the cases of direct impact, when applying a conservative Bonferonni Correction. Results obtained for the NOM exchanged (NOTO data set) are consistent with those obtained on the PHLX exchange (PHOTO data set), albeit weaker, with Market Makers still showing a significant annualized SR of around 2.5--3.

Another contribution of this paper was the introduction a new approach to tackling future returns regression, namely the P\&L regression, which was motivated by the needs of this particular research, but may also be of independent interest for other financial studies. This approach, while slightly more intricate than a simple OLS method, is problem-tailored for directional predictions, bypasses the requirement for linear effects, and is able to work under sparse data. We have, for the most part, avoided the usage of linear methods, and focused on the usage of metrics like the SR, providing a general framework for analyzing return prediction tasks.

This research could lead to further studies that examine option volumes from different data sets, and at different time horizons. This research could also be extended to an intraday analysis (given that data comes in batches of 10-minute bucket increments), which we consider in Section 3 of the Supplementary Material. 
This led to promising observations, which we have not been able to quantitatively approach, due to our limited access to intraday data. Another interesting angle would include a study of the interplay between option and stock volumes, and ultimately the prediction of future volumes (for both the option and stock markets), which would be of potential interest for market making and optimal execution.  

\section*{Acknowledgments}
This work was supported by the UK Engineering and Physical Sciences Research Council (EPSRC) grants EP/R513295/1 and EP/V520202/1. We would also like to thank Rama Cont and Stefan Zohren for their valuable and productive feedback.

\bibliography{bib_file}

%% file: input_app_ab.tex

\vspace{-2mm}

\section{Financial Definitions}
\label{appendix:definitions}
The following definitions are from NASDAQ's website \url{https://www.nasdaq.com/glossary}. The types of market participants used in the analysis are: 
\begin{itemize}
\item \textbf{Firm (Proprietary Trades):} \textit{Principal trading in which a firm seeks direct gain rather than commission dollars.}
\item \textbf{Broker:} \textit{An individual who is paid a commission for executing customer orders. Either a floor broker who executes orders on the floor of the exchange, or an upstairs broker who handles retail customers and their orders. Also, person who acts as an intermediary between a buyer and seller, usually charging a commission. A "broker" who specializes in stocks, bonds, commodities, or options acts as an agent and must be registered with the exchange where the securities are traded.}
\item \textbf{Market Makers:} \textit{One who maintains firm bid and offer prices in a given security by standing ready to buy or sell round lots at publicly quoted prices.}
\item \textbf{Customers} \textit{A person or entity that is not a broker or dealer in securities (and is not a Professional Customer as defined below).}
\item \textbf{Professional Customers:} \textit{A person or entity that is not a broker or dealer in securities and places more than 390 orders in listed options on average per day during a calendar month for its own beneficial account(s)}
\end{itemize}

For each market participant class other than Market Makers, the volumes are further decomposed into four types based on transaction intent: Opening Buy, Opening Sell, Closing Buy and Closing Sell, which are defined as follows
\begin{itemize}
\item \textbf{Opening Buy:} \textit{an option where the intention of the purchaser is to create or increase a long position.}
\item \textbf{Opening Sell:} \textit{An option where the intention of the writer is to create or increase a short position.}
\item \textbf{Closing Buy:} \textit{An option where the intention of the purchaser is to reduce or eliminate a short position in a stock.}
\item \textbf{Closing Sell:} \textit{An option where the intention of the writer is to reduce or eliminate a short position in a stock.} 
\end{itemize}

\vspace{-3mm}
\section{Approximation of trade flow between MPCs}
\label{appendix:tradeflow}
In Section 3 of the main text, we construct an approximation of the trade flow between different market participant classes. 

We approximate the trade flow, by inspecting the 10-minute intraday aggregate volume reports, and then interpreting and matching the flow. Within a 10-minute window, if there is exactly one MPC buying and/or selling a particular product, we can easily record the flow between the two classes. However, if several MPCs trade a specific product during the window, we can estimate the flow between the different participants, by interpreting the buy and sell volumes of each MPC as left and right nodes of a bipartite graph. The existence and volume of transactions between the two types of participants is represented by the existence and weight of an edge. Inspired by the Chung-Lu generative model for bipartite graphs \cite{benson2020simple}, we can then approximate the flow between two MPCs, for each asset, at each window, by assuming the flow is proportional to the product of the buyer and seller volume. 

Let $V_{i,j,d,m}^{\textrm{Buy},t}$ (resp. $V_{i,j,d,m}^{\textrm{Sell},t}$), correspond to the volume of option contract $j$ for underlying asset $i$ bought (resp. sold) during day $d$, during intraday time window $t \in \{ 1,\dots,39 \}$, by MPC $m$. Then, we denote the nominal flow between MPCs $m_1$ and $m_2$ on day $d$ as 
\begin{equation} \label{flow_sum}
     \textrm{NF}_{d,m_1,m_2} = \sum_{i=1}^{N_d} \sum_{j=1}^{C_{i,d}} \sum_{t=1}^{39}  P_{i,j,d} V_{i,j,d,m_1}^{\textrm{Buy},t} \times \frac{V_{i,j,d,m_2}^{\textrm{Sell},t}}{\sum_{m'} V_{i,j,d,m'}^{\textrm{Sell},t}}.
\end{equation} 
Next, we normalize the flow for each day so that 
\[ \sum_{m_1} \sum_{m_2}  \textrm{NF}_{d,m_1,m_2} = 1, \forall d \in \{ 1,\dots,D \},\]
and average across all $D$ days, to obtain the average nominal flow between MPCs $m_1$ and $m_2$
\begin{equation} \label{flow_fraction} \widetilde{\textrm{NF}}_{m_1,m_2} = \text{median} \left(  \frac{\textrm{NF}_{d,m_1,m_2}}{\sum_{m_1',m_2'} \textrm{NF}_{d,m_1',m_2'}} ; d=1,\dots,D \right). 
\end{equation}

%% file: input_app_cde.tex
\section{OVI of different Market Participant Classes: Additional Results}
In this part, we provide additional numeric results regarding Section 5 of the main paper. 

\begin{figure}[h]
\centering
\includegraphics[width=1\textwidth,center]{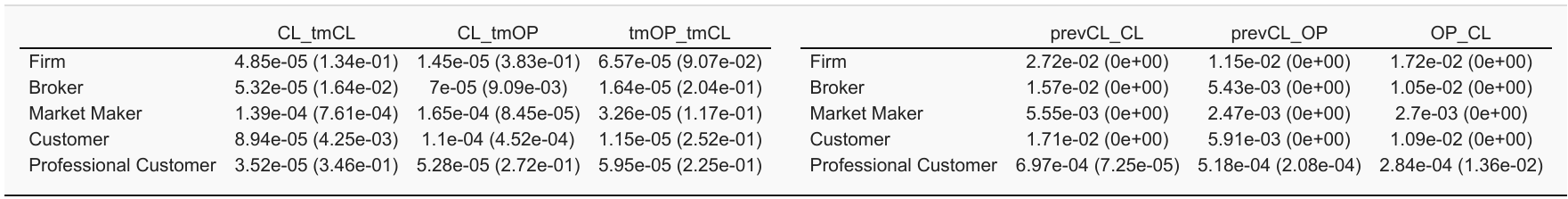}
\caption{Adjusted $R^2$-values obtained for the coefficient of the OVI in (\Cref{f_fut}) and (\Cref{f_cont}) respectively for different MPCs, for the Q3 QR groups, Uniform Weighting Scheme. The $p$-values for the regression are shown in brackets.}
\label{fig:contemporaneous}
\end{figure}

\paragraph{Contemporaneous Returns}
The main paper was focused on the usage of Option Volume Imbalance as a predictor for future returns. We are also interested in testing whether the OVI has explanatory power on contemporaneous returns. To this end, we test for the linear relationship 
\begin{equation} \label{f_cont}
    \tilde{f}_{i,d-1}^{\textrm{CL\_tmCL}}= \alpha + \beta_{i,d} \textrm{OVI}_{i,d,m} + \epsilon_{i,d}.  
\end{equation}
For comparison purposes, we also consider the relationship of OVI with future returns 
\begin{equation} \label{f_fut}
    \tilde{f}_{i,d}^{\textrm{CL\_tmCL}}= \alpha + \beta_{i,d} \textrm{OVI}_{i,d,m} + \epsilon_{i,d}.  
in\end{equation}
In a sliding window manner, with a window length of 200 days which gets shifted every 50 days, we regress both CL\_tmCL (\Cref{f_fut}) and prevCL\_CL (\Cref{f_cont}) returns against OVI, by filtering non-zero OVI values. In \Cref{fig:contemporaneous}, we show the median $p$ and $R^2$ value for contemporaneous and future returns by using the above method. 

The OVI is a significant explanatory variable for all MPCs, for both contemporaneous and future returns. However, we observe that the $p$-values for the contemporaneous returns are much lower than those corresponding to the future  returns. The $R^2$'s obtained are generally quite low across all cases, but slightly higher for contemporaneous returns.

\begin{figure}
\includegraphics[width=0.9\textwidth,left]{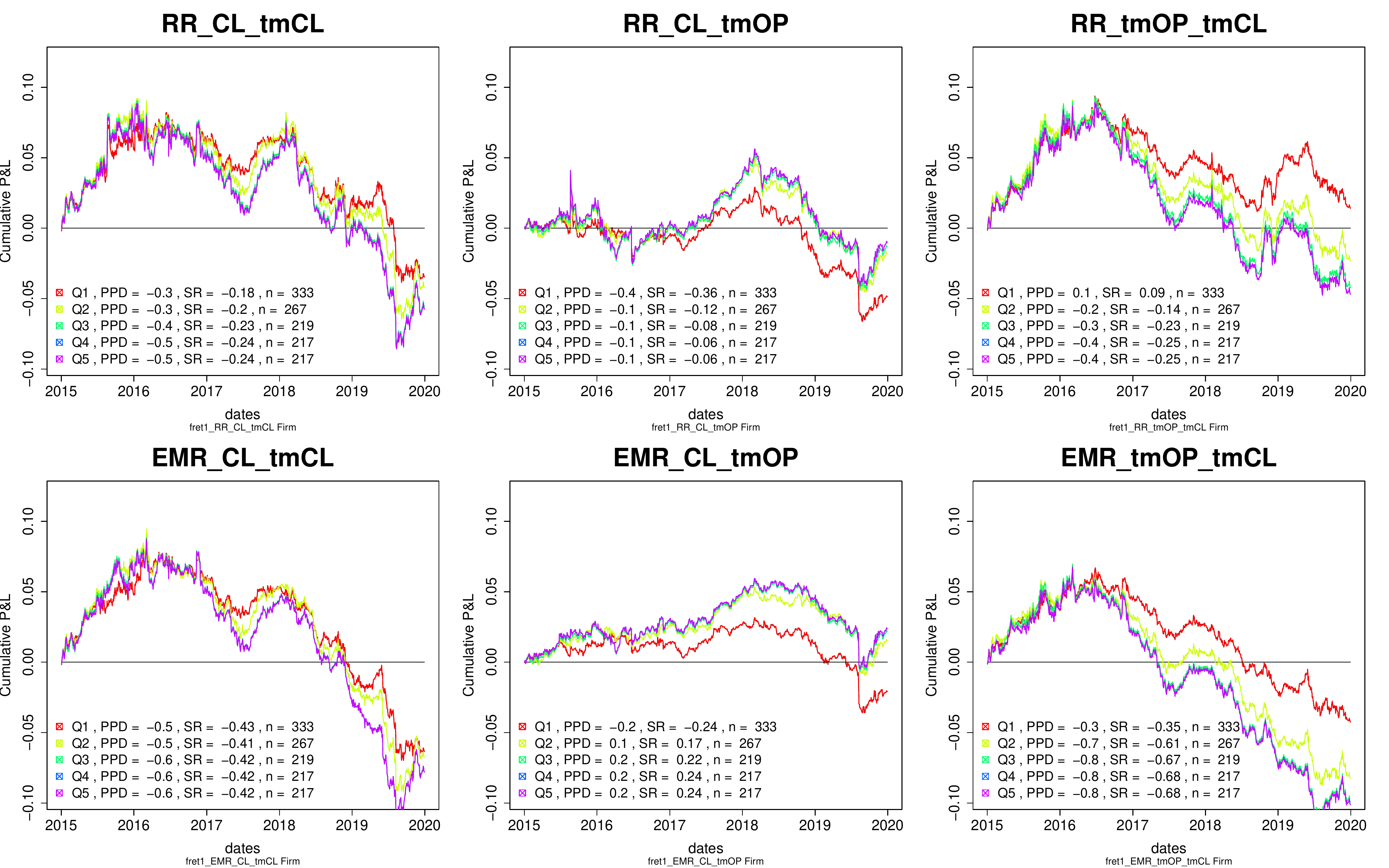}
\caption{Cumulative P\&L plot - Firm: This shows the cumulative P\&L from following a Uniform Weighting Scheme - one-day holding period strategy, for the five quantile rank groups Q1-Q5, using the OVI of Market Makers for six different types of returns. Also given, are SR (Annualized Sharpe Ratio), PPD (Profit per dollar in basis points), and $n$ (average number of instruments in the portfolios).}
\label{fig:cumulative_PnL_firms}
\end{figure}

\begin{figure}
\includegraphics[width=0.9\textwidth,left]{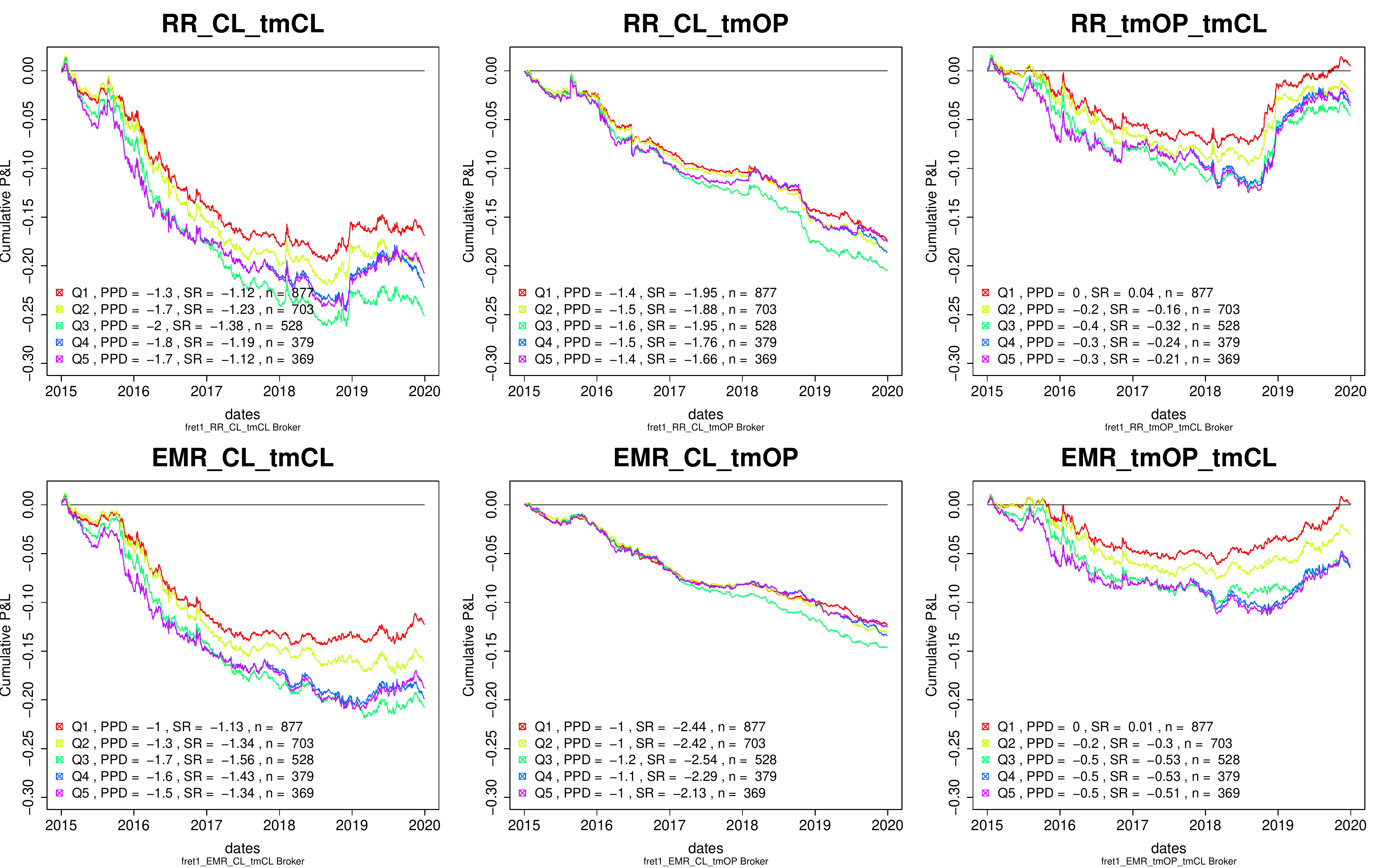}
\caption{Cumulative P\&L plot - Broker: This shows the cumulative P\&L from following a Uniform Weighting Scheme - one-day holding period strategy, for the five quantile rank groups Q1-Q5, using the OVI of Market Makers for six different types of returns. Also given, are SR (Annualized Sharpe Ratio), PPD (Profit per dollar in basis points), and $n$ (average number of instruments in the portfolios).}
\label{fig:cumulative_PnL_brokers}
\end{figure}

\begin{figure}
\includegraphics[width=0.9\textwidth,left]{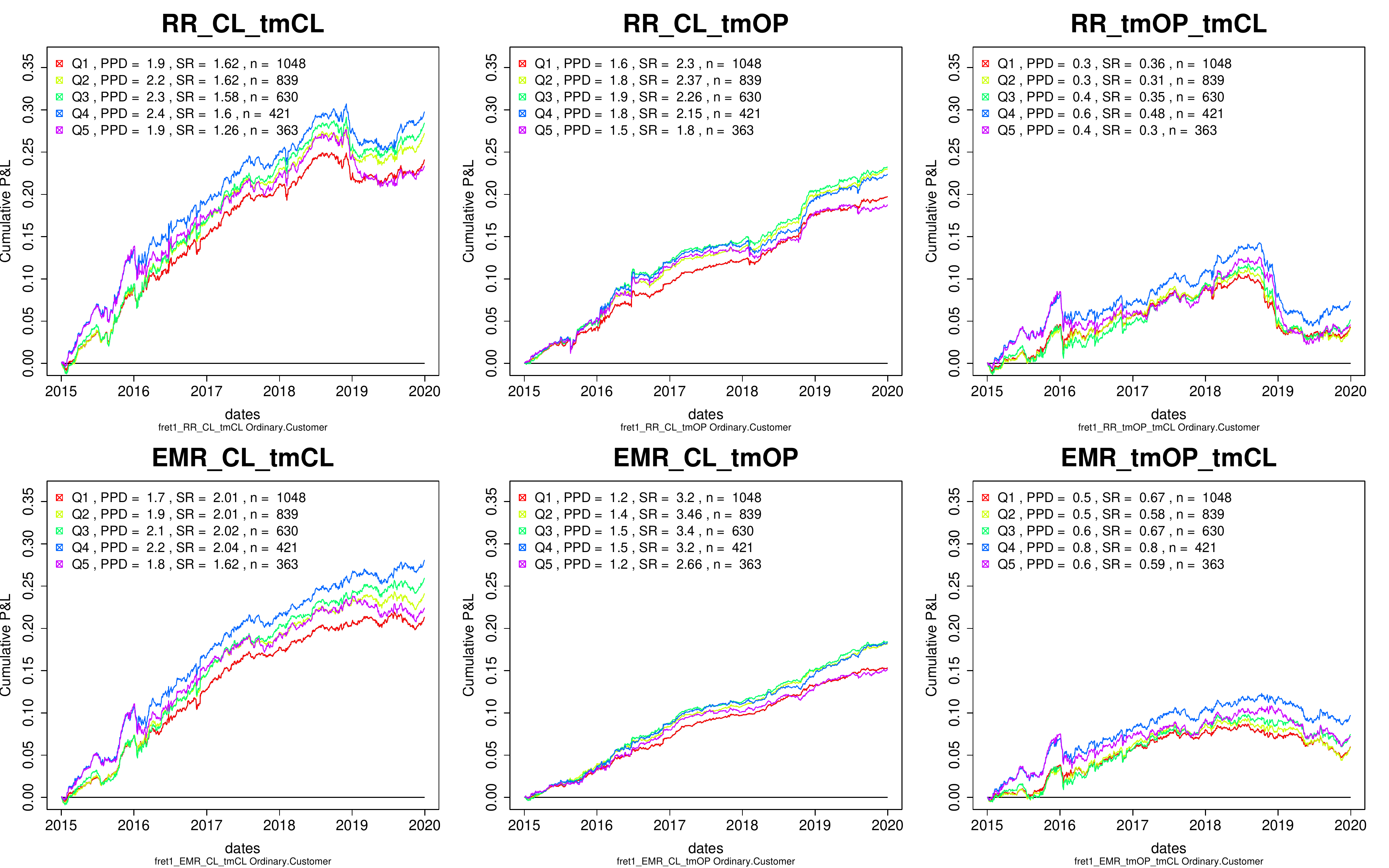}
  \caption{Cumulative P\&L plot - Customer: This shows the cumulative P\&L from following a Uniform Weighting Scheme - one-day holding period strategy, for the five quantile rank groups Q1-Q5, using the OVI of Market Makers for six different types of returns. Also given, are SR (Annualized Sharpe Ratio), PPD (Profit per dollar in basis points), and $n$ (average number of instruments in the portfolios).}
\label{fig:cumulative_PnL_customers}
\end{figure}

\begin{figure}
\includegraphics[width=0.9\textwidth,left]{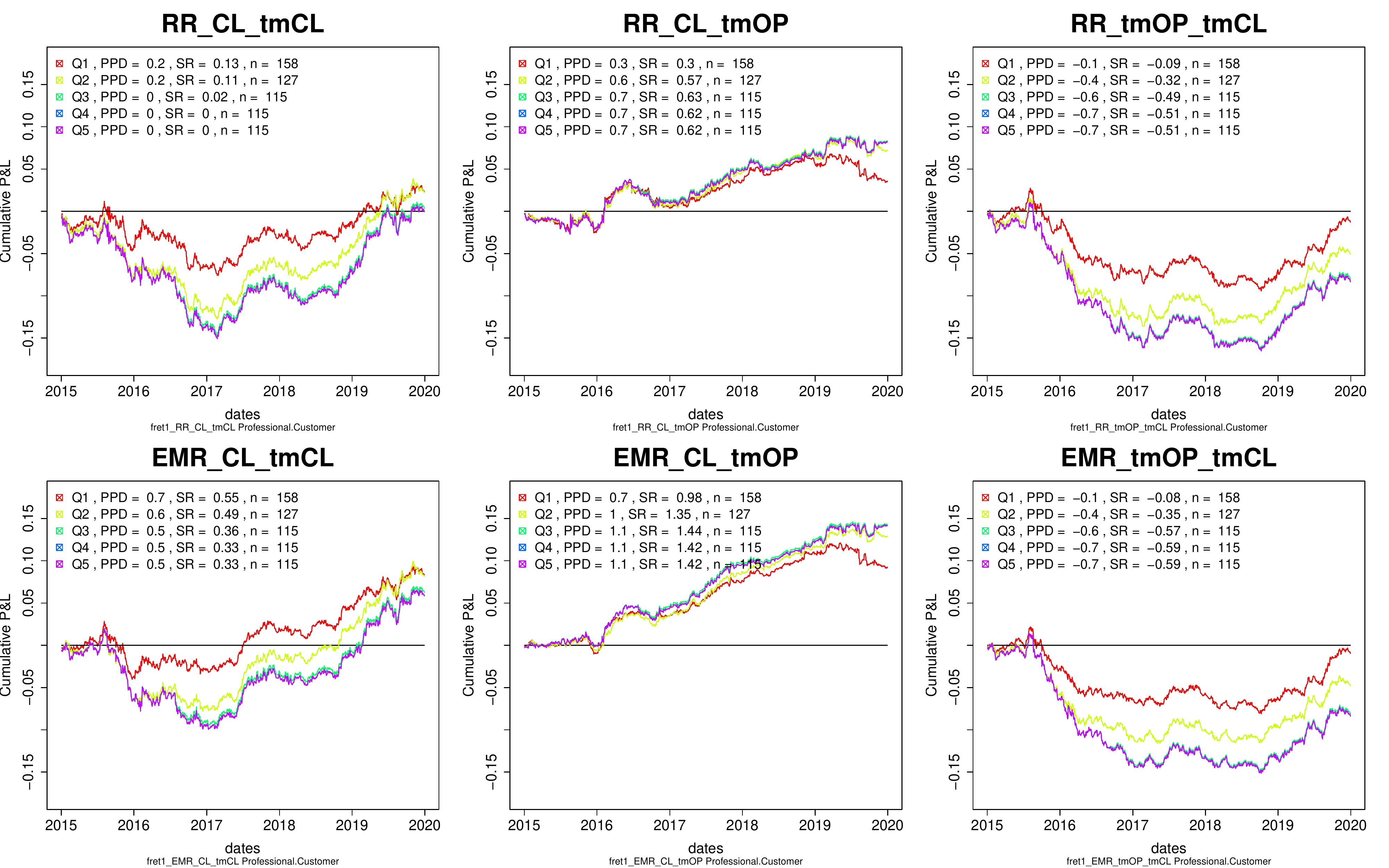}
\caption{Cumulative P\&L plot - Professional Customer: This shows the cumulative P\&L from following a Uniform Weighting Scheme - one-day holding period strategy, for the five quantile rank groups Q1-Q5, using the OVI of Market Makers for six different types of returns. Also given, are SR (Annualized Sharpe Ratio), PPD (Profit per dollar in basis points), and $n$ (average number of instruments in the portfolios).}
\label{fig:cumulative_PnL_professional_customers}
\end{figure}

\paragraph{Cumulative P\&L for non-market maker participants}
In Section 5 of main the paper, we showed the cumulative P\&L plots for the various return modes for Market Makers. Below, in Figures \ref{fig:cumulative_PnL_firms}-\ref{fig:cumulative_PnL_professional_customers}, we show the equivalent plots for Firms, Brokers, Customers and Professional Customers.

Similarly to Market Makers, the highest performance observed for Brokers and Customers corresponds to overnight EMR CL\_tmOP returns. OVI also seems to have predictive power on tmOP\_tmCL returns, especially for the years 2015-2018. However, in both cases, performance drops in 2019. The CL\_tmCL return data is  essentially a combination of the other two types of returns, thereby resulting in a cumulative P\&L plot with a clear trend, and a drop in performance for 2019. For Professional Customers, most of the OVI predictability corresponds to overnight returns, which follow a clear upward trend. For EMR tmOP\_tmCL, the cumulative P\&L seems to have a somewhat downward trend for 2015 and 2016 and a somewhat upward trend for 2018 and 2019. For Firms, the EMR tmOP\_tmCL returns correspond to a upward trend for 2015, which later reverses. On the other hand, Firm OVI seems to be uninformative for overnight returns, as the cumulative P\&L remains close to zero.   

\begin{figure}[h]
\centering
\includegraphics[page=1,width=1.1\textwidth,center]{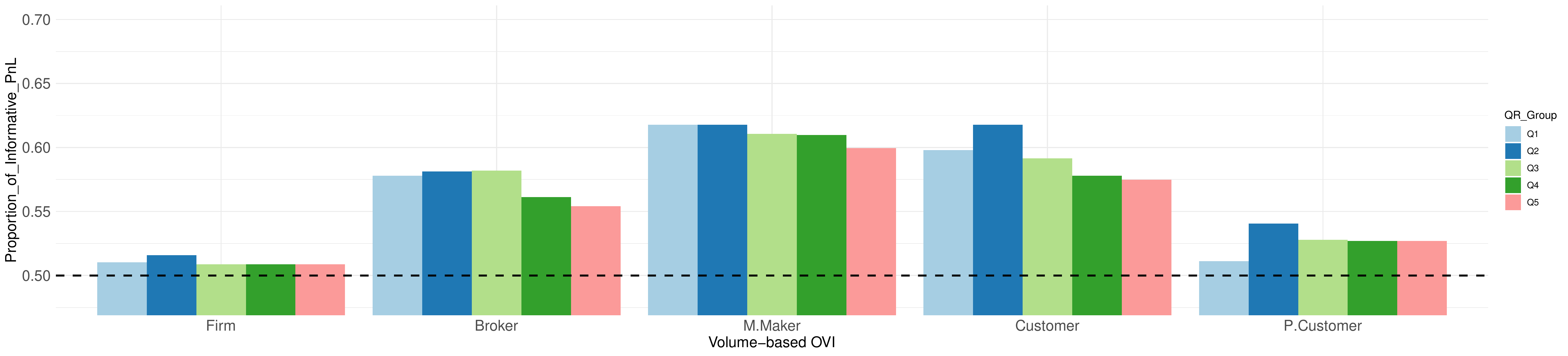}
\caption{P\&L accuracy: The ratio of correct predictions for the EMR CL\_tmOP returns across days is shown for each of the OVI measures, for the five quantile rank groups. Accuracy is defined as $\textrm{max} \{ \pi , 1-\pi \}$, where $\pi$ is defined as the ratio of positive daily P\&L.}
\label{fig:informative_PnL}
\end{figure}

\paragraph{Proportion of ``profitable" days}
An alternative way to evaluate the performance of the different OVI strategies, would be to consider the percentage of profitable days. Denote the proportion of positive daily P\&L's by $\pi$ corresponding to a particular OVI. In \Cref{fig:informative_PnL}, we show the value of $\textrm{max} \{ \pi , 1-\pi \}$ for different OVI signals, for a Uniform Weighting Scheme, evaluated for EMR CL\_tmOP returns. This is the proportion of P\&L's that are either positive (e.g by using the Customer OVI) or negative (e.g by using the Market Maker OVI). The OVIs of Firms and Professional Customers have values close to 0.5, showing how uninformative they are as predictors for future returns. The other three MPCs fare much better, with the Market Makers having a rate of informative P\&L's above 60\%, which is much higher than what would be expected at random (around 50\%). Depending on the quantile rank, the OVI of Customers and Brokers results in a proportion lying between 55 and 60\%. Note that using the total number of trades or the nominal volume results in similar results.

\FloatBarrier

\section{Broader Scope Analysis: Additional results}
In this section, we list additional results for Sections 6 and 7 of the main paper.
\paragraph{Decomposition based on Option Features}
As mentioned in Section 6 of the main paper, we consider how certain option features affect the predictability of the OVI. \Cref{tab:options}  shows a summary of the analysis, including results corresponding to Broker and Customer OVI, which were not shown in the main paper.
\phantom{~}\noindent
\begin{table}[h]
    \begin{tabular}{p{0.26\textwidth}>{\centering}p{0.34\textwidth}>{\centering}p{0.11\textwidth}>{\centering}p{0.11\textwidth}>{\centering\arraybackslash}p{0.11\textwidth}}
\hline
\\
\multirow{2}{*}{Option Feature} & \multirow{2}{*}{Definition} & \multicolumn{3}{c}{Best Performing Bucket} \\\cline{3-5}
& & Brokers & Market Makers & Customers \\
\hline \\
Delta ($\Delta$) & $
   \Delta = \partial P^{\textrm{Option}}/\partial S$ & $1^{st}$  and $3^{rd}$  & $1^{st}$ & $1^{st}$-$2^{nd}$  \\ \\
Gamma($\Gamma$) & 
    $\Gamma = \partial \Delta / \partial S = \partial^2 P^{\textrm{Option}} / \partial S^2$ & $2^{nd}$-$4^{th}$ & $4^{th}$ & $2^{nd}$ and $4^{th}$ \\ \\
 Theta ($\Theta$) & $\Theta = - \partial P^{\textrm{Option}} / \partial \tau$ & $2^{nd}$ and $4^{th}$ & $2^{nd}$-$4^{th}$ & $2^{nd}$ \\ \\
 Vega ($\mathcal{V}$) & $\mathcal{V} =\partial P^{\textrm{Option}} / \partial r$ & None & $2^{nd}$ & None \\ \\
Rho ($\rho$) & $\rho = \partial P^{\textrm{Option}}\partial r$  & $2^{nd}$ & $1^{st}$ -$2^{nd}$  & $1^{st}$ -$2^{nd}$  \\  \\
Moneyness (Mon) & $\textrm{Mon} = \left\{
\begin{array}{ll}
      +\log \left(S  e^{r \tau}/K \right)/(\sigma  \sqrt{\tau}) & \textrm{for call options} \\
      -\log \left( S  e^{r \tau}/K \right)/(\sigma  \sqrt{\tau})& \textrm{for put options}\\ 
\end{array}
\right. $  & $2^{nd}$ & $2^{nd}$-$4^{th}$ & $2^{nd}$-$4^{th}$ \\
    \\
Maturity & Number of days until expiration.  & None & None & None \\
    \\
Implied Volatility ($\sigma_I$)  & $\sigma$ that solves BS option price equation & $4^{th}$ & $4^{th}$ & $4^{th}$ \\
\hline \\
\end{tabular}
\caption{Summary information for option feature decomposition: For each option feature, its definition is given as well as the ``Best Performing Bucket" for each MPC, which is the bucket that displays the highest Sharpe Ratio.} 
\label{tab:options}
\end{table} 

\paragraph{Difference between Opening and Closing Transactions} 
For MPCs other than Market Makers, we have a distinction between opening and closing transactions. In both the cases of Brokers and Customers, the number of closing transactions is much lower than for opening transactions. Therefore, in both cases there is no significant difference between opening transactions and the complete data set, whereas the performance of closing transactions is quite low. 

\paragraph{Influence of underlying equity's features on OVI's predictive power}
Similar to the experiments for the influence of option features on the predictive power of OVI, we divide underlying equities into four buckets based on some characteristic of the underlying. For Market Cap, there is no significant difference across buckets. For Turnover (both the daily, as well as the median of the last 21 trading days), higher turnover seems to be linked with higher predictability. However, the differences in SR are not significant in either cases. 

\begin{figure}
   \centering
   \includegraphics[width=1\textwidth]{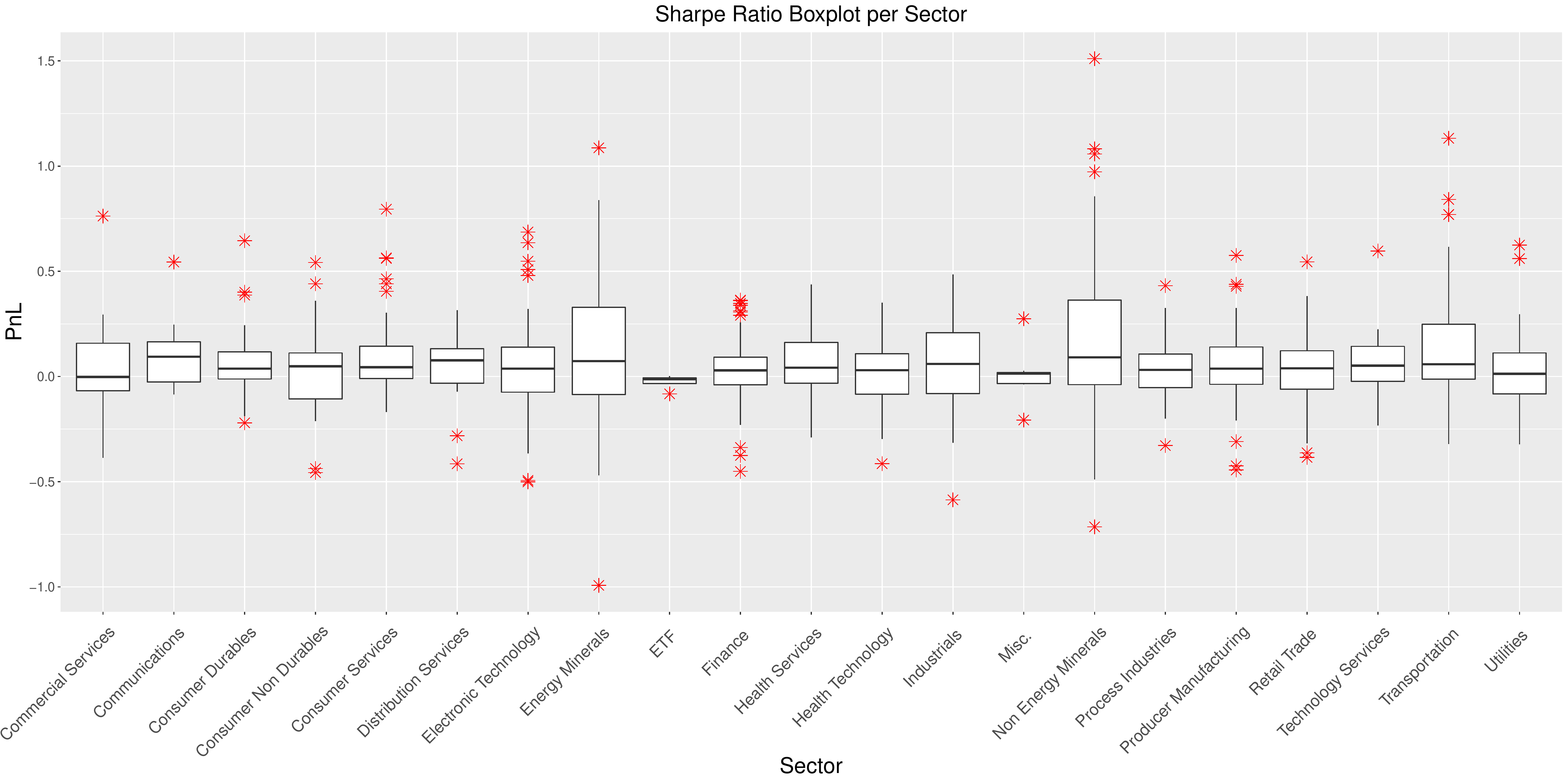}
   \caption{Boxplots of Sharpe Ratios for each instrument, obtained using the Market Maker Q3 strategy. Results are shown in the form of boxplots, broken down by sector. }
   \label{fig:sector}
\end{figure}

We also consider differences in performance between different sectors. In \Cref{fig:sector}, we show the boxplots for SRs obtained when running a Q3-Market Maker strategy, grouped by sector\footnote{Equity sectors are taken from https://www.marketbeat.com/.}. We observe that most of the sectors have a positive SR on average, even though the lower quantile of P\&L for most sectors falls below 0. However, the discrepancies across sectors are not statistically significant, as we do not get a rejection for an ANOVA test when testing for group differences. 

The results seem to suggest that predictability does not depend as much on the features of the underlying, but more so on the features of the option features, as shown in Section 6 of the main paper. Finally, note than when we test the SR obtained by following the Q3 Market Maker strategy (with a Uniform Weighting Scheme) for each individual instrument, 148 of instruments reject the SR significance test (SR = 0) at a 5\% level, while only one instrument rejects the hypothesis for a Bonferonni corrected test.

\paragraph{Cross-Impact and Networks}
We now turn our attention to the network produced by following a Q3 and row-level corrected Bonferonni level (i.e $\frac{0.05}{\# \textrm{Number of Instruments}}$) network, as discussed in Section 7 of the main paper. The resulting network has one large connected component. In the weak sense (dropping the directions of the edges), it includes all but two instruments, while in the strong sense, it includes 942 out of 1792 instruments, where the rest of the instruments form no other component. 

\begin{figure}
   \centering
   \includegraphics[width=1\textwidth]{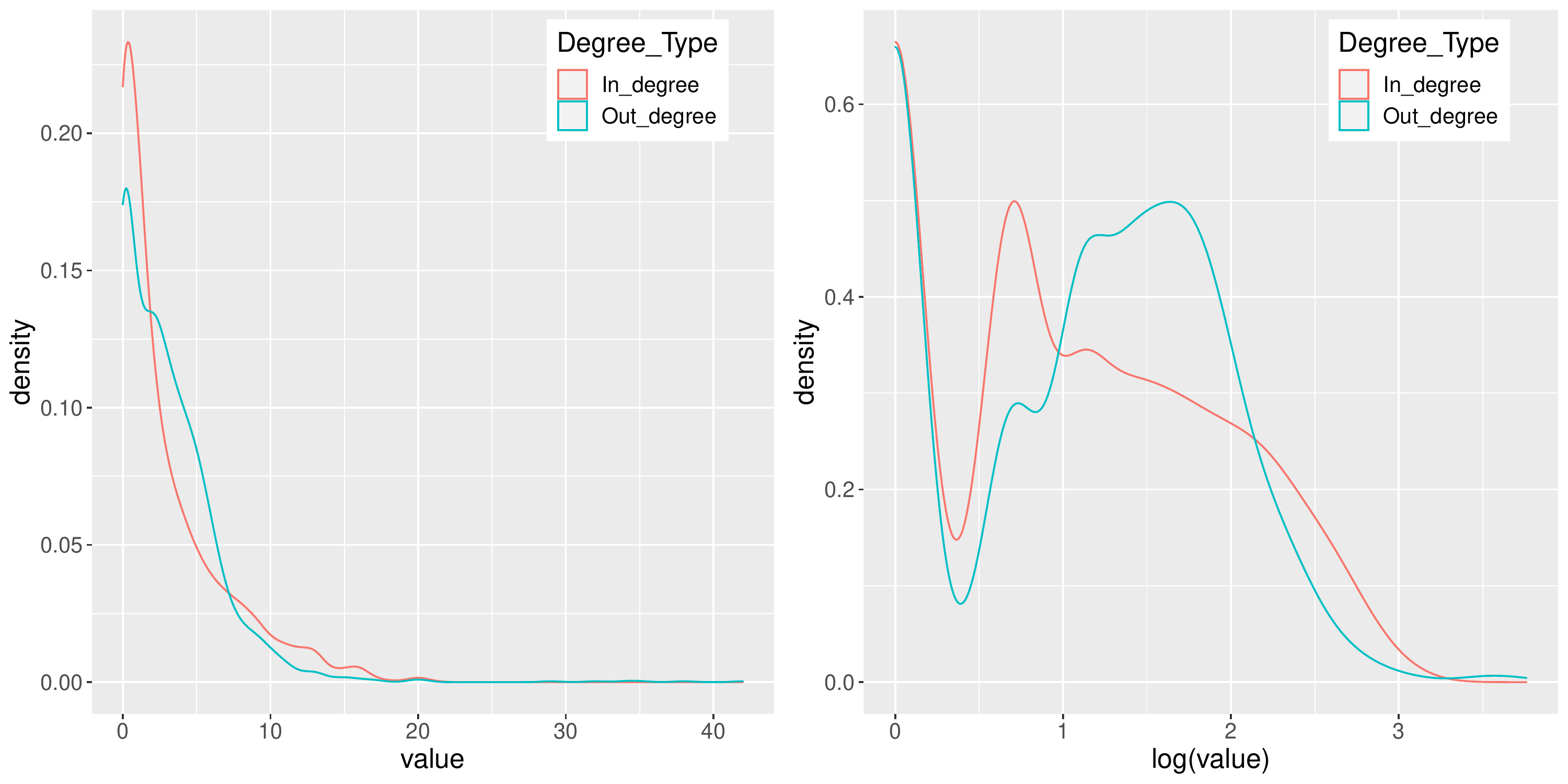}
   \caption{In- and Out-Degrees of Network. The left sub figure shows the degree density plot, while the right sub figure shows the density of the logarithm of the density, corresponding to a Q1 row-level Bonferonni Corrected network discussed in \Cref{section:networks}.}
   \label{fig:degrees}
\end{figure}

In \Cref{fig:degrees}, we show a histogram of the in- and out-degrees, in the normal and log-scale. The in-degrees give a $p$-value of 0 when testing the null hypothesis that they follow a power-law distribution. Out-degrees do not reject and give a coefficient of 3.75, but from \Cref{fig:summary_info}, it seems quite clear that the resulting network does not follow a power-law, as about half of the produced networks reject such a hypothesis. 

\begin{figure}
   \centering
   \includegraphics[width=1\textwidth]{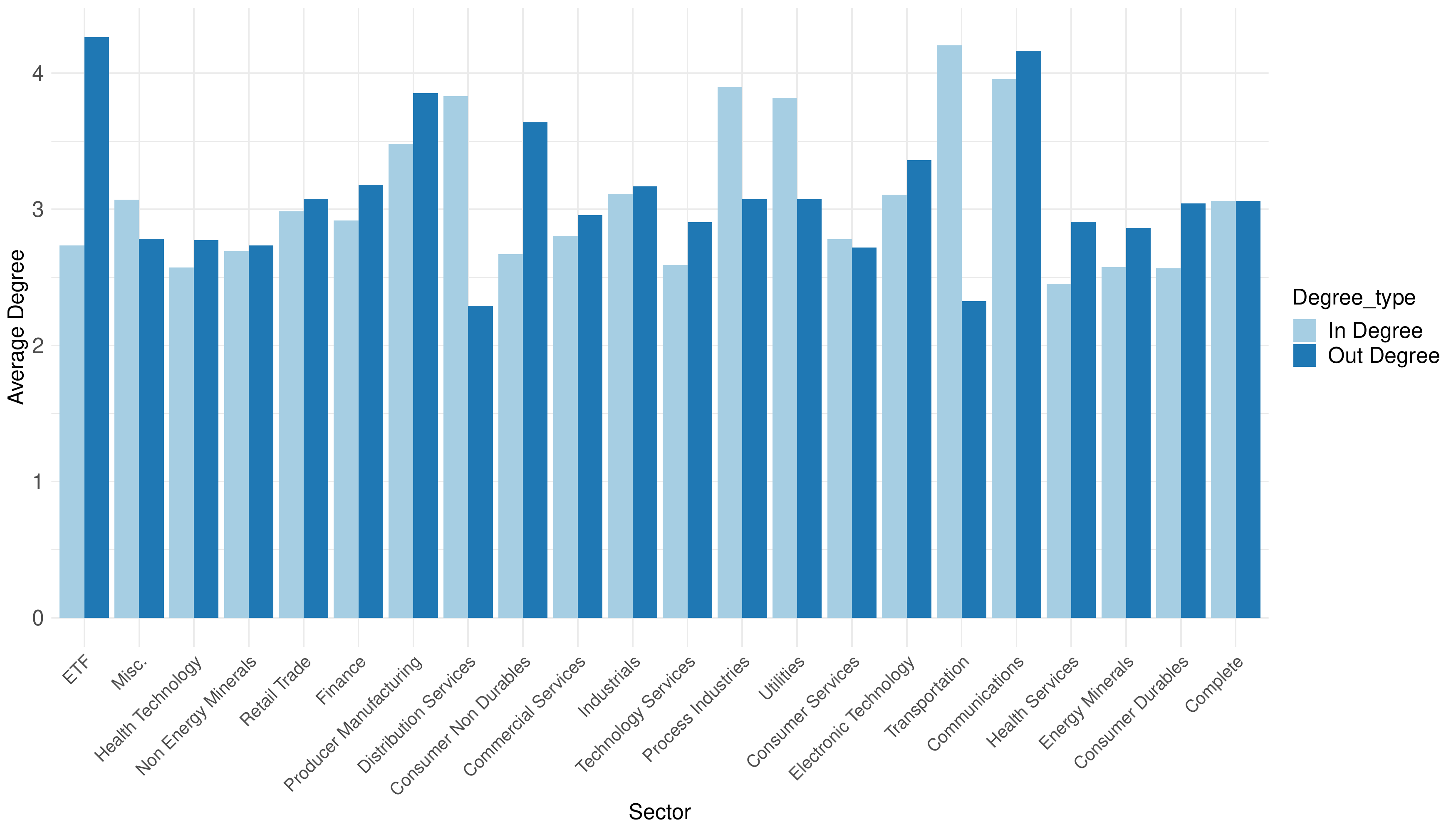}
   \caption{In and Out Degrees per Sector: This figure shows the average degree for nodes corresponding to each sector of instruments, for the Q1 row-level Bonferonni Corrected network discussed in \Cref{section:networks}. The right-most bar shows the average degree of the whole network.}
   \label{fig:degrees_sectors}
\end{figure}

We consider the average degrees based on each underlying sector in \Cref{fig:degrees_sectors}, based on the Q3-market maker network with row-level Bonferonni correction. Some interesting observations are: 
\begin{itemize}
    \item Transportation is the sector with the highest average in degree, and the lowest average out degree. 
    \item Distribution Services, Process Industries and Utilities sectors also have higher in-degree than the average node. On the other side, Health Services, Consumer-Durables and Energy Minerals have lower than average in-degrees. 
    \item The out-degrees seem to vary more than than in-degrees. The communication sector as well as ETF's have much higher out-degrees than average, while Consumer Non Durables and Producer Manufacturing sectors also have higher than average out-degrees. Reversely, the Distribution service has a lower than average out-degree.
\end{itemize}

We also consider a bipartite graph, under the Chung-Lu model similarly to the main paper, where weights are proportional to the out-degree of the outgoing node and the in-degree of the incoming node. No pair of sectors has a value significantly different from the actual number of links observed. In other words, there is no significant evidence for a sector-sector cluster structure. 

\begin{figure}
    \centering
   \includegraphics[width=1\textwidth]{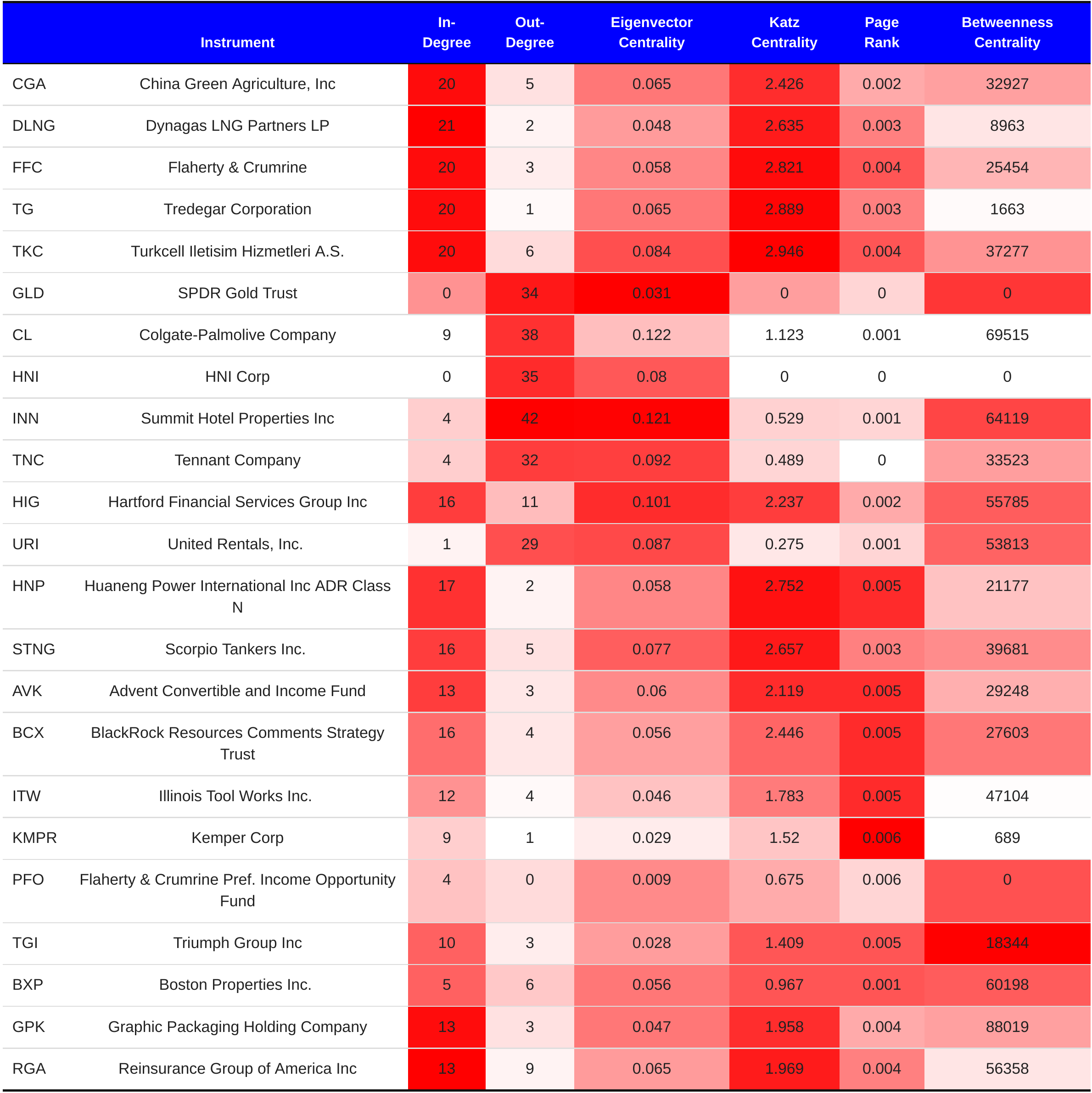}  
   \caption{Selection of Instruments with high centralities for different centrality measures. Red cells denote high centrality for the metric corresponding to the particular column.}
   \label{fig:centralities}
\end{figure}

Lastly, we consider the centrality of the nodes. Besides the in- and out-degrees, we also consider four more centrality measures: Eigenvector, Katz, Page-Rank and Betweeness Centrality. In \Cref{fig:centralities}, we show the instruments with the highest centralities (we choose the five instruments with the highest value for each centrality). Note that instrument with high in-degrees also tend to have high Katz centralities and vice versa, while instruments with high Katz Centrality also tend to have a high Page rank. 

\FloatBarrier 

\section{Intra-day patterns.}
The interplay between OVI and future returns in the main paper was conducted on a daily level. We also briefly consider an extension to the intraday case and report on some of the preliminary observed patterns. We consider the five MPCs, using three QR Groups defined similarly as before, and consider various types of intraday excess market returns for different time horizons ($\{1,5,10,30,60,90,180\}$-minute future returns, as well as overnight returns). We refer to the OVI used as ``open--$t$" OVI, as it captures all transactions conducted from the market's opening up to current time $t$. Note that a limitation of this analysis is that the universe of underlying assets is smaller and constrained only to 2018-2019 due to the availability of intraday data\footnote{Returns are calculated using LOB data, by considering the midpoint of the bid and ask price}. This renders any direct comparison between the intraday and daily returns not possible at this stage. Therefore, we limit ourselves to some interesting qualitative observations. Indeed, running a fully Bonferonni Corrected test taking into account all tests (correcting for the three QR Buckets, ten return modes, five MPCs and 39 values for $t$). In this case, none of the SRs comes out as significant; using the less conservative, Benjamini-Hochberg Procedure does not end up in any rejection either. 

\begin{figure}[h]
   \centering
   \includegraphics[width=0.88\textwidth]{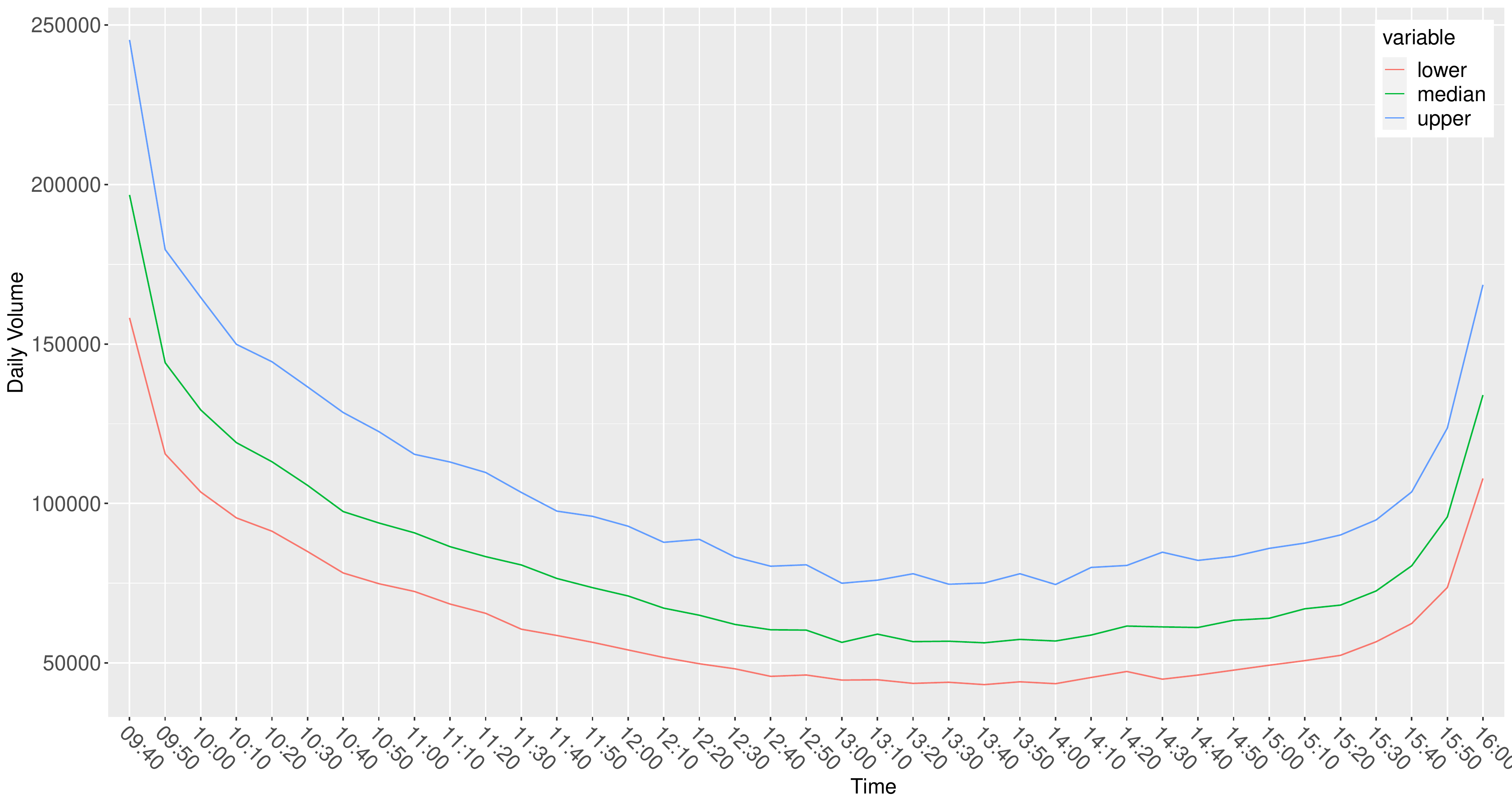}
   \caption{Intraday Volume: The median total option volume at different time points is shown with the black line, while the upper quantile (resp. lower quantile) volume is given by the red line (resp. blue line).}
   \label{fig:intraday_volumes}
\end{figure}

\begin{figure}[h]
   \includegraphics[width=0.85\textwidth]{PLOTS/intraday_time_series_firm.pdf}
   \caption{Intraday Returns for Firm open\_t OVI: This Figure shows the 90-minute market excess returns by following an open\_t (use all volumes from opening up to t) strategy against t. The quantities shown are given for 3 cumulative quantile rank groups. The top figure shows the (Annualized) SR corresponding to each strategy, the second figure shows the Profit Per Dollar (PPD) and the third figure shows the average number of instruments used for each portfolio. }
   \label{fig:firm_intraday}
\end{figure}

\begin{figure}[h]
   \includegraphics[width=0.85\textwidth]{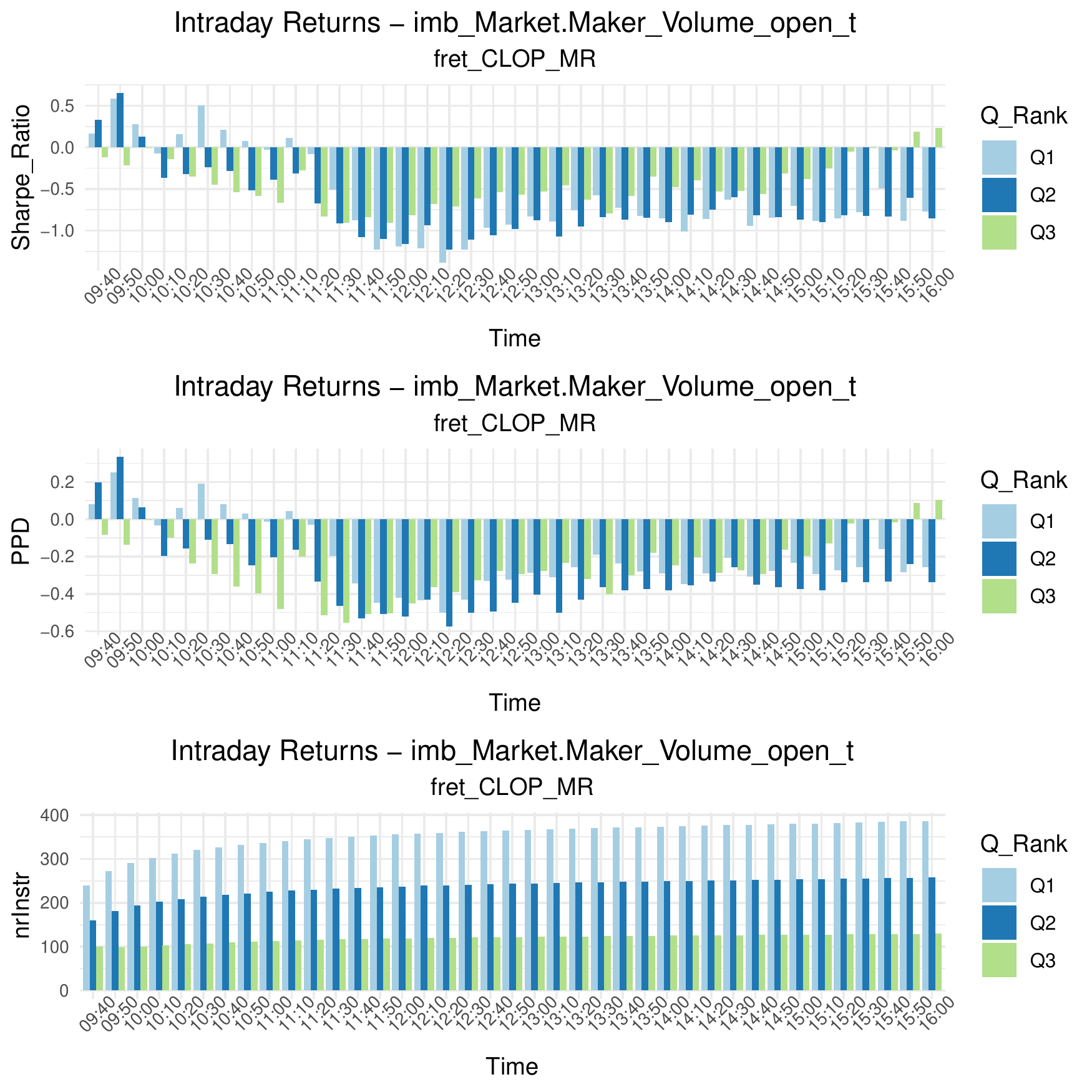}
   \caption{Intraday Returns for Market Maker open\_t OVI: This Figure shows the overnight (CL\_tmOP) market excess returns by following an open\_t (use all volumes from opening up to t)strategy against t. The quantities shown are given for 3 cumulative quantile rank portfolios. The top figure shows the (Annualized) Sharpe Ratio corresponding to each strategy, the second figure shows the Profit Per Dollar (PPD) and the third figure shows the average number of instruments used for each portfolio. }
   \label{fig:mm_intraday_open}
\end{figure}

In most cases, Market Makers attain the most significant and consistent performance. The open--t OVI at mid-day results in a moderately high SR (maximized at 13:10), when we consider returns for 30-180 minutes (see for example \Cref{fig:mm_intraday_open}). Furthermore, setting t near midday, results in positive CL\_tmOP returns. Thus, a holding period from midday to opening of next day seems to be the optimal strategy when using the Market Maker OVI. If the holding period is further extended to include overnight returns (i.e CL\_tmCL returns), then the SR remains positive but is decreased in magnitude. 

Across MPCs, for either open-t or t-close OVIs and a return window of 60, 90 or 180 minutes, a seasonal-like effect can be observed for SRs. For example, we observe this pattern in \Cref{fig:firm_intraday}, for the open--t Firm OVI used for 90 minute EMR. Setting $t$ to any time earlier than 10:30am, yields a consistently positive profit. For $t$ after 10:30am, we observe a decrease in predictability and this drop eventually results in a negative SR for $t$ later than 11:00am. Fluctuations continue with returns changing signs again at 12:30pm, 13:20pm and 15:10pm, displaying a somewhat periodic pattern. This seasonality of the time series appears in the other MPCs as well, albeit to a lesser extent. Across MPCs, these oscillations also appear to be dumped, as they usually decrease in magnitude (either in terms of SR or PPD). 

In \Cref{fig:intraday_volumes}, we observe that the option volumes follow a U-shape similar to what is observed for the spot market. Higher volumes should make OVI a more reliable signal, which could explain the highest performance observed when using the volumes from the early part of the day.

Due to the rather restrictive data sets employed in our analysis, the above intraday results are difficult to quantify. On the other hand, a lot of observed SRs exceed 1, and are considered statistically significant (when ignoring Bonferonni correction). The oscillations of the observed SR with respect to $t$, also explain why CL\_tmCL are worse performing than CL\_tmOP. The Market Maker OVI is the exception, as its corresponding SR does not revert sign as $t$ increases within the day, and remains mostly negative. This appears to be in line with the observation that the Market Maker's OVI has the highest predictability for longer horizon returns.   

%% file: main_arXiv.bbl
\begin{thebibliography}{}

\bibitem[Amin and Lee, 1997]{Amin&Lee}
Amin, K.~I. and Lee, C. M.~C. (1997).
\newblock Option trading, price discovery, and earnings news dissemination*.
\newblock {\em Contemporary Accounting Research}, 14(2):153--192.

\bibitem[Anthony, 1988]{anthony}
Anthony, J.~H. (1988).
\newblock The interrelation of stock and options market trading-volume data.
\newblock {\em The Journal of Finance}, 43(4):949--964.

\bibitem[Bailey and Lopez~de Prado, 2014]{bailey2014}
Bailey, D. and Lopez~de Prado, M. (2014).
\newblock The deflated sharpe ratio: Correcting for selection bias, backtest
  overfitting, and non-normality.
\newblock {\em The Journal of Portfolio Management}, 40:94--107.

\bibitem[{Bank for International Settlements}, 2021]{BIS}
{Bank for International Settlements} (2021).
\newblock Bis statistics explorer: Table d5.1 and d8.
\newblock Data retrieved from \url{https://stats.bis.org/statx}.

\bibitem[Bao, 2008]{bao}
Bao, Y. (2008).
\newblock {Estimation Risk-Adjusted Sharpe Ratio and Fund Performance Ranking
  under a General Return Distribution}.
\newblock {\em Journal of Financial Econometrics}, 7(2):152--173.

\bibitem[Benson et~al., 2020]{benson2020simple}
Benson, A.~R., Liu, P., and Yin, H. (2020).
\newblock A simple bipartite graph projection model for clustering in networks.

\bibitem[Bhattacharya, 1987]{Batcha}
Bhattacharya, M. (1987).
\newblock Price changes of related securities: The case of call options and
  stocks.
\newblock {\em The Journal of Financial and Quantitative Analysis},
  22(1):1--15.

\bibitem[Biais and Hillion, 2015]{Biais}
Biais, B. and Hillion, P. (2015).
\newblock {Insider and Liquidity Trading in Stock and Options Markets}.
\newblock {\em The Review of Financial Studies}, 7(4):743--780.

\bibitem[Black, 1975]{black}
Black, F. (1975).
\newblock Fact and fantasy in the use of options.
\newblock {\em Financial Analysts Journal}, 31(4):36--72.

\bibitem[Black and Scholes, 1973]{BS}
Black, F. and Scholes, M. (1973).
\newblock The pricing of options and corporate liabilities.
\newblock {\em Journal of Political Economy}, 81(3):637--654.

\bibitem[Cao et~al., 2005]{cao}
Cao, C., Chen, Z., and Griffin, J. (2005).
\newblock Informational content of option volume prior to takeovers.
\newblock {\em The Journal of Business}, 78(3):1073--1109.

\bibitem[Cao et~al., 2020]{cao_SR_opt}
Cao, H.~K., Cao, H.~K., and Nguyen, B.~T. (2020).
\newblock Delafo: An efficient portfolio optimization using deep neural
  networks.
\newblock In Lauw, H.~W., Wong, R. C.-W., Ntoulas, A., Lim, E.-P., Ng, S.-K.,
  and Pan, S.~J., editors, {\em Advances in Knowledge Discovery and Data
  Mining}, pages 623--635, Cham. Springer International Publishing.

\bibitem[Chan et~al., 2015]{kalok}
Chan, K., Chung, Y.~P., and Fong, W.-M. (2015).
\newblock {The Informational Role of Stock and Option Volume}.
\newblock {\em The Review of Financial Studies}, 15(4):1049--1075.

\bibitem[Clauset et~al., 2009]{clauset_powerlaw}
Clauset, A., Shalizi, C.~R., and Newman, M. E.~J. (2009).
\newblock Power-law distributions in empirical data.
\newblock {\em SIAM Review}, 51(4):661--703.

\bibitem[Copeland, 1976]{copeland1976}
Copeland, T.~E. (1976).
\newblock A model of asset trading under the assumption of sequential
  information arrival.
\newblock {\em The Journal of Finance}, 31(4):1149--1168.

\bibitem[Curme et~al., 2014]{curme2014emergence}
Curme, C., Tumminello, M., Mantegna, R.~N., Stanley, H.~E., and Kenett, D.~Y.
  (2014).
\newblock Emergence of statistically validated financial intraday lead-lag
  relationships.

\bibitem[Easley et~al., 1998]{ohara}
Easley, D., O'Hara, M., and Srinivas, P. (1998).
\newblock Option volume and stock prices: Evidence on where informed traders
  trade.
\newblock {\em The Journal of Finance}, 53(2):431--465.

\bibitem[Fama and French, 1993]{FF:1992}
Fama, E.~F. and French, K.~R. (1993).
\newblock Common risk factors in the returns on stocks and bonds.
\newblock {\em Journal of Financial Economics}, 33:3--56.

\bibitem[Ge et~al., 2016]{GE2016601}
Ge, L., Lin, T.-C., and Pearson, N.~D. (2016).
\newblock Why does the option to stock volume ratio predict stock returns?
\newblock {\em Journal of Financial Economics}, 120(3):601--622.

\bibitem[Goncalves-Pinto et~al., 2020]{GoncalvesPinto2020WhyDO}
Goncalves-Pinto, L., Grundy, B., Hameed, A., Heijden, T. V.~D., and Zhu, Y.
  (2020).
\newblock Why do option prices predict stock returns? the role of price
  pressure in the stock market.
\newblock {\em Manag. Sci.}, 66:3903--3926.

\bibitem[Grossman, 1977]{grossman}
Grossman, S.~J. (1977).
\newblock {The Existence of Futures Markets, Noisy Rational Expectations and
  Informational Externalities}.
\newblock {\em The Review of Economic Studies}, 44(3):431--449.

\bibitem[Hu, 2014]{Hu2014}
Hu, J. (2014).
\newblock Does option trading convey stock price information?
\newblock {\em Journal of Financial Economics}, 111(3):625--645.

\bibitem[Johnson and So, 2012]{JOHNSON2012262}
Johnson, T.~L. and So, E.~C. (2012).
\newblock The option to stock volume ratio and future returns.
\newblock {\em Journal of Financial Economics}, 106(2):262--286.

\bibitem[Kingma and Ba, 2014]{ADAM}
Kingma, D. and Ba, J. (2014).
\newblock Adam: A method for stochastic optimization.
\newblock {\em International Conference on Learning Representations}.

\bibitem[Ledoit and Wolf, 2008]{wolf2008}
Ledoit, O. and Wolf, M. (2008).
\newblock Robust performance hypothesis testing with the sharpe ratio.
\newblock {\em Journal of Empirical Finance}, 15(5):850--859.

\bibitem[Lin and Lu, 2015]{Lin2015}
Lin, T.-C. and Lu, X. (2015).
\newblock Why do options prices predict stock returns? evidence from analyst
  tipping.
\newblock {\em Journal of Banking \& Finance}, 52.

\bibitem[Lin et~al., 2013]{lin2013}
Lin, T.-C., Lu, X., and Driessen, J. (2013).
\newblock Why do options prices predict stock returns?
\newblock {\em Netspar Discussion Paper}, No. 07/2013-079.

\bibitem[Manaster and Rendleman, 1982]{mandle}
Manaster, S. and Rendleman, R.~J. (1982).
\newblock Option prices as predictors of equilibrium stock prices.
\newblock {\em The Journal of Finance}, 37(4):1043--1057.

\bibitem[Pan and Poteshman, 2006]{Pan_Poteshman}
Pan, J. and Poteshman, A.~M. (2006).
\newblock The information in option volume for future stock prices.
\newblock {\em Review of Financial Studies}, 19(3):871--908.

\bibitem[Roll et~al., 2010]{ROLL20101}
Roll, R., Schwartz, E., and Subrahmanyam, A. (2010).
\newblock O/s: The relative trading activity in options and stock.
\newblock {\em Journal of Financial Economics}, 96(1):1--17.

\bibitem[Schlag and Stoll, 2005]{SCHLAG200569}
Schlag, C. and Stoll, H. (2005).
\newblock Price impacts of options volume.
\newblock {\em Journal of Financial Markets}, 8(1):69--87.

\bibitem[Sharpe, 1994]{Shapiro1994}
Sharpe, W.~F. (1994).
\newblock The sharpe ratio.
\newblock {\em The Journal of Portfolio Management}, 21(1):49--58.

\bibitem[Stephan and Whaley, 1990]{stephan_whaley}
Stephan, J.~A. and Whaley, R.~E. (1990).
\newblock Intraday price change and trading volume relations in the stock and
  stock option markets.
\newblock {\em The Journal of Finance}, 45(1):191--220.

\bibitem[Tibshirani, 1996]{LASSO}
Tibshirani, R. (1996).
\newblock Regression shrinkage and selection via the lasso.
\newblock {\em Journal of the Royal Statistical Society: Series B
  (Methodological)}, 58(1):267--288.

\bibitem[Tumminello et~al., 2011]{tum2011}
Tumminello, M., Miccichè, S., Lillo, F., Piilo, J., and Mantegna, R.~N.
  (2011).
\newblock Statistically validated networks in bipartite complex systems.
\newblock {\em PLOS ONE}, 6(3):1--11.

\bibitem[{World Bank}, 2021]{worldbank}
{World Bank} (2021).
\newblock World bank: Gross domestic product 2019.
\newblock Data retrieved from \url{https://databank.worldbank.org/data}.

\bibitem[Xiaoyan et~al., 2008]{poteshman2005}
Xiaoyan, S., Pan, J., Poteshman, A., Chen, J., Liu, J., Pollet, J., Vayanos,
  D., Wang, J., and White, J. (2008).
\newblock The information in option volume for future stock volatility.
\newblock {\em Journal of Finance}.

\bibitem[Zhang et~al., 2020]{Zhang8}
Zhang, Z., Zohren, S., and Roberts, S. (2020).
\newblock Deep learning for portfolio optimization.
\newblock {\em The Journal of Financial Data Science}, 2(4):8--20.

\end{thebibliography}
